\documentclass[12pt]{article}

\usepackage{latexsym}
\usepackage{amsmath}
\usepackage{amssymb}
\usepackage[usenames]{color}
\usepackage{cite}
\usepackage{amssymb}
\def\parsmallskipn      {  \par\smallskip\noindent  }
\def\al{\alpha}
\def\ga{\gamma} \def\Ga{\Gamma}
\def\ep{\epsilon}

\def\Lam{\Lambda}

 \def\calH{{\cal H}} 
  
 \def\calN{{\cal N}} \def\calO{{\cal O}}

\def\atil{{\tilde{a}}}

\def\ftil{{\tilde{f}}}
\def\gtil{{\tilde{g}}}

\def\ttil{{\tilde{t}}}

\def\betil{\tilde{\beta}}

\def\gatil{\tilde{\gamma}}

\def\Lamtil{\widetilde{\Lambda}}

\def\phitil{\tilde{\phi}}

\def\pitil{\tilde{\pi}}

\def\Ehat{\hat{E}}

\def\behat{\hat{\beta}}
\def\gahat{\hat{\gamma}}

\def\pihat{\hat{\pi}}

\def\del        {  \partial }
\def\half       {  {1\over 2}  }
\def\rootof#1   {  \left( #1 \right)^{1/2}  }

\def\abs#1      {  \vert #1 \vert  }
\def\ie         {{\it i.e.}\,\,}
\def\evalat#1   {  \left\vert_{#1} \right. }

\def\comma          {\, ,}
\def\period         {\, .}
\def\lsim      {\lower .65ex \hbox{\ $\stackrel{<}{\sim}$\ } }
\def\gsim      {\lower .65ex \hbox{\ $\stackrel{>}{\sim}$\ } }

\def\bra#1{{\langle #1 | } }
\def\ket#1{{| #1 \rangle } }

\def\com#1#2{{ \left[#1, #2\right] } } 

\def\matel#1#2#3  {{\langle #1 | #2 | #3 \rangle } }

\def\lrvec#1    {\hbox{$\stackrel{\leftrightarrow}{#1}$}}
\def\lvec#1     {\hbox{$\stackrel{\leftarrow}{#1}$}}

\def\vecii#1#2      {  \left(\begin{array}{c}#1\\#2\end{array}\right)  }
\def\veciii#1#2#3   {  \left(\begin{array}{c}#1\\#2\\#3\end{array}
                     \right)  }
\def\veciv#1#2#3#4  {  \left(\begin{array}{c}#1\\#2\\#3\\#4
                                 \end{array}\right)  }
\def\vecfv#1#2#3#4#5 {  \left(\begin{array}{c}#1\\#2\\#3\\#4\\#5
                                 \end{array}\right)  }

\def\matrixii#1#2#3#4            {  }
\def\matrixiii#1#2#3#4#5#6#7#8#9 {  \left(\begin{array}{ccc}#1&#2&#3\\
                                     #4&#5&#6\\#7&#8&#9\end{array}
                               \right)  }
\def\mativ#1#2#3#4               {  \left(\begin{array}{cccc}
                                       #1\\#2\\#3\\#4\end{array}\right) }
\def\matv#1#2#3#4#5              {  \left(\begin{array}{ccccc}
                                     #1\\#2\\#3\\#4\\#5\end{array}
                              \right)  }
\def\eqabegin         {  \begin{eqnarray}  }
\def\eqaend           {  \end{eqnarray}  }

\def\bracetwo#1#2     {  \left\{ \begin{array}{l} #1 \\ #2 \end{array}
                         \right.  }
\def\bracetwocases#1#2#3#4  {   \left\{ \begin{array}{ll} #1 &
                                 \qquad #2 \\
                                 #3 & \qquad #4 \end{array} \right.  }
\def\bracebegin#1     {  \left\{ \begin{array}{#1}   }
\def\braceend         {  \end{array}\right.   }
\def\boxit#1#2      {  \vbox{\hrule\hbox{ \hskip -4.1pt \vrule\kern3pt 
                     \vbox
                    {  \hsize #1 \strut\kern3pt #2 \kern3pt\strut  }
                       \kern3pt  \vrule} \hrule  } }
\def\centerbox#1#2  {  \mbox{  }\par\bigskip  \hfil \boxit{#1}{#2} \hfil
                       \par\bigskip\noindent }
\def\rightbox#1#2   {  \hfill\boxit{#1}{#2}  }
\def\leftbox#1#2    {  \boxit{#1}{#2}  }
\def\fullbox#1      {  \boxit{\textwidth}{#1}  }

\newcommand{\nullify}[1]{}

\def\mpg#1#2{\begin{minipage}[t]{#1} #2  \end{minipage} }

\def\nxt{\parsmallskipn}

\def\epsfig#1#2#3{
{\lower #3 \hbox{
 \mpg{#1}{\begin{center} \includegraphics[width=#1,clip]{#2.eps} \\
 Fig. #2\end{center} }}}}

\definecolor{darkgreen}{rgb}{0,0.5,0}
\definecolor{darkblue}{cmyk}{0.9,0.9,0,0}
\definecolor{darkred}{rgb}{0.6,0,0.3}
\definecolor{MyRed}{cmyk}{0,1,1,0.15}
\definecolor{MyBlue}{cmyk}{1,1,0,0.25}

\newcommand{\beqa}{\begin{eqnarray}}
\newcommand{\eeqa}{\end{eqnarray}}
\newcommand{\bea}{\begin{array}}
\newcommand{\eea}{\end{array}}
\newcommand{\beqn}{\begin{equation}}
\newcommand{\eeqn}{\end{equation}}

\def\eqref#1{(\ref{#1})}
\definecolor{darkgreen}{rgb}{0,0.5,0}
\definecolor{darkblue}{cmyk}{0.9,0.9,0,0}
\definecolor{darkred}{rgb}{0.6,0,0.3}

\def\WR{W${}_{\rm R}$\,\,}
\def\WL{W${}_{\rm L}$\,\,}
\def\WF{W${}_{\rm F}$\,\,}
\def\WP{W${}_{\rm P}$\,\,}

\def\vacM{\ket{0}_{\rm M}}

\def\braM{{}_{\rm M}\bra{0}}
\def\SS{Schwarzschild }

\def\Htwo{H^{(2)}}
\def\Hone{H^{(1)}}

\def\ahat{{\hat{a}}}

\def\uhat{{\hat{u}}}

\def\that{{\hat{t}}}
\def\x1hat{{\hat{x}^1}}

\def\ahat{{\hat{a}}}

\def\uhat{{\hat{u}}}

\def\ttil{{\tilde{t}}}

\def\xonehat{{\hat{x}^1}}
\def\xonetil{{\tilde{x}^1}}
\def\yonehat{{\hat{y}^1}}
\def\yonetil{{\tilde{y}^1}}

\def\Ehat{{\hat{E}}}
\def\ponehat{{\hat{p}^1}}
\def\qonehat{{\hat{q}^1}}
\def\gatil{\tilde{\gamma}}
\def\betil{\tilde{\beta}}

\def\ahat{{\hat{a}}}
\def\Lamtil{\tilde{\Lambda}}

\def\atil{\tilde{a}}

\def\am{a_-}
\def\ap{a_+}

\def\m{{\mu}}
\def\w{{\omega}}

\def\n{{\nu}}

\def\ep{{\epsilon}}
\def\d{{\delta}}

\def\a{{\alpha}}

\def\SG{{\Sigma}}

\def\s{\sqrt}

\def\b{{\beta}}

\def\pp{\partial}

\def\be{\begin{equation}}
\def\ee{\end{equation}}
\def\ba{\begin{eqnarray}}
\def\ea{\end{eqnarray}}
\def\bal#1\eal{\begin{align}#1\end{align}}

\def\pic(#1,#2){\begin{figure}[h!]
\begin{center}
  \includegraphics[width=5cm]{#1.png}
  \caption{#2}
 \end{center}
 \end{figure}\\}
 
 \def\mpic(#1,#2){\begin{figure}[h!]
\begin{center}
  \includegraphics[width=7.5cm]{#1.png}
  \caption{#2}
 \end{center}
\end{figure}\\}
 
 \def\bpic(#1,#2){\begin{figure}[h!]
\begin{center}
  \includegraphics[width=10cm]{#1.png}
  \caption{#2}
 \end{center}
\end{figure}\\}
\def\bbpic(#1,#2){\begin{figure}[h!]
\begin{center}
  \includegraphics[width=13cm]{#1.png}
  \caption{#2}
 \end{center}
\end{figure}\\}

\def\spic(#1,#2){\begin{figure}[h!]
\begin{center}
  \includegraphics[width=3cm]{#1.png}
  \caption{#2}
 \end{center}
 \end{figure}\\}
\def\dg{\dagger}

\def\r{\rightarrow}

\def\LR{\Leftrightarrow}
\def\f {\frac}

\def\ddd{\cdot\cdot\cdot}
\def\no{\nonumber \\}

\def\ep{\epsilon}
\def\r{\rightarrow}

\def\q{\quad}
\def\qq{\quad\quad}
\def\qqq{\quad\quad\quad}

\def\bep{b_+}
\def\bem{b_-}
\def\calNtil{\widetilde{\calN}}

\def\matrixii#1#2#3#4            {  \left(\begin{array}{cc}#1&#2\\#3&#4
                                       \end{array}\right) }
\makeatletter
\def\section{\@startsection {section}{1}{\z@}{-3.5ex plus -1ex minus 
-.2ex}{2.3ex plus .2ex}{\large\bf}}
\makeatother
\makeatletter
\def\subsection{\@startsection {subsection}{1}{\z@}{-3.5ex plus -1ex minus 
-.2ex}{2.3ex plus .2ex}{\normalsize\bf}}
\makeatother
\usepackage{graphicx}

\def\picture#1#2{\text{\huge Figures}}
\def\papertitlepage{\baselineskip 3.5ex \thispagestyle{empty}}
\def\Title#1{\baselineskip 1cm \vspace{1.5cm}\begin{center}
 {\Large\bf #1} \end{center} 
\vspace{0.5cm}}
\def\Komabanumber#1#2#3{\hfill \begin{minipage}{6cm} {#1}
              \par\noindent {#2} 
              \par\noindent #3 \end{minipage}}
\def\Authors#1{\begin{center} {\it #1} \end{center}}
\def\Abstract{\vspace{1.0cm}\begin{center} {\large\bf Abstract} 
           \end{center} \par\bigskip}
\begin{document}
\papertitlepage
\renewcommand{\thefootnote}{\fnsymbol{footnote}}
\setcounter{page}{1}
\setcounter{footnote}{0}
\setcounter{figure}{0}
\Komabanumber{UT-Komaba 18-2, RUP-18-8}{ RIKEN-QHP-366}{} \\

\vspace{-1.5cm}

\Title{
On the observer dependence of the Hilbert space \\ \vspace{-0.2cm}  
near 
 the horizon of  black holes 
}
\Authors{\baselineskip 3ex
{\sc Kanato Goto$^{a}$\footnote[1]{{\tt kgoto@hep1.c.u-tokyo.ac.jp}} and 
 Yoichi Kazama$^{a,b,c}$\footnote[2]{{\tt yoichi.kazama@gmail.com}}
}
\vskip 3ex
${}^a$     Institute of Physics, University of Tokyo, \\
 Komaba, Meguro-ku, Tokyo 153-8902 Japan\vskip 0.3ex\noindent
${}^b$  Research Center for Mathematical Physics,  
Rikkyo University, \\  Toshima-ku, Tokyo  171-8501  
Japan 
  \vskip 0.3ex\noindent
${}^c$ Quantum Hadron Physics Laboratory, RIKEN  Nishina Center, \\
Wako 351-0198, Japan \vskip 0.3ex\noindent
  }
\vspace{-1cm}
\renewcommand{\thefootnote}{\arabic{footnote}}
\numberwithin{equation}{section}
\numberwithin{figure}{section}
\numberwithin{table}{section}
\parskip=0.9ex
\baselineskip 3.3ex
\Abstract
One of the pronounced  characteristics of gravity,  distinct from other interactions,  is that there are no local observables which are independent of the choice of the spacetime coordinates.  This property acquires crucial importance in 
 the quantum domain in that  the structure of the Hilbert space pertinent to  different observers can be drastically different. Such intriguing phenomena  as Hawking  radiation and the Unruh effect are all rooted in this feature. As in these examples, 
the quantum effect due to such observer-dependence is most conspicuous in the 
 presence of an event horizon and there are still many questions to be clarified  in such a situation. 
In this paper, we perform a comprehensive and explicit study of the 
 observer dependence of the quantum Hilbert space of a massless scalar field in the vicinity of the horizon of Schwarzschild black holes in four dimensions, both in the eternal (two-sided) case and in 
the physical (one-sided) case created by collapsing matter. 
Specifically, we compare and relate the Hilbert spaces of three types of 
 observers, namely (i) the freely falling observer, (ii) the observer who stays at 
 a fixed proper distance outside of the horizon and (iii) the natural observer inside of the horizon analytically continued from outside. The concrete results we obtain have a number of important implications on black hole complementarity 
pertinent to  the quantum equivalence principle and the related firewall phenomenon,  on the number of degrees of freedom seen by each type of observer, and on the ``thermal-type" spectrum of particles realized in a pure   state. 
\newpage
\baselineskip 3.5ex
\thispagestyle{empty}
\renewcommand{\contentsname}{\hrule {\flushleft{Contents}}}
\tableofcontents
\nxt\hrule
\newpage
\section{Introduction}
A quantum black hole  is a fascinating but as yet an abstruse object. 
Recent endeavors to identify it in a suitable class of conformal field theories (CFTs)  
in the AdS/CFT context \cite{Maldacenaa,GKP,W}
\cite{FKW,FKWn,Aspl,Anous}
or by an ingenious model such as the one proposed by Sachdev, Ye, and Kitaev \cite{PhysRevLett.70.3339,Kitaev,Kitaev2} have seen only a glimpse of it, to say the most. Unfortunately, the string theory,  at the  present stage of development,  does not seem to give us a useful clue either. 
This difficulty is naturally expected 
 since an object whose  profile  fluctuates  by quantum self-interaction 
would be hard to capture. We must continue our struggle to find an effective
 means to characterize it more precisely. 

Although the quantization of a  black hole itself is  still a formidable task, some analyses of quantum effects around a (semi-)classical black hole have been  
 performed since a long time ago and they have already uncovered various intriguing phenomena. Among them 
are the celebrated Hawking radiation\cite{Hawking1975}\cite{aHawking76,ISRAEL,
HartleHawking} and the closely related Unruh effect\cite{Unruh}\cite{ Fulling1973,Davies:1974th}. 
These effects revealed the non-trivial features of the quantization in curved 
 spacetimes, in particular in those with event horizons.  At the same time, they 
 brought out new puzzles of deep nature, such as the problem of information loss, the final fate of an evaporating black hole, and so on. 

More recently, further unexpected quantum effect in the black hole environment 
was argued to occur, namely that  a freely falling observer encounters 
 excitations of high-energy quanta, termed ``firewall", as he/she 
crosses the event horizon of a  black hole\cite{Almheiri,Almheiri2013}\cite{Marolf2013b}. This is clearly at odds with 
the equivalence principle, which is one of the foundations of  classical general relativity. An enormous number of papers have appeared since then, both for and against the assertion\footnote{It is practically impossible to list all such papers on this subject. We refer the reader to those citing the the basic papers \cite{Almheiri,Almheiri2013}.} . The various arguments presented have all been rather indirect, however, 
making use of the properties of the entanglement entropy, application of the no-cloning theorem, use of information-theoretic arguments, etc. 

At the bottom of these phenomena lies the strong dependence of the quantization 
on the frame of observers, which is one of the most 
characteristic features of quantum gravity. 
This is particularly crucial  when the 
spacetime of interest contains event horizons as seen by some observers and 
 leads to the notion of black hole complementarity\cite{Susskind:1993if}.

The main aim  of the present work  is to investigate 
 this  observer dependence in some physically 
 important situations  as explicitly as possible to gain some firm and direct understanding of the phenomena rooted in this feature. 
 For this purpose, we shall study   the quantization of a massless scalar field 
 in the vicinity of the horizon of the \SS black hole in four dimensions as  performed 
 by three typical observers.  They are  (i) the freely falling observer crossing the horizon,  (ii) the  stationary observer hovering at a fixed proper distance outside  the horizon (\ie the one under constant acceleration), and (iii) the natural analytically continued observer inside the horizon. 

Such an investigation, we believe, will be  important for at least two reasons. 
One is that we will deal directly with the states of the scalar fields as seen by 
 different observers and will not rely on any indirect arguments alluded to above. 
This makes  the interpretation of the outcome of our study quite transparent 
(up to certain approximations that we must make for computation).  Another role of our  investigation is that the concrete result we obtain should serve as the properties  of quantum fields in the background of a black hole, which should be 
 compared, in the semi-classical regime, to the results  to be obtained by other means of investigation, notably and hopefully by the AdS/CFT duality\footnote{As far as the vicinity of the horizon is concerned, the \SS black hole and the AdS black hole have the same structure.}. For  some progress and intriguing proposals in the related directions, see \cite{HKLL,HKLL2,HKLL3,Heemskerk2012,Papadodimas2013,Papadodimas2014b,Papadodimas2016,Jafferis}. This is important since, as far as we are aware, there has not been a serious attempt to understand how the observer dependence is described in the context of AdS/CFT duality. 

We will perform our study both for the case of two-sided eternal 
Schwarzschild black hole and for that of one-sided physical black hole modeled 
by a simple Vaidya metric produced by collapsing matter or radiation 
at  the speed of light\footnote{Actually, we shall make an infinitesimal regularization to make the trajectory of the matter slightly timelike in order to avoid a certain 
 singularity.}\cite{Vaidya1,Vaidya2,VAIDYA1953}. 
 What makes such an investigation feasible explicitly is 
 the well-known fact that near the horizon of the \SS black hole (roughly within 
 the \SS radius from the horizon; see Sec.~3.1 for more precise estimate) 
there exists a coordinate frame in which  the metric takes the form of the flat 
four-dimensional Minkowski spacetime $M^{1,3}$. Thus, one can make use of the knowledge of the quantization in the  flat space for  observers corresponding to the various Rindler frames.  As this will serve as the platform upon which we develop our picture and computational methods for the black hole cases, 
 we will give, in Sec.~2, a review of this knowledge together with  some further new information about the relations between the quantizations by the three aforementioned observers. 

In making use of this flat space approximation to the near-horizon region of 
 a black hole, an important care must be taken, however. 
 Although the scalar field and its canonical conjugate momentum are locally well-approximated by  those  in the flat space for the region of our interest and hence  the canonical quantization can be performed without any problem,  as we try to 
 extract the physical modes which create and annihilate the quantum states, 
such a local knowledge  is not enough in general.  This is because  the notion of {\it a quantum state}  requires the global information of the wave function.  Technically, this is reflected  in the fact that  the orthogonality relation needed 
for the extraction of the mode  is expressed by an integral over the entire spacelike surface at  equal time,  and depending on the region of interest such a surface may not be totally contained within the region where the flat space approximation is valid.  

One such problem, which, however can be easily dealt with, stems from the simple fact that the approximation by the four-dimensional flat space includes that of the 
 spherical surface of the horizon by a tangential plane around a point. Clearly, since the physical modes of the scalar field should better be classified by the angular momentum, not by the linear momentum,  we shall  use  $\mathbb{R}^{1,1} \times \mathbb{S}^2$,  instead of $M^{1,3}$,  as the more accurately  approximated  spacetime, where $\mathbb{R}^{1,1}$ stands for a portion of two-dimensional  flat spacetime realized near the horizon and $\mathbb{S}^2$ is the sphere at the \SS radius.  Various formulas reviewed and/or developed  in Sec.~2 for $M^{1,3}$ can be readily transplanted to this case by replacing the plane waves by the spherical harmonics. 

 The problem pointed out above of the extraction of the modes  within 
 the flat region is much more non-trivial in the near-horizon region of $\mathbb{R}^{1,1}$, since the flat region which extends to infinity is only along the direction of the lightcone. The problem about this situation  is that the use of the trajectory along the light cone leads to the quantization of a {\it chiral} boson, which is  known to be notoriously complicated.  In addition such a trajectory is not connected by a  Lorentz transformation to the trajectory of a general  observer, which is timelike.  This problem is particularly severe  when we deal with the one-sided black hole produced by a massless shock wave,  the effect of which will be treated by the imposition of an effective Dirichlet boundary condition on the scalar field along the trajectory of the shock wave. To solve this problem, we have made a careful regularization of taking the trajectory of  the shock wave to be {\it slightly timelike}\footnote{Evidently this corresponds to the case of a slightly massive falling matter, which is physically reasonable.}. Then 
we are able to treat the quantization for the observers freely falling with arbitrary velocity by making a  suitable Lorentz transformation. Such a proper analysis has not been performed  in the literature and this allowed us to obtain firm  results for the question of major  interest.

Although we cannot summarize here  all the results on how the different observers see their quanta and how they are related, let us list two that are of obvious interest: 
\begin{itemize}
	\item Under the assumption that the metric of the interior of a physical Schwarzschild black hole, in particular the one large enough so that the curvature at the 
 horizon is very small, can be described by a Vaidya type solution, our results  indicate that the equivalence principle still holds quantum mechanically near the horizon of the black hole, and the freely falling observer finds no surprise as he/she goes through the horizon. 
	\item For a physical (one-sided) black hole, the vacuum\footnote{The vacuum referred to here will be explained in Sec.~4.2.3.} $\ket{\hat{0}}_{-}$ for the freely 
falling observer is a pure state which is  not  the same as the usual Minkowski vacuum $\ket{0}_M$.  
Nevertheless the expectation value of the number operator for the observer in the 
frame of the right Rindler wedge 
 in $\ket{\hat{0}}_{-}$  has an Unruh-like  distribution, which contains 
 a ``thermal"  factor together with another portion depending on the assumed  interaction  between the scalar field and the collapsing matter, effectively expressed as a boundary condition.   This is in contrast to the case of the two-sided eternal black hole, where  tracing out of the modes of the left Rindler wedge must be performed and  the resultant mixed state density matrix produces the usual purely thermal form of the Unruh distribution. 
The effect  for the physical black hole occuring in the pure state described above 
 is essentially of the same origin as the Hawking radiation seen by the asymptotic 
 observer, who is a Rindler observer\footnote{For related work, though in a different setting, see \cite{Louko1998}.}. 
\end{itemize} 

The plan of the rest of the paper is as follows: In Sec.~2, we begin by describing the quantization of a massless scalar field in four-dimensional flat Minkowski space from the point of view of various observers, and provide explicit relations between them. Although this section is mostly a review, we also derive some useful relations that have not been discussed in the literature. This includes the construction of the explicit unitary transformation between the Minkowski mode operators and those of the future Rindler wedge and  how the Poincar\'e algebra is realized in various wedges. Next, in Sec.~3, this knowledge about  the quantization in flat spacetime will be utilized to discuss how the scalar 
 field is quantized by various observers in the vicinity of the event horizon of a two-sided \SS black hole, which by a suitable choice of coordinates can be approximated 
 by a part of $\mathbb{R}^{1,1}$ times $\mathbb{S}^2$.  In Sec.~4, we study the similar problem in the case of a 
Vaidya model of the physical one-sided black hole that is produced by a collapse of 
  matter with infinitesimal mass, introduced as a regularization. The effect of this collapse is treated as an effective  boundary condition  on the scalar field along a slightly timelike trajectory of such a shock wave. 
Even though  we focus on the flat region near the horizon, the quantum states, 
 which depend on the global situation, show different properties as compared with 
 the two-sided case studied in Sec.~3. In Sec.~5, we disucss the implications of 
 the results obtained in the previous sections on some  important questions, such 
 as the quantum equivalence principle,  the firewall phenomenon, and the Unruh effect near the horizon. Finally, in Sec.~6, after summarizing the results, we re-emphasize that the effect of 
 the observer dependence of quantization is one of the most crucial characteristics 
  of any theory of quantum gravity and it should be seriously investigated, in particular,  in the framework of the AdS/CFT approach.  Several appendices are provided to 
give  further useful details of the formulas  and calculations discussed in the main text. 
\section{Quantization of a scalar field in the Rindler wedges and the degenerate Kasner universes\label{sec:Rindler}}
\begin{figure}[h]
\hspace{4.5cm}
  \includegraphics[width=10cm]{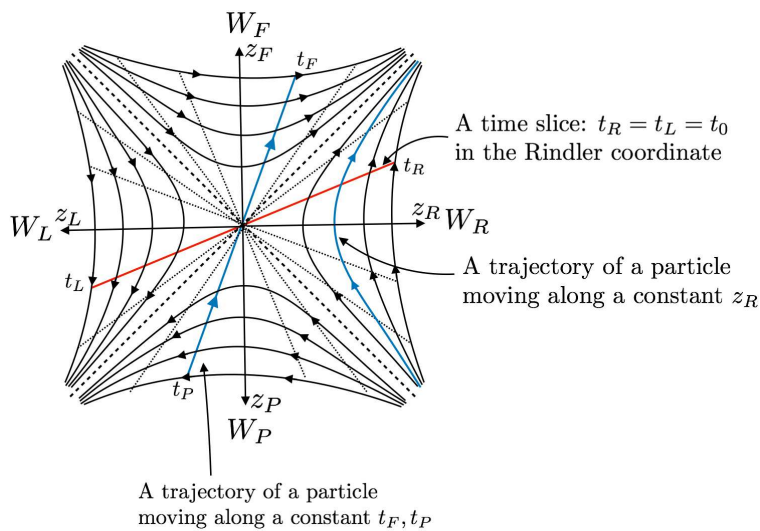}
\caption{Trajectories and equal-time slices of the Rindler observers in various wedges.  The boundaries of the wedges \WR\!\!, \WF\!\!, \WL and \WP are shown by dotted lines. The arrowed blue lines represent the trajectories of a particle, while the red line is a typical  time slice at $t_R=t_L$ for  \WR and \WL\!\!. \label{Rindlerwedges}}
\end{figure}

We begin by describing the quantization of a massless\footnote{Massive case can be treated in an entirely similar manner.} scalar field in the four dimensional  Minkowski space,  from the standpoint of  a uniformly  accelerated  Rindler 
 observers for the  right  and the left wedges \WR and \WL\!\!,  and their appropriate  analytic continuations  for the future and the past wedges \WF and \WP\!\!, which can be identified as degenerate  Kasner universes. In Figure\  \ref{Rindlerwedges}, we draw the trajectories of  the corresponding observers and the equal time slices in each wedge. 

The subject of the quantization by Rindler observers  has a long history\cite{Fulling1973,Boulware1975,Narozhny2002,Blasone2015} and hence the content of this  section is largely a review\footnote{For a review article closely related to this section, see \cite{Crispino:2007eb}.}.  However, a part of our exposition supplements the description 
 in the existing literature by providing some clarifying details and new relations.  
The results of this section will serve as the foundation upon which to discuss 
 the observer-dependent quantization around the horizon of \SS black holes, 
 both eternal (two-sided) and physical (one-sided), as will be performed  in 
 Sec.~3. 
\subsection{Relation between the Minkowski  and the Rindler coordinates}
Before getting to the quantization of a scalar field, we need to describe the relationship between the Minkowski coordinate  and the Rindler coordinates in various wedges.

 The $d$-dimensional Minkowski metric is described in the usual Cartesian coordinate as 
\ba
ds^2=-(dt_M)^2+(dx^1)^2+\sum_{i=2}^{d-1} (dx^{i})^2\period 
\ea
Since we will be mostly concerned with the first two coordinates and the roles of 
 the rest of the $d-2$ coordinates are essentially  the same, hereafter we will deal with the four dimensional case, \ie $d=4$. 

As for the Rindler coordinates, we begin with the one in the right wedge \WR shown  in Figure\  \ref{Rindlerwedges}. As is well-known, it is related to the coordinates 
 of the observer who is acclerated in the positive $x^1$ direction with a uniform 
 acceleration. The trajectory of the observer in the $(t_M,x^1)$ Minkowski 
 plane with a value 
 of acceleration $\kappa(>0)$  is given by 
\begin{align}
(x^1)^2 -(t_M)^2 = (1/\kappa)^2 = z_R^2 \period
\end{align}
Here the symbol $z_R$ is introduced  as a variable, meaning that  different values of  $z_R$ describes different trajectories. 
Thus  the Rindler 
 coordinate system is spanned by the proper time $\tau_R$
of the observer and the spatial coordinate $z_R$. 
The  relation to the Minkowski coordinate is given by 
\ba
t_M=z_R\sinh  t_R\comma \qquad  x^1=z_R\cosh t_R\comma 
\qquad (z_R >0) \comma \label{reltoMink}
\ea
where we introduced for convenience the rescaled time  $t_R$ defined  by 
\begin{align}
t_R &\equiv \kappa \tau_R \period
\end{align}
The metric in terms of these variables is 
\ba
ds^2=-z_{R}^2dt_{R}^2+dz_{R}^2+\sum_{i=2}^3(dx^{i})^2\period
\label{coordr}
\ea
Note that $z_R=0$ corresponds to the (Rindler) horizon, which consists of two dimensional planes along the  lightlike lines bounding the region \WR .  
It will often be  convenient to use  the following lightcone variables:
\begin{align}
x^\pm &\equiv x^1\pm t_M = z_R e^{\pm t_R} \period 
\end{align}
This shows that $t_R$ is nothing but the {\it rapidity-like variable} and gets simply translated by the Lorentz boost in the $x^1$ direction. 

The coordinates $(t_L, z_L)$ in the left wedge \WL can be obtained in an entirely similar manner and are related to the Minkowski coordinates by 
\ba
t_M=-z_L\sinh t_L\comma \qquad  x^1=-z_L\cosh  t_L \comma 
\qquad (z_L >0)\period
\ea
The  metric takes exactly the same form as  (\ref{coordr}),  with the subscript $R$ replaced by $L$.  Note that as $t_R$ increases from $-\infty$ to $\infty$, the 
 Minkowski time $t_M$ also increases, while when $t_L$ increases from $-\infty$ to $\infty$, $t_M$ decreases, as indicated by the arrows in Figure\   \ref{Rindlerwedges}. 

Next consider  the future and the past wedges, \WF and \WP\!\!. They describe the interior of the Rindler horizon. The relation to the Minkowski coordinate for \WF  is 
\ba
t_M=z_F\cosh t_F\comma \qquad  x^1=z_F\sinh  t_F\comma \qquad (z_F >0)
\comma  \label{relMWF}
\ea
and the metric takes the form 
\ba
ds^2=-dz_{F}^2+z_{F}^2dt_{F}^2+\sum_{i=2}^3(dx^{i})^2\period    \label{coordf}\ea
This means that in \WF\!\!, {\it $z_F$ is the timelike and $t_F$ is the spacelike coordinates.} As in the case of \WR\!\!, the following lightcone combinations are often useful:
\begin{align}
x^\pm\equiv x^1\pm t_M = \pm z_F e^{\pm t_F} \period
\end{align}
Just like $t_R$, under a Lorentz transformation the variable  $t_F$ undergoes a 
 simple shift. 

This interchange of the timelike and the spacelike natures  also occurs in the past 
 wedge \WP. In the entirely similar manner, we have 
\ba
t_M=-z_P\cosh t_P\comma \qquad  x^1=-z_P\sinh  t_P \comma \qquad (z_P >0)
\comma 
 \ea
with the form of the metric identical to (\ref{coordf}) with the subscript $F \r P$. 

In Sec.~3, where we discuss how  the similar  Rindler 
 wedges for a flat space appear in the vicinity of the horizon of a \SS black hole, 
we will see that the $z_R$ variable expresses the proper distance from the horizon 
 in the outside region and is related to the radial variable $r$ and the \SS radius $2M$ (where $M$ is the mass of the black hole), by $z_R \simeq  \sqrt{8M(r-2M)}$. 
Hence, as we go through the horizon from \WR into \WF\!\!, we must make an 
  analytic continuation by choosing a branch for the square-root cut. 
Similarly, an analytic continutation connects \WF and \WL\!\!,  and so on. Such a continuation process must be such that as we go once around all the wedges, we should come back  to the same branch for \WR\!\!.  A simple analysis for this consistency 
 yields the following continutation rules, with a sign $\eta=\pm 1$ which can be 
 chosen by convention, for the adjacent wedges:
\begin{align}
t_F &= t_R -i{\pi \over 2} \eta \comma \qquad \qquad z_F = e^{i(\pi/2)\eta} z_R \comma \\
t_L &= t_F  -i{\pi \over 2} \eta \comma \qquad \qquad
 z_L = e^{-i(\pi/2)\eta} z_F = z_R \comma \\
t_P &= t_L + i{\pi \over 2} \eta = t_F \comma \qquad \!
z_P = e^{-i(\pi/2)\eta} z_L \comma \\
t_R &= t_P + i{\pi \over 2} \eta \comma \qquad \qquad z_R = e^{i(\pi/2)\eta} z_P 
\comma \\
& z_R, z_F, z_L, z_P \ge 0 \period 
\end{align}
One can easily check that these relations are compatible with the relations 
 between the Minkowski variables and the Rindler wedge variables 
 given above. 
\subsection{Quantization  in the Minkowski spacetime}
We now discuss the quantization of a massless scalar field $\phi$ in various coordinates. 

In this subsection, just for setting the notation, we summarize the simple case for the Minkowski  coordinate. The action, the canonical momentum  and the equation of motion are given by 
\begin{align}
&S =-\f{1}{2}\int dt_Mdx^1  d^2 x \left(-(\pp_{t_M}\phi^M)^2+(\del_{x^1} \phi^M)^2 + \sum_{i=2}^3 (\del_{x^i} \phi^M)^2\right)\comma  \\
&\pi^M  \equiv\f{\pp{\cal L}}{\pp (\pp_{t_M}\phi^M)}=\pp_{t_M}\phi^M \comma \\
&\left(-\pp_{t_M}^2 + \del_{x^1}^2 +\sum_{i=2}^3 \pp_{x^i}^2\right)\phi^M=0 \comma 
\end{align}
where we denote the fields and the time in the Minkowski frame with the super(sub)script $M$. 
 Now $\phi^M$ can be 
 expanded into  Fourier modes as 
\bal
\phi^M(t_M, x^1_M,x)&=\int \f{dp^1}{\s{2\pi}\s{2E_{kp^1}}} \int \f{d^2k}{2\pi}  e^{ikx+ip^{1}x^1-iE_{kp^1}t_M} a^M_{kp^1} + \mbox{h.c.}
 \comma \label{expphiM} \\
E_{kp^1}  &\equiv \sqrt{(k)^2 + (p^1)^2} \period 
\eal
Here and throughout,  we often denote $(x^2, x^3)$ simply by $x$  and similarly for  the momenta for the corresponding dimensions by $k$,   and write 
 the inner product  $\sum_{i=2}^3 k_i x_i$  as  $kx$.  
Canonical quantization is performed by  demanding that\footnote{$\delta(x-y)$ of course means the two-dimensional delta-function $\delta^2(x-y)$. This abbreviation 
 will be used throughout.}
\ba
[ \pi^M(t_M, x^1, x), \phi^M(t_M ,y^1,y)]=-i\d(x^1-y^1)\d(x-y)
\period
\ea 
Using the orthogonality of the exponential function, we can easily extract out 
 the mode operators and check that they satisfy the  usual commutation relations:
\ba
[a^M_{k p^1},a^{M\dg}_{k' {p'}^1}]=\d(p^1-{p'}^1)\d(k-k')
\comma \qquad \mbox{rest}=0 \period
\ea 
\subsection{Quantization  outside the Rindler horizon}
Let us now begin the discussion of quantization in the Rindler coordinates in various wedges. 

We first consider  the Rindler wedges outside the horizon, namely  \WR  and \WL\!\!. Since the metrics  in these wedges take  the same form in the respective 
 variables, we will focus on \WR\!\!. 
 The action takes the form 
\bal
S&=-\f{1}{2}\int dt_Rdz_Rd^2x\s{-g} g^{\m\n}\pp_\m\phi^R\pp_\n\phi^R
\no
&=-\f{1}{2}\int dt_Rdz_Rd^2x\left(-\f{1}{z_R}(\pp_{t_R}\phi^R)^2+z_R(\pp_{z_R}\phi^R)^2+z_R \sum_{i=2}^3 (\del_{x^i}\phi^R)^2\right) \period
\eal
The canonical momentum is given by 
\ba
\pi^R \equiv\f{\pp{\cal L}}{\pp (\pp_{t_M}\phi^R)}=\f{1}{z_R}\pp_{t_R}\phi^R \comma 
\ea
which has  an extra factor of $1/z_R$ compared with the Minkowski case. 
Variation of the action yields the equation of motion 
\ba
\left(\pp_{z_R}^2+{1\over z_R}\pp_{z_R}+\sum_{i=2}^3 \pp_{x^i}^2-\frac{1}{z_R^2}\pp_{t_R}^2\right)\phi^R=0.
\ea
As it is a second order differential equation, there are  two independent solutions, 
 which can be taken to be the exponential function times the modified Bessel functions, namely   $e^{i(kx-\w t)}I_{i\w}(|k|z)$ and $e^{i(kx-\w t)}K_{i\w}(|k|z)$. The appropriate 
 solution is the one which damps at $z\r \infty$ and we write it as\footnote{Let us make a remark on the boundary condition
 at $z=0$ for \WR (and similarly for \WL).    For the region \WR,  the point $z_R=0$ corresponds 
 to the perpendicularly  bent line consisting of the light-like segments $t=-x$ and $t=x$ (in the Cartesian   coordinate)  for $x \ge 0$. This ``point" should be defined  as the limit $z_R \rightarrow 0$. From the completeness relation for the 
 solutions  $K_{i\omega}(y)$ (where we have set $y=|k|z_R$) of the equation of motion described in Appendix A.2, one easily sees that a function $f(y)$ can be 
 expanded in terms of  $K_{i\omega}(y)$ in the form 
$f(y) =\int_0^\infty d\omega \mu(\omega) C(\omega) K_{i\omega}(y)$, 
where the coefficient is given by 
$C(\omega) = \int_0^\infty {du\over u} K_{i\omega}(u) f(u) $. 
This  integral is convergent  near $u\simeq 0$  if and only if $f(u)\rightarrow 0$ 
 as $u \rightarrow 0$. Therefore, in order to be expandable into creation and annihilation parts in terms of $K_{i\omega}(|k|z_R)$ functions, the scalar field $\phi^R(t_R, z_R, x)$ should vanish as one  approaches the $z_R=0$ boundary. 
This is however automatically built in due to the main dependence  of $K_{i\omega}(|k|z_R)$ on $z_R$ near $z_R=0$,  which is the divergent phase $e^{\pm i \omega \ln _R}$. Then by the use of the Riemann-Lebegue lemma, the integral defining 
$\phi^R(t_R, z_R, x)$ vanishes as $z_R \rightarrow 0$. For more discussions 
see \cite{Narozhny:2001} and \cite{Richtmyer}. }
\ba
f_{k\w }^R(t_R,z_R,x)&=N_{\w }^RK_{i\w}(|k|z_R)e^{i(kx-\w t_R)} \comma 
\ea
where $N_{\omega }$ is a normalization constant  given below. 
Thus, the scalar field in the right Rindler wedge can be expanded 
as 
\bal
\phi^R(t_R,z_R,x )&=\int_0^{\infty}d\w\int d^2k N_{\w }^R\left[K_{i\w}(|k|z_R)e^{i(kx-\w t_R)} a^R_{k\w }+{\rm h.c.}\right] \comma \no
N_{\w }^R&=\f{\s{\sinh\pi\w}}{2\pi^2} \period  \label{NRomega}
\eal

Let us make some remarks on this formula: \nxt
(i)\ For the hermitian conjugate part, only the conjugation for the exponential part
 is needed since $K_{i\omega}(|k|z)$ is real.  \\
(ii)\  The normalization constant chosen here will lead to the canonical  form of the commutation relations, as explained in Appendix \ref{app:WR}.  \\
(iii)\ The variable $\w$ here is the{\it  energy conjugate to the time-like variable $t_R$}, and hence {\it its range is  $\w\ge 0$}. 

Canonical quantization is performed by imposing the following equal-time commutation relation:
\ba
[\pi^R(t_R,z_R,x), \phi^R(t_R,z'_R,x'),]=-i\d(z_R-z'_R)\d(x-x') \period
\ea 
Using the orthogonality relation for the modified Bessel functions explained in the Appendix \ref{app:orthogonality}, it is straightforward to obtain the commutation relations for the mode operators 
\ba
[a^R_{\w k},a^{R\dg}_{\w' k'}]=\d(\w-\w')\d(k-k') \comma 
\qquad \mbox{rest}=0 \period
\ea 
For some details of the calculations, see Appendix \ref{app:WR}. 

The quantization in \WL is essentially similar to the one in \WR above, 
 except for one point that one must be careful about.  Recall that as the Minkowski time $t_M$ (and also $t_R$) goes from $-\infty $
 to $\infty$, the time $t_L$ in \WL runs oppositely from $\infty$ to $-\infty$. 
This is due to the definition of $t_L$ by a smooth  analytic continuation 
 and does not of course mean that a physical particle moves  from the future to the past. After all 
\WL is a part of the Minkowski space and all the particles and waves must evolve 
along the positive direction in Minkowski time. This  applies to the \WL observer 
as well, who is under constant acceleration  in the negative $x^1$ direction. The time which increases along the trajectory of the \WL observer is {\it not} $t_L$ but $\ttil_L \equiv -t_L$.  Therefore, the quantization in this frame should be 
done with $\ttil_L$ regarded as time.  Then, all the formulas for the quantization 
 in the \WR frame hold for the \WL frame, with $t_R$ replaced by $\ttil_L$. 
This means that {\it if one wishes to use  the ``time" $t_L$} to write   the 
mode expansion of the  field $\phi^L$ and define its conjugate momentum $\pi^L$, we have 
\bal
\phi^L(t_L,z_L,x )&=\int_0^{\infty}d\w\int d^2k N_{\w }^L\left[K_{i\w}(|k|z_L)e^{ikx+i\w t_L} a^L_{k\w }+{\rm h.c.}\right] \comma \no
N_{\w }^L&=\f{\s{\sinh\pi\w}}{2\pi^2} \comma    \label{NRomega}
\eal
and 
\begin{align}
\pi^L(t_L, z_L, x) &\equiv -{1\over z_L} \del_{t_L}\phi^L \period
\end{align}
One can then check that the equal time commutation relation $\com{\pi^L(t_L, z_L, x)}{\phi^L(t_L, z'_L, x')} = -i \delta(z_L-z'_L)\delta(x-x')$ holds correctly. 
\subsection{Quantization  inside the Rindler horizon}
Next  consider the quantization in the Rindler wedges inside the horizon,
 \ie in \WF and \WP\!\!.  Again they can be treated in parallel and we focus on \WF\!\!.    Compared to the previous analysis for the outside region,  an important difference  arises due to the interchange of the timelike and the 
spacelike coordinates. 

The action in \WF region  is given by 
\bal
S&=-\f{1}{2}\int dt_Fdz_Fd^2x\s{-g} g^{\m\n}\pp_\m\phi^F\pp_\n\phi^F
\no
&=-\f{1}{2}\int dt_Fdz_Fd^2x\left(-z_F(\pp_{z_F}\phi^F)^2+\f{1}{z_F}(\pp_{t_F}\phi^F)^2+z_F\sum_{i=2}^3  (\del_{x^i}\phi^F)^2\right) \period  \label{actioninside}
\eal
From the sign of various  terms, it is clear that  {\it $t_F$ is the space coordinate and $z_F$ is the time coordinate}. Therefore,  the canonical momentum must be defined by 
\ba
\pi^F \equiv\f{\pp{\cal L}}{\pp (\pp_{z_F}\phi^F)}=z_F\pp_{z_F}\phi^F.
\ea
The equation of motion takes the form 
\ba
\left(\pp_{z_F}^2+\f{1}{z_F}\pp_{z_F}-\sum_i \pp_i^2-\frac{1}{z_F^2}\pp_{t_F}^2\right)\phi^F=0\period
\ea
Again there are two independent solutions, which can be taken as 
\bal
f_{k\w }(t_F,z_F,x)&=N^F_{\w }H^{(2)}_{i\w}(|k|z_F)e^{i(kx-\w t_F)}\comma \no
f^*_{k\w }(t_F,z_F,x)&=N^F_{\w }e^{-\pi\w}H^{(1)}_{i\w}(|k|z_F)e^{-i(kx-\w t_F)}.\eal
where $H^{(1)}_{i\omega}$ and $H^{(2)}_{i\omega}$ are the Hankel functions  of imaginary order and $N^F_{\w }$ is the normalization constant, to be specified
 shortly. 

To expand the scalar field in terms of these functions, 
 a care should be taken as to which function should be associated with the 
annihilation (resp. creation) modes. 
This is because, in contrast to the previous case, 
$\omega$ is conjugate to the spacelike variable $t_F$ and hence it is not the energy 
 but the usual momentum. Therefore the range of $\omega$ is $-\infty \le \omega \le \infty$ and we cannot determine the positive (resp. negative) frequency mode 
from the exponential part of the functions above. 

 To guess  which Hankel function should be taken as describing the 
 positive frequency part, it is physically natural to first look at the asymptotic behavior of   $H^{(1,2)}_\omega(|k|z_F)$ at late time, {\it i.e.} at very large positive 
$z_F$.  Such behaviors are given by 
\begin{align}
H^{(1)}_{i\w}(|k|z_F)&\sim e^{i|k|z_F-i\pi/4}e^{\pi\w/2}\s{\f{2}{\pi |k|z_F}}
\ \comma \label{Honelargez} \\
H^{(2)}_{i\w}(|k|z_F)&\sim e^{-i|k|z_F+i\pi/4}e^{-\pi\w/2}\s{\f{2}{\pi |k|z_F}} 
\ \period \label{Htwolargez} 
\end{align}
We see that  $H^{(2)}_{i\w}(|k|z_F)$ behaves like $e^{-i|k|z_F}$ which corresponds to the positive frequency with respect to ``energy'' $|k|$ (with an overall 
inessential damping behavior).  This tells us that the 
the correct expansion is 
\bal
\phi^F(t_F,z_F,x )&=\int_{-\infty}^{\infty}d\w\int d^2k N^F_{\w }\left[e^{i(kx-\w t_F)}H^{(2)}_{i\w}(|k|z_F)a^F_{k\w }+{\rm h.c.}\right]\comma \label{expphiF} \no
(N^F_{\w })^2&=\f{e^{\pi\w}}{8(2\pi)^{2}} \comma 
\eal
where the  factor $N^F_\omega$ is determined such that the commutator of the 
 modes take the canonical form as in (\ref{comaF}) below. 
In the literature  the modes $a^F_{k\w }$ are often called  the Unruh modes,  whereas the modes $a^R_{k\w }$ are referred to as the  Rindler modes.

To check that  such an association is actually the correct one,  one must compute the   ``equal-time" (\ie equal $z_F$) commutation relation. This indeed gives
the right relation 
\ba
[\pi^F(z_F,t_F,x), \phi^F(z_F,t'_F,x'), ]=-i\d(t_F-t_F')\d(x-x') \period
\ea 
Using the orthogonality of the Hankel functions, we get the canonical form of the commutation relations for creation/ annihilation operators, namely, 
\ba
[a^F_{k\w },a^{F\dg}_{k'\w' }]=\d(\w-\w')\d(k-k') \comma \qquad 
\mbox{rest}=0 \period \label{comaF}
\ea 
See Appendix \ref{app:WF} for some details of this computation. 
\subsection{Hamiltonian in the future wedge}
We have seen that in \WF and \WP the timelike and the spacelike variables 
 are swapped  compared to the usual situations  in \WR and \WL  and this has 
 made the identification of the positive and negative frequency modes somewhat 
non-trivial. In fact, this swapping makes the Hamiltonian in \WF and \WP 
 {\it time dependent}. In this subsection, we briefly discuss the form of the Hamiltonian  and its action as the proper time-development operator. 

From the action (\ref{actioninside}) for the \WF  region, the Hamiltonian is
 readily obtained as
\begin{align}
H_F &
=\half \int dt_F \left(  {1\over z_F}{(\pi^F)^2} + {1\over z_F}(\del_{t_F} \phi^F)^2 +  z_F(\sum_{i=2}^3 \del_{x^i} \phi^F)^2 \right) \comma \qquad \pi^F = z_F \del_{z_F} \phi^F \period
\end{align}
Since $z_F$ is the time variable, the Hamiltonian $H_F$ is clearly time-dependent. 
Therefore the time development of a state $\ket{\psi(z_F)}$ is accomplished  by the 
 unitary operator $U(z_F)$ in the manner 
\begin{align}
\ket{\psi(z_F)} &= U(z_F) \ket{\psi(0)} \comma \\
U(z_F) &= T \exp \left( -i \int_0^{z_F} H_F (z') dz' \right) \comma 
\end{align}
where $T \exp( **)$ denotes the time-ordered exponential.  Thus for general $z_F$ the time-development is quite non-trivial. 

We now wish to express $H_F$ in terms of modes given in (\ref{expphiF}) and
 see how it simplifies for large $z_F$. 
The necessary computation is straightforward: Substitute the expansion (\ref{expphiF}) and perform the space integral over $t_F$.  Since the intermediate expressions are lengthy, we omit them and display the final form. It is given by 
\begin{align}
H_F &= {\pi \over 8}\int_{-\infty}^\infty d\w d^2k \biggl[ \left( {\omega^2 \over z_F} + z_Fk^2\right) \Htwo_{i\w}(|k|z_F)  \Htwo_{-i\w}(|k|z_F)  \no
&+z_F\del_{z_F} \Htwo_{i\w}(|k|z_F)  \del_{z_F} \Htwo_{-i\w}(|k|z_F) \biggr]
a^F_{k \omega }a^F_{-k, -\omega} + {\rm h. c.}  \no
&+ {\pi \over 4}\int_{-\infty}^\infty d\w  d^2k \biggl[ \left( {\omega^2 \over {z_F}} + z_Fk^2\right) \Htwo_{i\w}(|k|z_F)  \Hone_{i\w}(|k|z_F)  \no
&+z_F\del_{z_F} \Htwo_{i\w}(|k|z_F)  \del_{z_F} \Hone_{i\w}(|k|z_F) \biggr]
a^{F\dagger}_{k\omega }a^F_{k\omega} \comma 
\end{align} 
where we have discarded, as usual, an infinite constant coming from the 
normal ordering of the last term. 

Now let us consider the limit of large time, $z_F\rightarrow \infty$. In this limit, since $t_M =\sqrt{z_F^2+(x^1)^2}$,  the line of equal time will approach that of equal Minkowski time $t_M$ and hence we expect that $H_F$ will take the form for the free scalar field. Using the formulas (\ref{Honelargez}) and (\ref{Htwolargez}) for large $z$, we can drastically simplify the expressions for $H_F$. The leading term which does not vanish as $z_F\rightarrow \infty$ takes the form
\begin{align}
H_F\big|_{z \rightarrow \infty} &= \int_{-\infty}^\infty d\omega\, d^2 k\, |k|a^{F\dagger}_{k\omega }a^F_{k\omega} \period
\end{align}
This is independent of $z_F$ and indeed coincides with the form for the free scalar field in Minkowski space. 
\subsection{Relation between the quantizations  in \WR\!\!, \WL\!\!, \WF  and the Minkowski frames }
We are ready to  discuss the relation between the quantizations  in \WR\!\!, \WF and  
the Minkowski frames. 
\subsubsection{Minkowski and \WR  frames}
First, since \WR  is contained in the Minkowski space, it should be possible to 
express the modes in the \WR  frame in terms of the modes in the Minkowski 
frame. Using the Klein-Gordon inner product  for \WR  defined in  Appendix \ref{app:KG}, we obtain the expression for the annihilation operator $a^R_{k\omega}$ in the \WR  frame as 
\bal
a^{R}_{k\w }&=(f^R_{k\w },\phi^M)^R_{\rm KG}\no
 &=i\int_{0}^{\infty}\f{dz_R}{z_R}\int dx^{2}(f^{R*}_{k\w }\overleftrightarrow{\pp_{t_R}}\phi^M)
 \no&=\int_{-\infty}^\infty  \f{dp^1}{\s{4\pi E_{kp^1}}} \f{1}{\s{\sinh\pi\w}}\left(\f{E_{kp^1}-p^1}{E_{kp^1}+p^1}\right)^{\f{-i\w}{2}}\left[e^{\pi\w/2}a^M_{kp^1}+e^{-\pi\w/2}a^{M\dg}_{-kp^1}\right] \comma 
\label{WRM}
\eal
where $\omega \ge 0$. 
Some details of the calculations are given in Appendix \ref{app:relaWR}. 

Actually, this expression for $a^R_{k\w}$ can be simplified rather drastically by introducing 
 the rapidity variable $u$ defined by 
\begin{align}
u \equiv \half \ln \left(\f{E_{k p^1}+p^1}{E_{k p^1}-p^1}\right) \period
\label{rapidityu}
\end{align}
Then we can immediately solve this relation for  $E_{k p^1}$ and $p^1$ in terms of $u$ and obtain 
\begin{align}
E_{k p^1} &= |k|\cosh u\comma \qquad p^1 = |k|\sinh u \period
\end{align}
Furthermore, the integration measures are related as 
\begin{align}
dp^1 = |k|\cosh u du = E_{k p^1} du \comma 
\end{align}
with the identical range of integration $[-\infty, \infty]$ for both $p^1$ and $u$. 
Further, if we define the annihilation operator in the rapidity variable as 
\begin{align}
a^M_{ku} &\equiv  \sqrt{|k|\cosh u}\,  a^M_{k p^1} =\sqrt{E_{kp^1}}\, a^M_{k p^1}
\comma  \label{defaMku}
\end{align}
the commutation relation with its conjugate is 
\begin{align}
[a^M_{ku}, a^{M\dagger}_{k'u'}] &= |k|\sqrt{\cosh u\cosh u'} \delta(p^1-{p'}^1) \delta (k-k') =\delta(u-u') \delta (k-k') \comma 
\end{align}
where we used $\delta(p^1 -{p'}^1) = \delta( |k|\sinh u  - |k|\sinh u') = \delta(u-u')/ (|k|\cosh u)$. 

Using these definitions, the relation (\ref{WRM}) can be written as   
\begin{align}
a^R_{k\w} = \int_{-\infty}^\infty {du \over \sqrt{4\pi \sinh \pi \w} } e^{i\omega u} 
\left[e^{\pi\w/2}a^M_{ku}+e^{-\pi\w/2}a^{M\dg}_{-ku}\right]  \period \label{aRaM}
\end{align}
Note that, as is well known,  the annihilation operator $a^R_{k\omega}$ 
is composed  both of the annihilation and the creation operators of the Minkowski 
frame.  Another important fact is that there is no negative frequency 
modes,   $a^R_{k\w } ({\rm for\ } \w<0)$,  in the \WR  frame since $\omega$ 
 is the energy conjugate to $t_R$. 
 Consequently it is {\it not possible to invert the relation above }
 to express the Minkowski annihilation/creation operators in terms of the ones in the \WR  frame only.  This means that {\it  the number of degrees of freedom that \WR  observer sees  is half as many as seen by the Minkowski observer. } 
Therefore, even when the \WR and the Minkowski observers\footnote{To avoid any confusion,  let us stress that what we mean by a ``Minkowski observer in $W_R$'' is an observer who is traveling along a constant $x_M$ line (=along the flow of the Minkowski time $t_M$) and happens to be in the $W_R$ region at some time $t_M$.} are within the same 
 \WR region, \WR observer cannot recognize half of the excitation modes that 
 the Minkowski observer sees. 
\subsubsection{Minkowski and \WF frames related by a Fourier transform} 
The situation is different for the quantization in the \WF frame. By 
 using the Klein-Gordon inner product for \WF\!\!,  we can obtain the 
 relation between the annihilation 
operator $a^F_{\w k}$ in the \WF frame and the mode operators in the Minkowski frame.  This time, what we obtain is the relation 
\bal
a^{F}_{k\w }&=(f^F_{k\w },\phi^M)^F_{\rm KG}
 =i\int_{-\infty}^{\infty}zdt_F dx^{2}(f^{F*}_{k\w }\overleftrightarrow{\pp_{z_F}}\phi^M)\no
 &=i\int_{-\infty}^\infty \f{dp^1}{\s{2\pi E_{k p^1}}}\left(\f{E_{k p^1}-p^1}{E_{k p^1}+p^1}\right)^{-\f{i\w}{2}}a^M_{kp^1} \comma  \label{WFM}
\eal
which requires only the annihilation operator in the Minkowski frame. 
Furthermore,  since $\w$ is conjugate to the {\it spacelike}  coodinate $t_R$ in this case, we {\it do} have negative frequency modes for  $a^F_{k\w }, \w <0$ and  hence {\it the number of degress of freedom of the modes that the \WF observer sees is the same as those for the Minkowski observer}. 

As in the case of $a^R_{k\w}$, the relation (\ref{WFM}) above can be 
 simplified  by the use of the  rapidity variable $u$. It can be written as 
\begin{align}
a^{F}_{k \w}&= i \int{du\over \sqrt{2\pi}} e^{i\omega u}   a^M_{k u}
\period \label{FtransFM} 
\end{align}
Apart from a factor of $i$, this is nothing but the Fourier transformation. 
Therefore the  inverse relation is trivial to obtain and we get 
\begin{align}
a^M_{ku} &= -i \int {d\omega \over \sqrt{2\pi}} e^{-i\omega u} a^F_{k \omega} \comma \label{FtransMF}\\
\Leftrightarrow \qquad a^M_{kp^1} &= -i\int \f{d\w}{\s{2\pi E_{k'p^1}}}\left(\f{E_{kp^1}-p^1}{E_{kp^1}+p^1}\right)^{\f{i\w}{2}}a^F_{k\w} \period 
\label{aMitoaF}
\end{align}
The fact that $a^M_{kp^1}$ and $a^F_{k\w}$ are in one to one correspodence 
with no mixing of the creation and the annihilation operators tells us that the 
vacuum state of the two observers are the same, namely\footnote {The vacuum  $\ket{0}_F$ is called  the ``Unruh vacuum".}  
\begin{align}
\ket{0}_M = \ket{0}_F \period
\end{align}
The important difference, however,  is that the entities recognized as ``particles" by  the two observers are quite distinct and their wave functions have ``dual" profiles.
\subsubsection{Fourier transform as a unitary transformation}
We now make a useful observation  that the Fourier transform exhibited above can be realized by a unitary transformation, in the sense to be described below\footnote{For related references, see\cite{Namias1980,Narozhny2002}.}.

Define the fourier transform $\gtil(p)$ of a function $g(x)$ as 
\begin{align}
\int {dx \over \sqrt{2\pi}} e^{ipx} g(x) &= \gtil(p) \period
\end{align}
The {\it  functional forms} of  $g(x)$ and $\gtil(p)$  are in general different. 

Let us look for a special class of functions for which the functional forms of their Fourier  transform are the same up to a proportionality constant.
 The simplest such function is 
 obviously the following Gaussian for which the proportionality constant is unity:
\begin{align}
f_0(x) \equiv \pi^{-1/4} e^{-x^2/2} \comma \qquad \ftil_0(p) =\pi^{-1/4} e^{-p^2/2} = f_0(p^2)
\period
\end{align}
We know that such a function is the coordinate representation of  the ground state of the one-dimensional  harmonic oscillators $\{a, a^\dagger\}$  
\begin{align}
f_0(x) &= \langle x | 0\rangle\comma 
\end{align}
where $\ket{0}$ denotes the  oscillator ground state 
defined by 
\begin{align}
a\ket{0} &=0 \comma \qquad \com{a}{a^\dagger} = 1 
\comma \qquad \langle 0|0\rangle =1 \period
\end{align}
and $\ket{x}$ is,   as usual,   the 
eigenstate of the operator $\hat{x}$ with the eigenvalue $x$, \ie 
 $\hat{x} \ket{x} = x \ket{x}$.

In what follows, we take the coordinate representations of $a$ and $a^\dagger$ 
as 
\begin{align}
a &= {1\over \sqrt{2}}(x+ip) = {1\over \sqrt{2}} \left( {d \over dx} + x \right)  = {i \over \sqrt{2}} \left({d \over dp} + p\right) \comma \label{axp}\\
a^\dagger &={1\over \sqrt{2}}(x-ip) = {1\over \sqrt{2}}\left(-{d \over dx} + x\right) 
 = {(-i)\over \sqrt{2}}\left(-{d \over dp} + p\right) \period   \label{adaggerxp}
\end{align}
Now, as is well-known,  the $x$-representation of the excited states of the oscillator system 
\begin{align}
\ket{n}&\equiv{ (a^\dagger)^n \over \sqrt{n!}} \ket{0} 
\comma \qquad \langle m |n \rangle = \delta_{m,n}
\end{align}
 is given by 
\begin{align}
f_n(x) &\equiv  \langle x | n\rangle  = {1\over \sqrt{n!}} \left[  {1\over \sqrt{2}}\left(-{d \over dx} + x\right) \right]^n \langle x | 0\rangle ={1\over \sqrt{n!}}\left[  {1\over \sqrt{2}}\left(-{d \over dx} + x\right) \right]^n  f_0(x) \period
\end{align}
Inserting the unity $\int (dp/\sqrt{2\pi}) |p\rangle \langle p|$ and using 
 $\langle x | p \rangle = e^{ipx}$, this can be 
written as the Fourier transform 
\begin{align}
 f_n(x) &=  \langle x | n\rangle = \int {dp \over \sqrt{2\pi}}\bra{x} p\rangle 
\langle p | n\rangle 
= \int {dp \over \sqrt{2\pi}}\bra{x} p\rangle 
(-i)^n  {1\over \sqrt{n!}} \left[  {1\over \sqrt{2}}\left(-{d \over dp} + p\right) \right]^n \langle p | 0\rangle \nonumber\\
&= \int {dp \over \sqrt{2\pi}}  e^{ixp} (-i)^n  f_n(p) \period
\end{align}
Thus  the functional form of the Fourier transform $\ftil_n(p)$ is the same 
 as the original up to a constant, namely $\ftil_n(p)  =(-i)^n f_n(p)$. 

Let us consider the number operator  $\calN=a^\dagger a$, for which  $\calN  \ket{n}  = n   \ket{n}$.  By using the $p$-representation of $a$ and $a^\dagger$, as exhibited in (\ref{axp}) and (\ref{adaggerxp}),    this is written as 
\begin{align}
 \calN_p f_n(p) &= \half \left( -{d^2\over dp^2} + p^2 -1\right)  f_n(p) =nf_n(p)  \period
\end{align}
Therefore, we can express the Fourier transform $(-i)^n f_n(p)$ as 
\begin{align}
e^{-i {\pi \over 2} \calN_p } f_n(p) &= (-i)^n f_n(p)  \label{Fourierfn} \period
\end{align}
Note that here the terminology ``Fourier  transform" refers to the transform  of the  {\it form of the function}, with the argument taken to be the same. 

Exactly the same formulas hold for $p$ replaced by $x$. Thus as far as the 
 set of functions $\{f_n(p)\}$ are concerned, the Fourier transform is 
realized by the operation on  the LHS of (\ref{Fourierfn}). 

Up to a constant, $f_n(x)$ is nothing but the Hermite polynomial $H_n(x)$ times 
 the Gaussian $e^{-x^2/2}$. More precisely, 
\begin{align}
f_n(x) &= {1\over (2^n n! \sqrt{\pi})^{1/2}} H_n(x) e^{-x^2/2} \comma 
\end{align}
where $H_n(x)$ is defind by\footnote{There are different conventions for 
 the normalization of the Hermite polynomials. Our definition is the most 
 standard one.}
\begin{align}
H_n(x) &\equiv  e^{x^2/2} \left(-{d \over dx} + x\right)^n e^{-x^2/2} \period
\end{align}

Now in order to apply this formalism to the oscillators, such as $a^M_{ku}$ and $a^F_{k\w}$, we consider a set of oscillators depending on a continuous variable and  satisfying the following commutation  relations
\begin{align}
\com{a(x)}{a^\dagger(y)} &= \delta(x-y) \period
\end{align}
Since so far we  have realized the Fourier transform as a differential operation 
 on the set of functions $f_n(x)$ only, 
in order to define the Fourier transform of such an oscillator function, we should first express  $a(x)$ and $a^\dagger(x)$ in terms of  $f_n(x)$.  This can be done 
 due to  the following  completeness relation 
\begin{align}
\delta(x-y) &=  \sum_{n=0}^\infty f_n(x) f_n(y)  \period
\end{align}
Thus, expanding 
\begin{align}
a(x) &= \sum_{m=0}^\infty b_m f_m(x) \comma  \qquad 
a^\dagger(y) = \sum_{n=0}^\infty b^\dagger_n f_n(y)  \comma 
\end{align}
the commutation relation can be reproduced as 
\begin{align}
 \com{a(x)}{a^\dagger(y)} = \sum_{m,n} [b_m, b^\dagger_n] f_m(x) f_n(y) 
=\delta(x-y)\comma 
\end{align}
provided we take $[b_m, b^\dagger_n]  \equiv \delta_{m,n}$.
Therefore, since the Fourier transform is a linear operation, we can apply the formula (\ref{Fourierfn}) to the operators $a(x)$ and $a^\dagger(x)$ as well. This can be implemented formally by the unitary transformation of the form
\begin{align}
\atil(x) &= U^\dagger a(x) U \comma \\
U &= \exp\left( -{i\pi \over 2}\int dy a^\dagger(y) \calN_y a(y) \right) \period
\end{align}
In fact one can easily verify 
\begin{align}
U^\dagger a(x) U &= a(x) + \com{{i\pi \over 2} \int dy a^\dagger(y) \calN_y a(y) }{a(x)} + \cdots 
= e^{-{i\pi \over 2} \calN_x } a(x) \period
\end{align}
So the Fourier transform for the form of the operator is  indeed reproduced. 

Applied to the oscillators  $a^M_{ku}$ and $a^F_{k\w}$, we have the 
relations 
\begin{align}
ia^M_{ku} &= U_F a^F_{k\omega} U_F^\dagger \big|_{\omega =u} \comma\q  \no a^F_{k\w} &= U_M^{\dagger}ia^M_{k u} U_M \big|_{u=\omega}\comma\label{unitaryaM}
\end{align}
where we defined
\bal
U_{\cal I} =\exp\left( {i\pi \over 2}\int d{\omega'} a^{{\cal I}\dagger}_{k\omega'}\calN_{\omega' }a^{{\cal I}}_{k\omega'} \right)\comma\q {\cal I}=F, M. 
\eal
In using the operators $U_{\cal I}$, one must make sure to act the  
 differential operator $\calN_\omega$ on any $\omega$-dependent quantity,  be it  a function or  an operator, to the right of it. 
Transformations using $U_{\cal I}$ are useful in converting various quantities  in  the Minkowski and the \WF frames, as will be demonstrated for the 
 Poincar\'e generators in Appendix \ref{app:Poincare}. 
\subsubsection{Relations between \WR\!\!,  \WL\!\!, \WF  and Minkowski frames}
Finally, let us relate the modes in \WR and \WL frames with those 
in the \WF frame. 
Combining   (\ref{WRM}) and (\ref{WFM}) (and their hermitian conjugates) and eliminating the Minkowski modes, we can obtain a simple {\it algebraic} relationship between $a^R_{k\w}$ and $a^F_{k\w}$: 
\ba
a^{R}_{k\w}=-\f{i}{\s{2\sinh\pi\w}}    \left[e^{\pi\w/2}a^{F}_{k\w}-e^{-\pi\w/2}a^{F\dg}_{-k,-\w}\right]\comma  \qquad \w \ge 0 
\period \label{WRF} 
\ea 
Again since $a^R_{k\omega}$ exists only for $\omega \ge 0$, 
{\it this relation cannot be inverted. }

However, recall that  the ``full"  Rindler spacetime has the left wedge \WL  
in addition to the right wedge \WR\!\!.  By  similar arguments we can obtain the relationship between $a^L_{k\w}$ and $a^F_{k\w}$ as
\ba
a^{L}_{k\omega }=-\f{i}{\s{2\sinh\pi\w}} \left[e^{\pi\w/2}a^{F}_{-k,-\w}-e^{-\pi\w/2}a^{F\dg}_{k\w}\right] \comma \qquad \w \ge 0 \period \label{WLF}
\ea
Note that the right-hand side  contains $a^F_{k,-\omega}$ instead of $a^F_{k\omega}$,  in contrast to the expression of $a^R_{k\omega}$ given in (\ref{WRF}). Therefore, combining (\ref{WRF}) and (\ref{WLF}), one can express 
the modes in the future wedge \WF  in terms of the modes in \WR and \WL in the following 
 combinations:
\ba
a^F_{k\w}=-\f{i}{\s{2\sinh\pi\w}} \left[e^{\pi\w/2}a^{R}_{k\w}-e^{-\pi\w/2}a^{L\dg}_{k\w}\right]\comma \label{relpFRL}\\
a^F_{-k,-\w}=\f{i}{\s{2\sinh\pi\w}} \left[e^{-\pi\w/2}a^{R\dg}_{k\w}-e^{\pi\w/2}a^{L}_{k\w}\right] \period \label{relmFRL}
\ea
Intuitively, this is the reflection of the fact that  the region \WF  can be reached both from  \WR  by  left-moving waves and  from \WL  by right-moving waves. Note that the right hand sides  contain both the creation and the annihilation operators and hence these relations constitute the  Bogoliubov transformations between the Rindler modes and the Unruh modes.

As an application of the formulas (\ref{relpFRL}) and (\ref{relmFRL}), let us 
 express the \WF vaccum $\ket{0}_F$, which is the same as the Minkowski vacuum $\ket{0}_M$,  in terms of the states in the 
 \WR and \WL frames.  The condition that $\ket{0}_F$ must be annihilated by $a^F_{k\omega}$ and $a^F_{k, -\omega}$ can be expressed in the form
\begin{align}
a^L_{k\w} \ket{0}_F &= e^{-\pi \w} a^{R\dagger}_{k\w} \ket{0}_F \comma \qquad 
a^R_{k\w} \ket{0}_F = e^{-\pi \w} a^{L\dagger}_{k\w} \ket{0}_F \period
\end{align}
The solution is 
\begin{align}
\ket{0}_F &=\ket{0}_M= \calN  \exp \left(\int d^2k \int_0^\infty  d\omega e^{-\pi \w} a^{L\dagger}_{k\w} a^{R\dagger}_{k\w} \right) \ket{0}_L \otimes \ket{0}_R\comma  \label{vacFM}
\end{align}
where $\calN$ is a normalization factor\footnote{The normalization constant $\calN$ 
is divergent as it stands. To make it finite, one must discretize $k$ and $\w$ and 
regularize the infinite sum. } and $\ket{0}_{L,R}$ are the vacua for the  \WL and \WR frames 
 defined by $a^L_{k\w} \ket{0}_L =0, a^R_{k\w}\ket{0}_R=0$, for all $k$ and positive $\w$. They are   known as the Rindler vacua.  

Let us make a few remarks  on the relation between the field expressed in the Minkowski frame and in the combined \WR and \WL frame. 
\begin{itemize}
	\item In the context of the discussions of  the entanglement and the entropy thereof, 
 instead of the expression (\ref{vacFM}) for the Minkowski vacuum, a simpler formula of the form 
\begin{align}
\ket{0}_M &=  \sum_{n=0}^\infty e^{-\omega \pi n} \ket{n}_L \otimes \ket{n}_R 
\end{align}
is often quoted in the literature. This of course is an expression for the two dimensional toy model with only one frequency kept. The full expression (\ref{vacFM}) for 
 four dimensions can be written in a form similar to the above  after 
 discretizing the energy $\w$ and the momenta $k$ and expanding the exponential. \item By using the relations  (\ref{aMitoaF}), (\ref{relpFRL}) and (\ref{relmFRL}), 
 one can express $\phi^M(t_M, x^1, x)$ in terms of the modes of \WL and \WR\!\!. 
An important check is if  $\phi^M(t_M, x^1, x)$ so constructed depends only on the modes of \WL(\WR\!\!) when $x^1<0\  (x^1>0)$.  In Appendix \ref{app:phiMWLWR}, we shall sketch  a proof \footnote{In the basic literature such as \cite{Fulling1973} and \cite{Unruh}, this property appears to be put in by  hand rather than derived.},  which turned out to require  a careful treatment of the proper analytic continutation. 
\end{itemize}
\subsection{Representation of Poincar\'e algebra for  various observers\label{sec:Poincare}} 
\subsubsection{Poincar\'e algebra for the $1+1$ dimensional subspace }
Evidently, the Poincar\'e symmetry of the flat Minkowski space is a fundamental symmetry  governing,  above all,  the structure of  correlation functions. 
Although the quantum generators of the Poincar\'e algebra are well-known in the ordinary  Minkowski frame, their forms are non-trivial in terms of the modes of 
 the observers in the \WF\!\!, \WR  and \WL wedges and have not been 
discussed in the literature.  In this subsection, we shall construct them by using the 
 relations among the modes of the various observers established in the previous 
subsections. 

As will be described in the next section,   the vicinity of the horizon of the four dimensional \SS black hole of our interest has the structure of the $1+1$ dimensional flat space $\mathbb{R}^{1,1}$. 
For that reason, in what follows we shall focus exclusively on the generators and the algebra  pertaining to such subspace of the four-dimensional  Minkowski space. 
In terms of the coordinates of the aforementioned observers, the metric of the 
 subspace $\mathbb{R}^{1,1}$ is expressed as  
\bal
ds^2&=-(dt_M)^2+(dx^1)^2
=z_F^2dt_F^2-dz_F^2=-z_R^2dt_R^2+dz_R^2=-z_L^2dt_L^2+dz_L^2
\period
\eal
As usual, the Poincar\'e generators can be constructed in terms of the 
 energy-momentum tensor, which for a scalar field takes the form 
\bal
T^\mu_{\nu}&=\f{\d{\cal L}}{\d(\pp_\mu\phi)}\pp_\nu\phi-\d^\mu_\nu
{\cal L}
=-\pp^{\mu}\phi\pp_\nu\phi+\f{1}{2}\d^{\mu}_\nu\pp^\rho\phi\pp_\rho\phi \period
\eal
Here $\mu,\nu, \rho$ refer to  general coordinates. Then, the generators of the 
 Poincar\'e algebra of $\mathbb{R}^{1,1}$ are the energy $H$ and the momentum $P_1$ 
 in the first direction 
\bal
H&\equiv P_0=\int d^2x \int dx^1 T^0_0 \comma
\qquad 
P_1=\int d^2x \int dx^1 T^0_1 \comma 
\eal 
and the boost generator along the first direction $M_{01}$
\bal
M_{01}&=\int d^2x \int dx^1 (x^{0}T^{01}-x^{1}T^{00}) \period
\eal
The subscripts $0$ and $1$ here refer of course to the directions in the Minkowski frame and when quantized the normal-ordering prescription for the modes 
is taken for granted.  Then, in terms of the Minkowski modes, these generators are  given by 
\bal
H&=\int dk^2 \int dp^1 \ E_{k p^1} a^{M\dg}_{kp^1}a^M_{kp^1}
\comma \\
P_1&=\int dk^2 \int dp^1 p_1 a^{M\dg}_{kp^1}a^M_{kp^1}\comma \\
M_{01}&=i\int dk^2 \int dp^1 E_{kp^1}\ a^{M\dg}_{kp^1}\f{\pp}{\pp p_{1}}a^M_{kp^1} \comma \label{Mzeroone}
\eal
and can be checked to form the $1+1$ dimensional Poincar\'e algebra
\ba
[H,M_{01}]=iP_1\comma \q [P_1, M_{01}]=iH \comma \q  [H, P_1] =0 \period
\label{oneonePalg}
\ea
\subsubsection{Poincar\'e generators for \WF observer} 
Recall that the  relation between the Minkowski modes $a^M_{kp^1}$ and the (Unruh) modes $a^F_{k\w}$ for the \WF observer have been worked out in (\ref{aMitoaF})  (which is reproduced for convenience below)
\bal
 a^{M}_{kp^1}&=-i\int \f{d\w}{\s{2\pi E_{k',p^1}}}\left(\f{E_{kp^1}-p^1}{E_{kp^1}+p^1}\right)^{\f{i\w}{2}}a^F_{k\w}
 =\f{-i}{\s{2\pi E_{k',p^1}}}\int d\w e^{-i\w u} a^F_{k\w} \comma 
\eal
where $u\equiv \half \ln {E_{kp^1} +p^1 \over E_{kp^1} -p^1 }$ is the 
 ``rapidity" variable. 
Substituting this into (\ref{Mzeroone}),  $M_{01}$ can be 
 rewritten in terms of the \WF modes as 
\ba
M_{01}^F=\int d^{2}k\int d\w\ \w a^{F\dg}_{k\w}a^F_{k\w} \period
\label{MF}
\ea
Note that this is diagonal in $\w$ and hence interpretable as the  ``momentum" 
operator.  In Appendix \ref{app:Unitary}, we show explicitly that by the unitary 
 transformation constructed in (\ref{unitaryaM}), $M_{01}$ and $M_{01}^F$ 
are transformed into each other. 

Next, let us rewrite  the Hamiltoian operator in terms of the Unruh modes. 
Using the rapidity representation, with $E_{kp^1} = |k|\cosh u$, we get 
\bal
H&=\int dk^2 \int dp^1\, E_{k'p^1} a^{M\dg}_{kp^1}a^M_{kp^1}\no
&=\int d^{2}k |k|  \int d\w d\w'\int \f{du}{2\pi}  \cosh u e^{i(\w'-\w)u}a^{F\dg}_{k\w'}a^F_{k\w} \period
\eal
Since the integral over $u$ is divergent and behaves as $\sim e^{|u|}$ at large $|u|$, 
we should define this integral with a suitable regularization.  We adopt the definition \ba
\int \f{du}{2\pi}  \cosh u e^{i(\w-\w')u}\equiv \lim_{\ep\r +0}\int \f{du}{2\pi}e^{-\ep u^2}  \cosh u e^{i(\w-\w')u} \period
\ea
Then, expanding $\cosh u$ in powers, we can rewrite this integral as 
\bal
&\lim_{\ep\r +0}\int \f{du}{2\pi}e^{-\ep u^2}  \cosh u e^{i(\w'-\w)u} \no
&\qquad =\lim_{\ep\r +0}\int \f{du}{2\pi}e^{-\ep u^2}  \sum_{n}\f{u^{2n}}{(2n)!} e^{-\ep u^2}e^{i(\w'-\w)u}
\no&\qquad =\lim_{\ep\r +0}\int \f{du}{2\pi}e^{-\ep u^2}  \sum_{n}\f{(-1)^n}{(2n)!}\left(\f{\pp}{\pp\w}\right)^{2n} e^{i(\w'-\w)u}\no
&\qquad =\lim_{\ep\r +0}\int \f{du}{2\pi}e^{-\ep u^2}  \cos\left(\f{\pp}{\pp\w}\right) e^{i(\w'-\w)u}\period
\eal
The remaining Gaussian integral produces a $\delta$ function $\delta(\omega'-\omega)$ and hence the Hamiltonian can be written as 
\bal
H^F
&=\int d^{2}k |k|  \int d\w d\w'  \cos\left(\f{\pp}{\pp\w}\right)\d(\w'-\w) a^{F\dg}_{k\w'}a^F_{k\w} \no
&= \int d^{2}k |k|\int d\w \  a^{F\dg}_{k\w}\cos \left(\f{d}{d\w}\right)a^F_{k\w}\period \label{HF}
\eal
In an entirely similar manner, $P_1$ operator is expressed as 
\bal
P_1^F&=-i\int d^{2}k|k| \int d\w \  a^{F\dg}_{k\w}\sin \left(\f{d}{d\w}\right)a^F_{k\w}\period \label{PoneF}
\eal
These operators are understood to be used within a matrix element such that the 
object is infinitely differentiable with respec to $\w$. 

In  Appendix \ref{app:PoincareWF}, we demonstrate that these operators $M_{01}^F, H^F$ and $P_1^F$ do satisfy the $1+1$ dimensional Poincar\'e algebra  of (\ref{oneonePalg}). 


\subsubsection{On the Poincar\'e generators for \WR and \WL observers }
Having derived the expression of the generators in terms of \WF oscillators, 
we can now write them in terms of the \WR and \WL mode operators 
 using the relations (\ref{relpFRL}) and (\ref{relmFRL}), that is, 
\ba
a^F_{k\w}=-\f{i}{\s{2\sinh\pi\w}} \left[e^{\pi\w/2}a^{R}_{k\w}-e^{-\pi\w/2}a^{L\dg}_{k\w}\right]\comma \\
a^F_{-k,-\w}=\f{i}{\s{2\sinh\pi\w}} \left[e^{-\pi\w/2}a^{R\dg}_{k\w}-e^{\pi\w/2}a^{L}_{k\w}\right] \period 
\ea

As for the generator of the angular momentum, it  cleanly separates  into the \WR part and the \WL part:
\begin{align}
M_{01} &= M_{01}^R + M_{01}^L\comma 
\end{align}
where 
\begin{alignat}{2}
 M_{01}^R &= \int d^{2}k \int^{\infty}_0d\w\ \w a^{R\dg}_{k\w}a^R_{k\w} \comma \qquad  && M_{01}^L =- M_{01}^R\bigr|_{a^R_{k\w}\rightarrow a^L_{k\w}} \period
\end{alignat}
Two remarks are in order. \nxt
(i)\  First, $M_{01}^R$ is 
 diagonal in $\omega$, which in this case has the meaning of the energy  conjugate to the Rindler time $t_R$. This clearly shows that the boost generator $M_{01}^R$ is the Hamiltonian for the \WR observer. \\
(ii)\ Second, the relative minus sign between $M_{01}^R$ and $M_{01}^L$ simply means that the ``time" flows in opposite directions in \WR and \WL\!\!. 
\nxt
These remarks are expressed by the following simple relations:
\bal
e^{i\xi M^R_{01}}a^R_{k\w}e^{-i\xi M^R_{01}}
&=e^{-i\w\xi}a^R_{k\w}\comma \\
e^{i\xi M^L_{01}}a^L_{k\w}e^{-i\xi M^L_{01}}
&=e^{i\w\xi}a^L_{k\w} \period
\eal
Thus, acting on the field, the boost generator indeed  induces  the Rindler time evolution in each wedge as shown below:
\bal
e^{i\xi M^R_{01}}\phi^R(t_R, z_R, x)e^{-i\xi M^R_{01}}&=\int_0^{\infty}d\w\int d^{2}kN^R_{\w}[K_{i\w}(|k|z_R)e^{ikx}e^{-i\w (t_R+\xi)}a^R_{k\w}+{\rm h.c.}] \comma \no
e^{i\xi M^L_{01}}\phi^L(t_L, z_L, x)e^{-i\xi M^L_{01}}&=\int_0^{\infty}d\w\int d^{2}kN^L_{\w}[K_{i\w}(|k|z_R)e^{-ikx}e^{i\w (t_L+\xi)}a^L_{k\w}+{\rm h.c.}]\period \nonumber
\eal

In contrast,  the gerators $H$ and $P_1$ turned out not to factorize  into the \WR part and the \WL part. They can be written as 
\bal
H&=\int_{\w>0} d\w \int_{\w'>0} d\w'\int\f{du}{2\pi}  \f{\cosh u e^{i(\w-\w')u}}{\s{\sinh\pi\w\sinh\pi\w'}}\left[e^{-\f{\pi(\w+\w')}{2}}\d(\w-\w')\right. \no& +
\left.\cosh\f{\pi(\w+\w')}{2}\left(a^{R\dg}_{k\w}a^{R}_{k\w'}+a^{L\dg}_{k\w'}a^{L}_{k\w}\right)-\cosh\f{\pi(\w-\w')}{2}\left(a^{L\dg}_{k\w'}a^{R\dg}_{k\w}+a^{L}_{k\w}a^{R}_{k\w'}\right)\right]\, ,\no
P_1&=\int_{\w>0} d\w \int_{\w'>0} d\w'\int\f{du}{2\pi}  \f{\sinh u e^{i(\w-\w')u}}{\s{\sinh\pi\w\sinh\pi\w'}}\left[e^{-\f{\pi(\w+\w')}{2}}\d(\w-\w')\right. \no& +
\left.\cosh\f{\pi(\w+\w')}{2}\left(a^{R\dg}_{k\w}a^{R}_{k\w'}+a^{L\dg}_{k\w'}a^{L}_{k\w}\right)-\cosh\f{\pi(\w-\w')}{2}\left(a^{L\dg}_{k\w'}a^{R\dg}_{k\w}+a^{L}_{k\w}a^{R}_{k\w'}\right)\right]\, .
\eal
Note that the last term in both $H$ and $P_1$ contains the mixture of 
the oscillators of \WL and \WR of the form $a^{L\dg}_{k\w'}a^{R\dg}_{k\w}+a^{L}_{k\w}a^{R}_{k\w'}$, which prevents the factorization. The basic reason 
is simple. As we already emphasized, while the oscillators of \WF observer (and 
 the Minkowski observer) can be expressed in terms of those of \WR and \WL 
observers, these relations are not invertible, reflecting the fact the number of 
 the degrees of freedom in the \WR and the \WL frame each is half that of 
 the \WF (or the Minkowski) frame. Thus, the full Poincar\'e algebra does not exist 
 in \WR and \WL frame separately. 
\newpage
\section{Quantization in an eternal Schwarzschild black hole by various observers}
As was  emphasized in the introduction, the main aim of this paper is to study the 
structure of the Hilbert spaces of the scalar field near the horizon of the Schwarzschild black hole  quantized in the frames of different observers. 
This is made possible largely because such  near-horizon geometry  has the structure of the flat Minkowski space, to be recalled shortly. This allows us to make use of the knowledge of the quantization in various 
 frames which has been  reviewed,  with some additional new information,  in the previous section.  As we shall discuss, however, we must take due care that our 
computations should be performed in such a way that the  approximation used 
 is legitimate. 

Now,  in studying the quantization 
around the horizon of a black hole,  it will be  important to distinguish 
 the two cases,  namely the case of the eternal (\ie two-sided) 
black hole and the more physical one  where the (one-sided) black hole is 
 produced by a collapse of matter (or radiation). There are essential differerences between the two. 

In this section, we analyze   the  simpler case of the eternal  Schwarzschild black hole. 
\subsection{Flat geometry around the event horizon of a \SS black hole}
Let us first recall how the flat geometry emerges in the vicinity of the event horizon  of a  Schwarzschild black hole. 

We denote the  metric for the four-dimensional \SS of mass $M$ in the asymptotic  coordinate in the  usual  way:
\begin{align}
ds^2 &= -\left( 1-{2M \over r}\right) d\ttil^2 + \left( 1-{2M \over r}\right)^{-1} dr^2 + r^2 d\Omega^2 \period 
\end{align}
The notations are standard, except that we set the Newton constant $G$ to one and used  ``$\ttil$" to 
 denote the \SS time.  This  will be later  rescaled to ``$t$" to denote the Minkowski time. 

First we consider the region \WR  outside the horizon. 
It is convenient to introduce a positive coordinate ``$z$" which 
 measures the proper radial distance from the horizon:
\begin{align}
z &\equiv \int_{2M}^r \sqrt{g_{rr}(r')} dr' = \int_{2M}^r {1\over \sqrt{1-{2M \over r'}}}dr' \period \label{defz} 
\end{align}
Near the horizon at $r=2M$, we write  $r$ as $r=2M +y$, expand $z$ in powers of $y$  in the form $z=ay^{1/2} (1+by + \cdots)$ and then solve for $y$ in terms of $z$. After a simple calculation we obtain
\begin{align}
r &=2M +  {M\over 8}\left(  z\over M\right)^2 -{M\over 384} \left( {z \over M}\right)^4 + \calO((z/M)^6) \period  \label{expandr}
\end{align}
Now if keep up to the second term of this expansion, the \SS 
 the metric becomes 
\begin{align}
ds^2 &\simeq -z^2 (dt)^2 + dz^2 + r^2(z) d\Omega^2  \\
t &\equiv {\ttil \over 4M} \comma \qquad d\Omega^2 =  d\theta^2+\sin\theta^2 d\phi^2 \period \label{ttil}
\end{align}
Further, focusing on the small  two dimensonal region perpendicular to 
 the radial direction around $\theta=0$, we can parametrize it by the coordinates
\begin{align}
x^2 &= 2M \theta \cos\phi \comma \qquad x^3 = 2M \theta \sin \phi \period  \label{xtwoxthree}
\end{align}
Then in this region the metric further simplifies and becomes  identical to the Rindler metric for \WR  given in (\ref{coordr})
\begin{align}
ds^2 &= -z^2 dt^2 + dz^2 + (dx^2)^2 + (dx^3)^2 
\comma 
\end{align}
which expresses (a portion of) the  flat spacetime\footnote{The approximation of taking $r$ to be the fixed value $2M$ in (\ref{xtwoxthree}) is admissible, since in the expression 
 $(dx^2)^2 + (dx^3)^2$ the radial coordinate appears in the forms 
$r^2 d\theta^2, r^2 d\phi^2, dr^2$ and $rdr d\theta$. For these expressions, the order $\calO(z^2/M^2)$ terms are safely neglected. }. 

To see the region of validity of this approximation, let us find out the condition  under which we can neglect the third term of the expansion (\ref{expandr}) compared to the second. A simple calculation shows that the 
 condition is 
\begin{align}
{z \over M} \ll 4\sqrt{3} \sim 7 \comma 
\end{align}
showing that the flat space approximation is good for $z$ up to the order of the \SS radius $\calO(M)$ out from the horizon. 

For the other regions \WL\!\!, \WF\!\!, and \WP\!\!, by appropriate  analytic continutations, we obtain similar flat space form of the metric of appropriate signature, as already displayed in Sec.~2. In particular, we should remember that as we go  from \WR  to \WF  the role of time and space variables are interchanged. 
\subsection{Exact treatment for the transverse spherical space }
The approximation of the vicinity of the horizon as a four-dimensional flat Minkowski  space is certainly a great advantage, as long as we are interested only in 
 the quantities determined by the local properties of the fields.  However, 
as we have repeatedly emphasized,  in quantum treatment the concept of states
 created by the mode operators is a global one, and that is precisely what we 
 are interested in.  It turns out that the inadequacy of the flat approximation is particularly troublesome for the two-dimensional transverse space,  since the orthogonality relation needed to  extract the modes from the fields requires integration over the entire range of $(x^2, x^3)$ expressing the flat 2-space, which is  unjustified  for large values of these coordinates. 

The obvious cure for this part of the problem is to replace the expansion in terms of  the plane waves by the spherical harmonics $Y_{lm}(\theta, \varphi)$. Thus, instead of $M^{1,3}$, we will be dealing with the spacetime  $\mathbb{R}^{1,1} \times \mathbb{S}^2_{2M}$, where the subscript $2M$ denotes the radius of the sphere. 

Explicitly, we can write the general expansions of a massless scalar and its conjugate in the vicinity of the horizon in the form 
\begin{align}
\phi(t,x^1, \Omega) &=\sum_{l=0}^\infty \sum_{m=-l}^l  \int_{-\infty}^\infty {dp^1\over \sqrt{4\pi E_{k_lp^1}}}   e^{ip^1 x^1 -iE_{k_lp^1} t} Y_{lm}(\Omega) a_{ lm p^1} + {\rm h.c.}  \comma \label{phiYlm}\\
\pi(t, y^1, \Omega') 
& = -i \sum_{l'=0}^\infty \sum_{m'=-l'}^{l'} \int_{-\infty}^\infty {dq^1 \over \sqrt{4\pi E_{k_{l'}q^1}}} E_{k_{l'}q^1} e^{iq^1 y^1 -iE_{k_{l'}q^1} t} Y_{l'm'}(\Omega') a_{ l'm' q^1} + {\rm h.c.} 
\end{align}
where $\Omega = (\theta, \varphi)$.  The energy $E_{k_l p^1}$ is 
determined in terms of $p^1$ and $l$ by the equation of motion as 
\begin{align}
E_{k_lp^1}^2  &= (p^1)^2 + k_l^2 \comma \qquad 
k_l \equiv  {\s{l(l+1)} \over 2M} \period 
\end{align}
The equal time canonical commutation relation takes the form
\begin{align}
\com{\pi(t, x^1, \Omega) } {\phi(t,y^1, \Omega')}
&= -i \delta(x^1-y^1) \delta(\cos\theta -\cos\theta') \delta(\varphi- \varphi')
\period \label{comphipi}
\end{align}
The orthogonality  for $Y_{lm}(\Omega)$ is 
\begin{align}
\int_0^{2\pi} d\varphi \int_0^\pi \sin\theta d\theta Y^\ast_{lm}(\theta, \varphi) Y_{l'm'}(\theta, \varphi) &= \delta_{ll'} \delta_{mm'} \comma 
\end{align}
while the completeness reads 
\begin{align}
\sum_{l=0}^\infty \sum_{m=-l}^l Y_{lm}^\ast(\theta, \varphi) Y_{lm}(\theta', \varphi') &= \delta(\cos\theta -\cos\theta') \delta (\varphi -\varphi') \period
\end{align}
Using the orthogonality relation, we can extract the modes as 
\begin{align}
a_{lm p^1}&=\int {dx^1 \over \sqrt{4\pi E_{k_lp^1}} } \int d\Omega Y^\ast_{lm}(\Omega) e^{-ip^1 x^1 + iE_{k_lp^1}t}
i \overleftrightarrow{\del_t} \phi(t, x^1, \Omega) \comma \\
a^\dagger_{lm p^1}&=\int {dx^1 \over \sqrt{4\pi E_{k_lp^1}} } \int d\Omega Y_{lm}(\Omega) e^{ip^1 x^1 - iE_{k_lp^1}t}
{1 \over i} \overleftrightarrow{\del_t} \phi(t, x^1, \Omega) \period
\end{align}
From the canonical commutation relation (\ref{comphipi}), the modes satisfy 
\begin{align}
\com{a_{lm p^1}}{a^\dagger_{l'm' q^1}} &= \delta_{ll'} \delta_{mm'} \delta(p^1-q^1)\comma \qquad  \mbox{rest}=0 \period
\end{align}
In summary, the expansion in flat space described in Sec.~2 can be converted to
 the present case by the simple replacements 
\begin{align}
\int {d^2k  \over (2\pi)^2} e^{ikx}(\ast\ast) a_{kp^1}\quad &\longrightarrow \quad 
\sum_{l=0}^\infty \sum_{m=-l}^l Y_{lm}(\Omega) (\ast\ast) a_{lm p^1}\comma  \\
E_{kp^1} =\sqrt{ k^2 + (p^1)^2}\quad &\longrightarrow \quad E_{k_l p^1}
=\sqrt{k_l^2 + (p^1)^2 } \comma \quad  k_l^2\equiv {l(l+1) \over (2M)^2} \comma \\
a_{kp^1} \quad &\longrightarrow \quad a_{lmp^1} \period
\end{align}

As the behavior of the scalar field on the transverse spherical surface near the horizon is treated exactly as above, we need only be concerned with the dependence on  the remaining two dimensions $(t_M, x^1)$.  Thus  from now on, we will use expressions such as ``flat approximation" or ``flat space"  to refer only to the two dimensional part near the horizon within $\mathbb{R}^{1,1}$. 

\subsection{Quantization in the frame of freely falling observer near the 
 horizon }
Among the many interesting questions  that stem from the observer dependence 
 of the quantization around a black hole, perhaps the 
 most provocative one is whether the freely falling observer, hereafter 
abbreviated as FFO,  sees a different Hilbert space structure  for the quantized  scalar field before and after he/she passes through the horizon.  In other words, 
whether the equivalence principle for the field is affected by the quantum effects or not. 

In this subsection we will perform some preparatory computations in the frame of an FFO,  who  crosses  the horizon along various directions in the Penrose diagram, \ie  with various  velocities.  

First, let us  briefly describe how the geodesic of a massive classical particle (which represents a FFO) near the horizon maps to the motion in the flat coordinate system obtained by  the non-linear transformation of the previous subsection. Although the final answer should be  a straight line in the flat coordinate system, as the geodesic should map to a geodesic, it is instructive to see the physical meaning of this mapping. 

Consider first the motion in \WR\!\!. The geodesic equation in the radial direction of a massive particle (with mass set to unity) in the 
\SS spacetime in the region \WR  takes the form
\begin{align}
\half \left({dr \over d\tau}\right)^2 + \half \left( 1-{2M \over r}\right) = \half E^2 \comma 
\end{align}
where $E$ is a constant of motion given by 
\begin{align}
E &= \left( 1-{2M \over r}\right){dt \over d\tau} \period \label{eqE}
\end{align}
Here $t$ is the asymptotic time and $\tau$ is the proper time. 
Restricting to the region near the horizon, we can approximate $r$ by 
$r \simeq 2M + (z_R^2/8M)$ as worked out in (\ref{expandr}). Then, from (\ref{eqE}) one can express $d\tau$ in terms of $t$ and $z_R$ and rescaling $t$ like 
 $t=4Mt_R$  as in (\ref{ttil}), we can easily rewrite the geodesic equation 
as 
\begin{align}
&\left( {1\over z_R} {dz_R \over dt_R} \right)^2 + b^2 z_R^2 =1 \comma 
\qquad b \equiv {1\over 4ME} \period 
 \label{massiveeqR} 
\end{align}
This differential equation for $z_R$ as a function of $t_R$ is easily solved to give\footnote{Actually, there is another solution with the sign in front of $t_R$'s flipped. 
But since they are related simply by changing the sign of $t_R$, we  deal with  the 
one displayed below without loss of generality.}
\begin{align}
z_R &= {2c \over c^2 e^{ t_R} + b^2 e^{-t_R} } \comma 
\qquad c >0 \comma \label{zgeodesic}
\end{align}
where $c$ is a positive  integration constant.  This shows that $z_R$ vanishes as $t_R \rightarrow \pm \infty$, meaning that the trajectory starts and ends at the horizon. 
The physical picture is that, due to the gravitational attraction of the black hole, 
the trajectory which starts out at  the horizon at $t_R=-\infty$ 
with some initial velocity goes out to a certain maximum distance (actually $z_R=4ME$) away from the horizon where it stops and then gets pulled  back to the horizon at $t_R =\infty$. 

Now let us rewrite  this motion (\ref{zgeodesic}) in terms of  the flat Minkowski coordinates $(t_M, x^1)$ 
related to $(z_R, t_R)$ by $t_M = z_R \sinh t_R, x^1=z_R \cosh t_R$ as 
in (\ref{reltoMink}). Then, we get 
\begin{align}
t_M &= -{1\over \beta}  x^1 + X  \comma \label{tMX} \\
\beta &\equiv {c^2-b^2 \over c^2 +b^2} \comma \qquad X \equiv {2c \over c^2 -b^2} \period 
\end{align}
As expected, this describes  a family of timelike straight line trajectories,  with the velocity $\beta$. To construct an orthogonal coordinate system with the trajectories above as specifying the time direction, we must supply spacelike lines  perpendicular (in the  Lorentz sense) to them.  Clearly they are of the form 
\begin{align}
t_M &= -\beta x^1 + T \comma \label{tMT}
\end{align}
where $T$ is a parameter. Thus by changing the values of  $X$ and $T$ we span
(a part of)  the 
 Minkowski space. In other words, $(T,X)$ serve as  new orthogonal coordinates. In fact  better coordinates are  the rescaled ones  $(t_\beta, x^1_\beta)$ defined in the following way:
\begin{align}
t_\beta &\equiv \ga T \comma \qquad x^1_\beta \equiv \ga \beta X \comma 
\qquad 
\ga \equiv {1\over \sqrt{1-\beta^2}} \comma \\
ds^2 &= -dt_\beta^2 + (dx^1_\beta)^2 \period
\end{align}
Then the relation to the canonical Minkowski variables $(t_M, x^1)$ are  
obtained from (\ref{tMX}) and (\ref{tMT}) and can be written  as 
\begin{align}
\vecii{t_\beta}{x^1_\beta} &= \ga \matrixii{1}{\beta}{\beta}{1} \vecii{t_M}{x^1} \period
\end{align}
This is nothing but the Lorentz boost by the velocity $\beta$ in the negative $x^1$ direction. 

One can perform a similar analysis of the geodesic  in the \WF  region 
sharing the horizon as the boundary with \WR\!\!.  The outcome of the study is that the geodesics which hit the same point on this common boundary from the inside and the outside with the same velocity $\beta$ are actually one and the same straight line 
which is obtained by the Lorentz boost of the trajectory along the time axis in the 
canonical Minkowski coordinate.  Physically this must be the case since the FFO  must be able to go through the horizon freely due to the {\it classical} equivalence 
 principle. 

\begin{figure}[h]
\begin{center}
  \includegraphics[width=14cm]{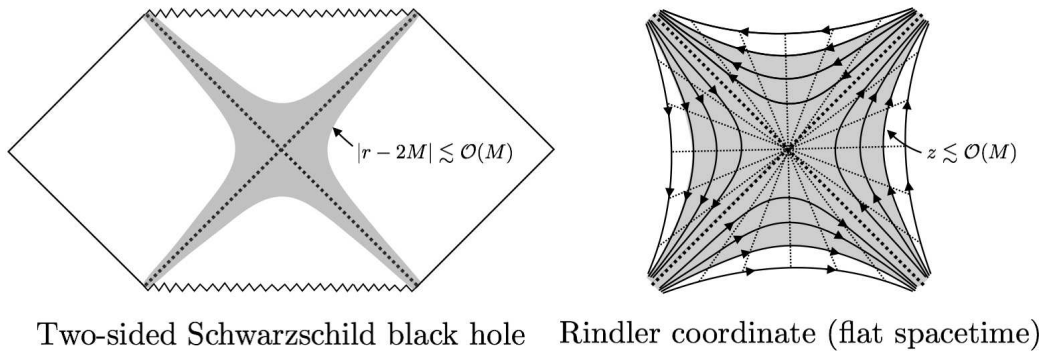}
\caption{ Approximately flat regions, shown in gray, near the horizon of the two-sided  \SS black hole and the corresponding region in the Rindler coordinate of the flat spacetime.   \label{flat-twosided}}
 \end{center}
\end{figure}

With this preparation, let us now  discuss the quantization and the mode expansion of the free scalar field by an FFO in the vicinity of the horizon where the flat space approximation for the dependence on $(t_M, x^1)$  is valid.  In the case of 
 the two-sided eternal \SS black hole studied in this section,   we have both \WR and \WL  regions and approximately flat region near the horizons 
can be dipicted as the shaded region in Figure\  \ref{flat-twosided}. 

\noindent
  In this region the general solution for a scalar field as seen by an FFO is 
\begin{align}
\phi^M(t_M, x^1, \Omega) &=\sum_{l, m}  \int_{-\infty}^\infty {dp^1 \over  \sqrt{ 2E_{k_l p^1 }}} 
\left( e^{ip^1 x^1 -iE_{k_l p^1 }t_M}Y_{lm}(\Omega) a^M_{lm p^1 } + {\rm h.c.} \right) 
\period
\end{align}
This expression is perfectly valid {\it locally} but as we try to extract 
the mode operators $a^M_{lm p^1}$ and their conjugate and check that they 
 obey the usual commutation relations, we encounter a problem of the same nature as occurred when using the flat coordinates $(x^2, x^3)$.  Namely, 
 such an extraction requires orthogonality relations for the plane waves, which 
involves integration over an infinite range for the spatial variable $x^1$. 
 As such a range is not within the flat space region, it appears to be quite difficult 
 to solve this problem. 

The observation which allows us to overcome this obstacle is that 
regions of infinite range do exist  around the horizon along the light cones in 
the $(t_M, x^1)$ space. Technically, however,  the quantization using the exact light cone variable as the time is rather singular.  Therefore, we shall make a very large (but not infinite)  two-dimensional Lorentz boost  so that the $x^1$-axis is rotated to the direction which is almost lightlike yet still slightly spacelike. Then assuming the usual regularization that the scalar field vanishes at $t_M=\pm \infty$,  we can integrate along this new $x^1$ axis, which is practically contained in the flat region,  and extract  the modes. Since what we used  here is a Lorentz transformation, 
the exponent is invariant, while 
 the modes are transformed in a well-known simple way, namely 
\begin{align}
{a'}^M_{lm {p'}^1 } \sqrt{E'_{k_l {p'}^1} }&= {a}^M_{lm {p}^1 } \sqrt{E_{k_l {p}^1 }}\comma \qquad \mbox{(same for the conjugates)} \comma \label{LtransaM}
\end{align}
where prime signifies  the Lorentz-transformed quantities.  The mode operators 
 satisfy the usual commutation relations, \ie 
$[a'_{lm{p'}^1}, {a'}^\dagger_{l'm' {q'}^1}] = \delta_{ll'}\delta_{mm'}\delta({p'}^1 -{q'}^1)$ etc. 
This immediately tells us 
that  the number of degrees of freedom observed by the FFO in the horizon region is exactly the same as that of the scalar field in the usual Minkowski space, and the structure of the Hilbert space is unchanged across the horizon. 
In this sense, {\it the equivalence principle is still valid quantum mechanically} 
around the eternal black hole. 
\subsection{Relation between the quantization by a freely falling  observer and the stationary observers  in \WF and \WP}
As the argument for \WP is the same as that for \WF  we will concentrate on 
 the case of a \WF observer. 

In the approximately flat region near the horizon, the scalar field $\phi^F$ 
 can be expanded in modes simply like 
\begin{align}
\phi^F(t_F, z_F, \Omega) &= \sum_{l,m} \int_{-\infty}^\infty d\omega N^F_\omega \left( e^{-i\omega t_F} H^{(2)}_{i\omega}( k_l z_F) Y_{lm}(\Omega) a^F_{lm \omega} + {\rm h.c.} \right) \period
\end{align}
Contrary to the case of the Minkowski frame discussed above, the extraction 
 of the modes in \WF is straightforward. This is because the equal-time spacelike 
 lines near the horizon are entirely contained in the approximately flat region, as 
 is clear from the Figure\  \ref{flat-twosided}. Therefore orthogonality relations 
 for the Hankel functions can be used just as in the case of the entire Minkowski space, described in Sec.~2.5. 

This means that  in the flat region around the horizon,  the number of modes 
is the same between a \WF observer and the FFO.  More explicitly,  the 
 relations between the mode operators are just as in the case of 
 the Minkowski space (with $k$ replaced by $lm$ taken for granted). This is particularly clear in the rapidity representation 
given in  (\ref{FtransFM}) and (\ref{FtransMF}).  Since $|k|\cosh u$ in the definition 
 (\ref{defaMku}) is the energy $E$, the operator $a^M_{ku}$ is  Lorentz invariant  as seen from (\ref{LtransaM}), which means ${a'}^M_{ku'} = a^M_{ku}$, where  $u' = u+\xi$, where $\xi$ is the rapidity for the boost. On the other hand,  the invariance of $\phi^F$ and $z_F$ and $t_F \rightarrow t_F + \xi$ under the Lorentz transformation in (\ref{expphiF})  dictates that we should have  ${a'}^F_{k\omega} =e^{i\omega \xi} a^F_{k\omega}$.  With the angular-momentum indices explicitly implemented, we have, under the Lorentz transformation, 
\begin{align}
{a'}^M_{lm,u+\xi} = a^M_{lmu}\comma \qquad {a'}^F_{lm\omega} =e^{i\omega \xi} a^F_{lm\omega} \period
\end{align}
It is easy to see that this is indeed compatible with the Fourier transform relation (\ref{FtransFM}) with $k$ replaced by $lm$. 
\subsection{Relation between the quantization by a freely falling  observer and the stationalry observers in \WR and \WL}
We now come to the more difficult situation of the quantization from the viewpoints of the \WR (and \WL) observer in the flat region. Expansion of the general 
solution into modes using the $K_{i\omega}$ functions is the same as in the Minkowski space and  the canonical quantization condition for the fields can be imposed.  But the extraction of the mode operators $a^M_{lm \omega}, a^{M\dagger}_{lm \omega}$ and verifying that they satisfy the canonical commutation relations cannot be performed explicitly. In contrast to the case of \WR discussed in the previous subsection,  there is no set of spacelike lines covering \WF, such as described by $t_R =$constant, that are contained entirely within the 
 flat region, and we cannot use the flat space orthogonality relation to 
express the mode operators in terms of the fields.  

What we can check easily is that,   if we assume the canonical form of the commutation relations for the modes as in the flat space, then by using the completeness relation, {\it which is a local relation},    the correct canonical commutation relations for the fields are reproduced.  This shows the self-consistency of the assumption. 

Actually, we can argue that the relation between $a^R$ and $a^M$ should be 
 the same as in the full Minkowski space in the following two ways:
\nxt
(i)\ In the flat region, using completeness, we can re-expand the field $\phi^R$, 
which contains $a^R$ and $a^{R\dagger}$ in terms of the plane waves, 
\ie in terms of the modes of $\phi^M$. In this calculation, we only need to use integration over the momenta. Now,  as described in Sec.~3, we can use the orthogonality of the plane waves along the contour which by a suitable Lorentz transformation is brought  within the flat region extending to infinity near the horizon, and extract the
 $a^M$ modes. Along such a line, we can relate the $a^M$ with the $a^R$ as in 
 the full Minkowski space. Then Lorentz transforming back this relation, we should 
 be able to express the $a^R$ in terms of the  $a^M$ in any flat region around the horizon. 
\nxt
(ii)\ Another argument goes as follows: For simplicity, consider the case where  we try to use the orthogonality integral along the spacelike straight line at $t_M=0$ extending from  $x^1=-\infty$ to $x^1=\infty$.  This  passes both  \WL and \WR, and only a portion of the contour is within the flat region. Outside the flat region, the 
eigenfunctions $f_{k_l \omega}(z_{R,L})$ satisfying the equation of motion starts to differ {\it continuously} from the modified Bessel functions $K_{i\omega}(k_lz_{R,L})$.   But since the differential equation expressing the equation of motion 
 does not acquire any new singularities,  one expects that such deformed eigenfunctions continue to satisfy appropriate forms of orthogonality relations. Then, using them, one can extract  the $a^{R,L}$ from the fields and compute the commutation relations among them. These relations should reduce (continuously) to the usual commutation  relations in the flat region, as they must lead, using the completeness  relation, to the correct canonical commutation relations for the fields expandable  in terms of the modified Bessel functions in such region. 

These arguments indicate that, as far as the flat region near the 
 horizon is concerned,  the relations between the modes for the FFO and the  observers in various Rindler frames should be the same as those already exhibited 
in Sec.~2 for the fully flat case, with the replacement of the linear momentum label $k$ by the angular momentum label $lm$. 
\section{Quantization in a  Vaidya model of a physical black hole by various 
 observers}
Black holes of more physical interest are the ones formed by a collapse of matter as actually occurs in nature. They are ``one-sided" and have rather different 
spacetime structures  compared with the two-sided eternal black holes discussed 
 in the previous section. 

In this section, we investigate how the observers  in various frames quantize
 a massless scalar field 
in the simplest model of \SS black hole of such a type,  namely  the so-called Vaidya spacetime\cite{Vaidya1,Vaidya2,VAIDYA1953}\footnote{For a review, see 
for example \cite{Griffiths-Podolsky}.},  created  by the collapse of a thin  spherical shell of matter at the speed of light, often referred to as a shock wave. 
 
\subsection{Vaidya model of a physical black hole and the effect of  the shock wave on the field as a boundary condition}
\subsubsection{Vaidya model of a physical black hole}
\begin{figure}[h]
\begin{center}
  \includegraphics[width=5cm]{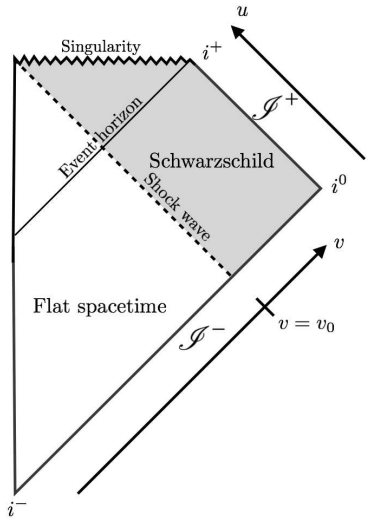}
  \caption{ Penrose diagram of the simplest Vaidya spacetime. It consists of two parts. One is the flat region  inside the  matter shell ($v<v_0$) shown in white. The other is the 
Schwarzschild spacetime outside the matter shell ($v>v_0$) shown in gray.\label{Vaidya}}
 \end{center}
\end{figure}

Let us begin by recalling the basics of such a Vaidya spacetime. 
After a black hole is formed by the spherical collapse, by the Birkohoff's theorem, the metric outside the horizon is always that of the \SS black hole. On the other hand, for the simplest situation above, the metric inside is isomorphic to a part of the flat Minkowski space. Thus the Penrose diagram of the entire spacetime 
 is obtained by gluing these two types of geometries along the light-like line 
 representing the falling shell, as shown in Figure\  \ref{Vaidya}. 

The Vaidya metric is a solution of the Einstein equation
\ba
R_{\mu\nu}-\f{1}{2}g_{\mu\nu}R=8\pi  T_{\mu\nu}\comma 
\ea
 with the energy momentum tensor
\ba
T_{vv}=\f{M}{4\pi r^2}\d(v-v_0).
\ea
The delta function at $v=v_0$ represents a shock wave induced by the matter collapsing along the lightlike direction $u$.
The metric of the Vaidya spacetime is described by
\ba
ds^2=-\left(1-\f{2m(v)}{r}\right)dv^2+2drdv+r^2d\Omega^2\comma 
\ea
where
\begin{align}
\bracetwocases{m(v)=0}{\mbox{for}\  v<v_0}{m(v)=M}{\mbox{for}\  v>v_0}
\period
\end{align}
The metric above consists of two parts, one of which corresponds to the region inside the shock wave $v<v_0$ and the other describes the outside, \ie the region $v>v_0$.
The solution  inside the shock wave is actually a flat spacetime described by 
\bal
ds^2&=-dv^2+2drdv+r^2d\Omega^2
=-dt^2+dr^2+r^2d\Omega^2\comma \label{MinkVai}
\eal
where $v=t+r$ is a lightcone coordinate. This is expected from the spherical symmetry of the matter shell. 
The solution for the region $v>v_0$ is the Schwarzschild black hole in the Eddington-Finkelstein coordinate 
\bal
ds^2=-\left(1-\f{2M}{r}\right)dv^2+2drdv+r^2d\Omega^2\comma 
\eal
as dictated by  the Birkohoff's theorem.
This metric can be transformed into the  Schwarzschild form 
\bal
ds^2=-\left(1-\f{2M}{r}\right)d\tilde{t}^2+\f{dr^2}{1-\f{2M}{r}}+r^2d\Omega^2 
\eal
by the coordinate transformation $v=\tilde{t}+r^*$. Here $\tilde{t}$ is the Schwarzschild time  and $r^*$ is the tortoise coordinate which is defined by 
\ba
r^*=\int \f{dr}{1-\f{2M}{r}}=r+2M\ln \left(\f{r}{2M}-1\right)\period
\ea
Notice that the two time coordinates,  $t$ inside the shell and $\tilde{t}$ outside, are different.
They are related as 
\bal
t+r&=\tilde{t}+r^*\qquad  {\rm at }\ v=v_0 \comma \no
\LR\quad \tilde{t}&=t-2M\ln \left(\f{v_0-t}{2M}-1\right)\period
\eal
In the subsequent sections, we mainly fucus on the special limit of the Vaidya spacetime,  for which the collapse of the matter has taken place very long time 
 ago so that the flat space region in Figure 4.1 is almost negligible.  This is 
  for the technical simplicity, as will be explained  more explicitly 
 in subsection 4.2.1. below Eq. (4.13).
\subsubsection{Effect of the shock wave on the scalar field as a regularized boundary condition}
In order to be able to study the quantization of a scalar field in an explicit manner, in what follows we shall (i) make a reasonable assumption about the effect of the 
 matter shock wave on the field, and (ii) implement it in a well-defined way by 
making a regularization which replaces the lightlike trajectory by a slightly timelike one. 

As for (i), since we focus on the \SS region outside of the locus of the 
shock wave, the effect  of the shock wave on  the scalar field $\phi(x)$  should be 
taken into account by an imposition of {\it an effective  boundary condition} on $\phi(x)$ along the trajectory of the shock wave, which must be  consistent with the bulk equation of motion.  Such boundary conditions are either Dirichlet or Neumann\footnote{Actually there is another  possibility that  the scalar field does not interact with the shock wave. In such a case, one needs to smoothly connect the solutions in the different spacetimes 
inside and outside the matter shell and in $1+3$ dimensions this is a difficult 
 task. For this reason, in this work we will not consider such a non-interacting case. }. 
This depends on the nature of the interaction between the shock wave and the 
scalar field,  and  for definiteness in this work we adopt the 
 Dirichlet condition  and demand that  $\phi(x)$ vanishes\footnote{For a massless scalar, by using the invariance of the action under a constant shift, we can do so without loss of generality.}  along the boundary\footnote{
 The boundary condition we introduce here should not be confused with the one considered  in the so-called moving mirror model in the two-dimensional gravity theory discussed in the literature ( see,  for example,  sec.~4.3 and 4.4 of \cite{Birrel-Davies}). In the moving mirror model,  the boundary condition is  imposed on the field at the origin $r=0$ of the Minkowski spacetime  (\ref{MinkVai}) inside the matter shock wave of the Vaidya spacetime, and  the field  outside the shock wave ($v>v_0$)  is smoothly connected to the one  inside ($v<v_0$).  On the other hand, 
in our treatment the  interaction of the shock wave and the field is represented by an effective boundary condition  imposed along  the shock wave at  $v=v_0$. }.

Next, let us elaborate on  the point (ii). 
If we take the boundary to be strictly  lightlike, \ie along $t_M = -x^1$, 
there is a complication for the spherical mode with zero angular momentum,  for which  $k_l=0$. Thus this component of the scalar field becomes massless 
in two dimensions and the future-directed massless field satisfying the Dirichlet condition along the lightlike line above  can only be right-moving and hence chiral. As is well known, quantization of a chiral scalar in two dimensions is notoriously 
troublesome and we would like to avoid it. 
 A physically natural regularization is to  endow the falling matter with an infinitesimal mass so that the trajectory is slightly timelike. Then the boundary condition can be treated in a non-singular manner by the standard canonical quantization procedure. 

Another advantage  of  such a regularization is the following. As it will 
 become evident, the effect of the boundary condition on the quantization 
can easily be taken into account in the frame of FFO moving in the direction 
 of the shock wave. When this direction is slightly timelike,  we can change 
it  by a Lorentz transformation into the case for a general FFO moving with any 
 velocity.  On the other hand, even if we could manage to treat the case of the strictly lightlike shock wave and an FFO moving along such a direction, we cannot relate such an observer by a Lorentz boost to a general FFO moving with a finite velocity.  
\subsection{Quantization of the scalar field with a boundary condition by a freely falling observer } 
In this section, we explicitly perform the quantization of a scalar field with the boundary condition imposed along a slightly timelike line from the point of view of  FFO's traversing the horizon with various velocities.  
\subsubsection{Three useful coordinate frames  and the imposition of a boundary condition }
\begin{figure}[h]
\begin{center}
  \includegraphics[width=5cm]{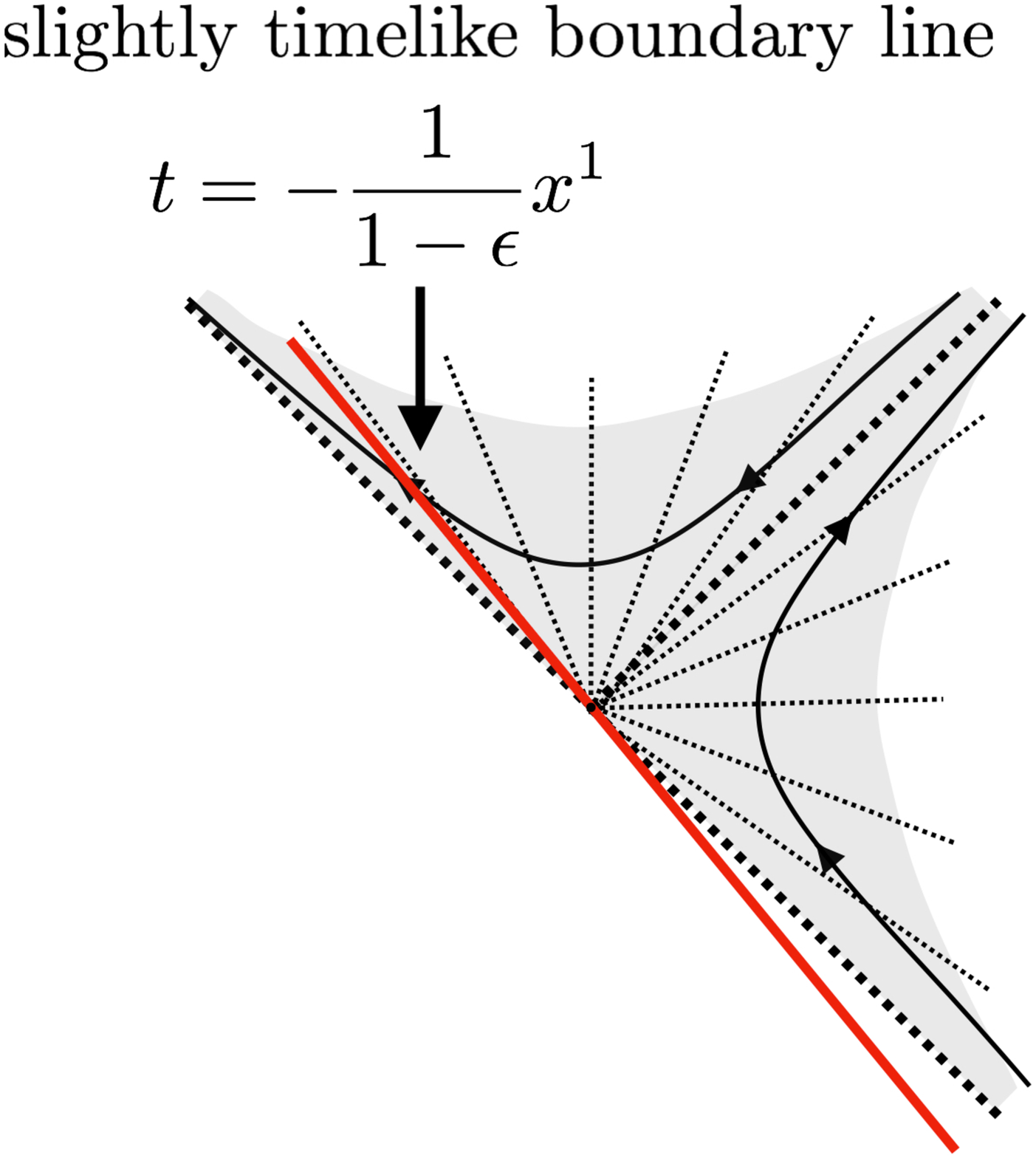}
\caption{Slightly timelike boundary line (shown in red).\label{boundaryline}}
 \end{center}
\end{figure}

In what follows, we will concentrate on the flat two dimensional portion 
in $\mathbb{R}^{1,1}$ and introduce three flat coordinates related 
 by Lorentz transformations. One is the canonical coordinates $(t, x^1)$, 
(where we use  $t$  for $t_M$ for simplicity in this subsection) for which $t$ and $x^1$ axes respectively run vertically and horizontally. The second is the coordinates $(\that, \xonehat)$, where the $\that$ axis runs almost lightlike but  slightly timelike direction. To go from $(t,x^1)$ to $(\that, \xonehat)$, we make a large Lorentz transformation of the form 
\begin{align}
\vecii{\that}{\xonehat} &= \Lam_\ep \vecii{t}{x^1} \comma \qquad 
\Lam_\ep = \gahat  \matrixii{1}{\behat}{\behat}{1} \comma \label{Lamep}
 \\
\behat &= 1-\ep \comma \qquad \gahat = {1\over \sqrt{1-\hat{\beta}^2}} \simeq {1\over \sqrt{2\ep}} \comma 
\end{align}
where  $\ep(>0)$ is an infinitesimal parameter. Thus, the explicit 
 transformations are 
\begin{align}
\that &= {1\over \sqrt{2\ep}}( (1-\ep) x^1+ t) \comma \qquad 
\xonehat = {1\over \sqrt{2\ep}}( (1-\ep)t + x^1)\period
\end{align}
We shall  take the boundary line to be the one expressed by  (see Figure\  \ref{boundaryline})
\begin{align}
\xonehat &= 0 \quad \Leftrightarrow \quad t = -{1\over 1-\ep}x^1 \period 
\end{align}
and demand that $\phi(\xonehat =0)=0$. 

\noindent
We should  remark that this 
 corresponds to the case where the shell of matter  collapses 
along the line for which the so-called  tortoise light-cone coordinate 
$v^\ast = t + r^\ast$, where $r^\ast \equiv \int dr/(1-(2M/r))$, 
takes a very large negative constant value compared to the scale of the \SS radius $2M$. An example is the case where $t \rightarrow -\infty$.   The reason for this rather special choice is strictly for technical convenienece: Such a trajectory is contained entirely within the region where 
 the flat space approximation is valid and hence the computations can be done 
explicitly and reliably.  We can deal with a general FFO later by making a Loretz transformation, as explained below. 
As far as the qualitative conclusions are concerned, 
a  constant shift in $v^\ast$  should not affect the quantum property 
 of the scalar field drastically,  because the Dirichlet 
condition $\phi=0$ along the matter trajectory,  as we shall see, 
 will act just like a reflecting wall for the scalar field.  

The third set of coordinates to be introduced is  $(\ttil, \xonetil)$, where $\ttil$ is the axis along which 
 a FFO travels with a general velocity $\betil$, which can be positive or negative.  He/she quantizes the scalar field with $\ttil$ as the time. This frame is defined to be related to the canonical frame by a 
 Lorentz transformation 
\begin{align}
\vecii{\ttil}{\xonetil} &= \Lamtil  \vecii{t}{x^1} = \gatil \matrixii{1}{\betil}{\betil}{1} \vecii{t}{x^1} \comma \label{Lamtil}\\
\ttil &= \gatil  (t + \betil x^1) \comma \qquad 
\xonetil = \gatil(x^1 + \betil  t) \period  
\end{align}
It will  also be convenient to relate 
the frame $(\ttil, \xonetil)$ with $(\that, \xonehat)$ directly. We shall write this 
 relation as 
\begin{align}
\vecii{\that}{\xonehat} &= \Lam \vecii{\ttil}{\xonetil} \comma \qquad 
\Lam =\ga \matrixii{1}{\beta}{\beta}{1} \comma \label{Lorentzhattil}\\
\that &= \ga(\ttil + \beta \xonetil) \comma \qquad \xonehat = \ga (\xonetil + \beta \ttil) \comma 
\end{align}
where $\Lam$, in terms of the Lorentz transformations  already introduced in (\ref{Lamep}) and (\ref{Lamtil}),  is the combination $\Lam = \Lam_\ep \Lamtil^{-1}$. For infinitesimal $\ep$, the  relations between $(\beta, \ga)$ and $(\betil, \gatil)$ can be approximated as 
\begin{align}
\beta&  \simeq 1-{1+\betil \over 1-\betil} \ep \comma 
\qquad 
\ga  \simeq {\gatil (1-\betil) \over \sqrt{2\ep}}   \period
\end{align}
%

\subsubsection{Quantization of the scalar field satisfying the boundary condition 
by a FFO  in the  $(\that, \xonehat)$ frame }
We begin with the quantization in the $(\that, \xonehat)$ frame. 
Since the boundary condition is imposed along the line $\xonehat=0$, obviously 
the quantization is easiest in such a frame.  More importantly, 
(regularizing the scalar field to vanish at infinity, as usual)  the trajectory 
of the FFO along the $\that$ axis is contained in  the region where the flat space 
 approximation is valid. Therefore, the following procedure is justified. 

The quantized 
scalar field which vanishes for $\xonehat=0$ is obtained by simply imposing such a 
 condition on the one without the boundary condition, namely the expression given in (\ref{phiYlm}) with unhatted variables replaced by hatted ones. Explicitly, 
 setting the coefficient of $\cos \ponehat\xonehat$, which does not vanish 
 for $\xonehat=0$,  in the expansion $\exp(i\ponehat \xonehat)=\cos \ponehat\xonehat + i \sin \ponehat\xonehat$, we obtain 
 the relation between the modes 
\begin{align}
\ahat_{lm, -\ponehat} 
 = -\ahat_{lm \ponehat} \period \label{relahat}
\end{align}
This clearly shows that the mode with negative $\ponehat$ is directly related to the mode with positive $\ponehat$ and hence {\it the number of independent  modes 
 is halved} by the imposition of the boundary condition.  
 Intuitively  the wave as seen in the hatted frame is reflected perpenticularly by the boundary line. 
Therefore, the scalar field as quantized by a FFO moving in the direction of the $\that$ axis and its conjugate momentum are of the form 
\begin{align}
\phi(\that,\xonehat, \Omega) &= \sum_{l=0}^\infty \sum_{m=-l}^l  \int_{-\infty}^\infty {d\ponehat\over \sqrt{4\pi E_{k_l\ponehat}}}  \left( e^{ -iE_{k_l\ponehat} \that} Y_{lm}(\Omega) (\ahat_{ lm \ponehat}-\ahat_{lm -\ponehat})  + {\rm h.c.} \right) \sin \ponehat\xonehat \period  \label{phihat} \\
\pihat(\that,\xonehat, \Omega) &= \del_{\that} \phi(\that,\xonehat, \Omega)  \nonumber \\
&= -i\sum_{l=0}^\infty \sum_{m=-l}^l  \int_{-\infty}^\infty {d\ponehat\s{E_{k_l\ponehat}} \over \sqrt{4\pi}}  \left( e^{ -iE_{k_l\ponehat} \that} Y_{lm}(\Omega) (\ahat_{ lm \ponehat}-\ahat_{lm -\ponehat}) - {\rm h.c.} \right) \sin \ponehat\xonehat \comma 
\end{align}
where the notation $\pihat$ reminds us  that the conjugate momentum in this frame  is defined using the derivative with respect to $\that$. 
By using the commutation relation $\com{\ahat_{lm \ponehat}}{\ahat^\dagger_{l'm'\qonehat}} = \delta_{ll'}\delta_{mm'} \delta(\ponehat-\qonehat)$ 
 and the formula $\int_0^\infty d\ponehat \sin\ponehat \xonehat \sin \ponehat \yonehat = (\pi/2)( \delta (\xonehat -\yonehat) -\delta(\xonehat + \yonehat))$, one can verify  that  the commutator of  the conjugate fields takes  the canonical 
form\footnote{Since $\xonehat$ and $\yonehat$ are both positive, we can discard $ -\delta(\xonehat + \yonehat)$.} 
\begin{align}
[\pihat(\that, \xonehat, \Omega), \phi(\that, \yonehat, \Omega')]   
= -i \delta(\xonehat -\yonehat) 
  \delta(\cos\theta-\cos\theta') \delta(\varphi -\varphi') \period  \label{comhatN}
\end{align}
%

\subsubsection{Comparison of Hilbert space of  FFO and the genuine Minkowski Hilbert space}
Let us define for convenience the following  combinations of the mode operators:
\begin{align}
\ahat^{\pm}_{lm \ponehat} &\equiv \ahat_{lm\ponehat} \pm \ahat_{lm, -\ponehat} \comma \label{ahatpm}\\
\ahat^{\pm \dagger}_{lm \ponehat} &\equiv \ahat^\dagger_{lm\ponehat} \pm \ahat^\dagger_{lm, -\ponehat}
\end{align}
Cearly, the  operators with plus and minus superscripts commute with each other. 
Then, from the discussion above, the Hilbert space $\calH_{\rm FFO}$ of the FFO  in the  $(\that, \xonehat)$ frame is constructed upon the vacuum $\ket{\hat{0}}_{-}$,  defined by 
\begin{align}
\ahat^{-}_{lm \ponehat}\ket{\hat{0}}_{-}& =0  \comma 
\end{align}
by applying the operators $\ahat^{- \dagger}_{lm \ponehat}$ repeatedly.  
In contrast, 
 the genuine Minkowski Hilbert space $\calH_{\rm M}$ is built  upon 
the vacuum $\ket{0}_M$, which is defined to be annihilated by $\ahat_{lm \ponehat}$  for all values of $l, m$ and $\ponehat$, by the (repeated) applications of $\ahat^\dagger_{lm \ponehat}$'s . 
This means that $\calH_{\rm M}$ can be  written as the tensor product 
\begin{align}
\calH_{\rm M} &= \calH^{-} \otimes \calH^{+} \comma 
\end{align}
where $\calH^{-}$ stands for $\calH_{\rm FFO}$ and the other half $\calH^{+}$ is constructed in the entirely similar manner as for $\calH^-$,  using the $a^+$ type operators. From the point of view of FFO, $\calH^+$ is unphysical but it is needed for the  construction of  $\calH_{\rm M}$. Note that this decomposition is 
completely {\it different} from the left-right decomposition $\calH_{\rm M} = \calH_{\rm W_L}  \otimes \calH_{\rm W_R}$. 

This structure will be important in the discussion of the Unruh-like effect 
 near the horizon of a physical \SS black hole, to be discussed  in Sec.~5. 
\subsubsection{Quantization by a FFO in a general frame  $(\ttil, \xonetil)$  with the boundary condition  \label{sec:onesidedFFO}}
We now consider the quantization by a FFO in a general frame $(\ttil, \xonetil)$ 
 with the {\it same boundary condition} $\xonehat =0$ along the shock wave.  Since this boundary condition is simplest to describe  in the $(\that, \xonehat)$ frame, 
 the most efficient way to quantize in the $(\ttil, \xonetil)$ frame with such a boundary condition is to express  the new conjugate momentum 
 $\pitil\equiv \del_{\ttil} \phi$ in terms of the quantities  in the $(\that, \xonehat)$ frame by applying the relation 
\begin{align}
{\del  \over \del \ttil} &= {\del \that \over \del \ttil} {\del \over \del \that}
 +  {\del \xonehat \over \del \ttil} {\del \over \del \xonehat} 
= \ga {\del \over \del \that} + \ga\beta  {\del \over \del \xonehat}
\label{delttil}
\end{align}
to  $\phi(\that, \xonehat, \Omega)$,  which we already have. Because $\del_{\ttil}$ contains the spatial derivative $\del_{\xonehat}$ as well, this leads to an important non-trivial change in the conjugate momentum, however.  
The result is\footnote{To get the explicit form of $\pitil(\ttil, \xonetil, \Omega)$, 
 we must rewrite all the hatted quantities in terms of the tilded ones obtained by 
 the Lorentz transformation (\ref{Lorentzhattil}). This produces a rather 
 involved expression. The procedure adopted here can avoid this complication.}
\begin{align}
\pitil(\ttil,\xonetil, \Omega) &=-i\widetilde{\calN} \sum_{l=0}^\infty \sum_{m=-l}^l  \int_0^\infty {d\ponehat\over \sqrt{4\pi E_{k_l\ponehat}}} \biggl[  \left((-i\ga\Ehat_{k_l  \ponehat}) e^{ -iE_{k_l\ponehat} \that} Y_{lm}(\Omega) a_{ lm \ponehat} + {\rm h.c.} \right) \sin \ponehat\xonehat \nonumber\\
&+ \ga\beta \ponehat \left( e^{ -iE_{k_l\ponehat} \that} Y_{lm}(\Omega) a_{ lm \ponehat} + {\rm h.c.} \right) \cos\ponehat \xonehat \biggr] \comma 
\label{pitilfield}
\end{align}
where we have denoted the normalization constant in this frame as $\widetilde{\calN}$. 

Now let us compute the  {\it equal $\ttil$} canonical commutation relation 
between $\pitil$ and $\phitil$.  In this process we need to take into account  the following two points: \nxt
(i)\ \ Since $\ttil = \ga(\that -\beta \xonehat)$, 
equal $\ttil$ is equivalent to equal $\that -\beta \xonehat$. In other words, if we 
 denote the hatted time that appears in  $\pitil$ given by (\ref{pitilfield}) 
by $\that$ and the one in  $\phi$ by 
 $\that'$, then the equal $\ttil$ can be expressed as 
$\that -\beta \xonehat = \that' -\beta \yonehat$, where $\xonehat$ and 
 $\yonehat$ are the spatial coodinates that appear in  $\pitil$ and $\phi$ respectively. 
Therefore we have the important relation 
\begin{align}
\that -\that' = -\beta (\xonehat -\yonehat)  \quad \mbox{at equal $\ttil$. } \label{equalttil1}
\end{align}
 (ii)\ \ The second point to keep in mind is that from the Lorentz transformation 
we easily find 
\begin{align}
\xonehat -\yonehat &= \ga(\xonetil -\yonetil) \quad \mbox{at equal $\ttil$, } 
\label{equalttil2}
\end{align}
so that the difference in the spatial coodinates  in the hatted frame can be 
 rewritten as the rescaled  difference in the tilded frame. 

With these facts in mind, the equal $\ttil$ commutator $\com{\pitil}{\phi}$ is 
 given by 
\begin{align}
&\com{\tilde{\pi}(\ttil, \xonetil, \Omega)}{\phi(\ttil, \yonetil, \Omega')} \nonumber\\
&\qquad =-i\calNtil^2\sum_{l,m}\sum_{l'm'} 
\int_0^\infty  d\ponehat \int_0^\infty d\qonehat{ Y_{lm}(\Omega)Y^\ast_{l'm'}(\Omega')  \over 4\pi  \sqrt{\Ehat_{k_l\ponehat} \Ehat_{k_{l'}\qonehat}}} \nonumber\\
&\qquad  \biggl[(-i\Ehat_{k_l\ponehat} \ga)\left( e^{-i\Ehat_{k_l\ponehat} \that +i\Ehat_{k_{l'}\qonehat} \that'} 
+{\rm h.c.} \right)\com{\ahat_{lm \ponehat}}{\ahat^\dagger_{l'm'\qonehat}}\sin\ponehat\xonehat\, \sin\qonehat \yonehat 
 \nonumber\\
&\qquad  +\ga\beta \ponehat  \left( e^{-i\Ehat_{k_l\ponehat} \that +i\Ehat_{k_{l'}\qonehat} \that'} 
-{\rm h.c.} \right)
 \com{\ahat_{lm\ponehat}}{\ahat^\dagger_{l'm'\qonehat}} 
\cos\ponehat\xonehat\, \sin\qonehat \yonehat \biggr] \nonumber\\
&\qquad = C_1+C_2  \comma  \label{compitilphitil}
\end{align}
where 
\begin{align}
C_1 &= -\gamma \calNtil^2 \sum_{l=0}^\infty \sum_{m=-l}^l \int_0^\infty {d\ponehat \over 4\pi }  Y_{lm}(\Omega)Y^\ast_{lm}(\Omega') 
\left(e^{i\beta\Ehat_{k_l \ponehat} (\xonehat-\yonehat)} +{\rm h.c.}\right)\sin\ponehat\xonehat\, \sin\ponehat \yonehat \comma \label{Cone} \\
C_2 &= -i\gamma\beta\calNtil^2 \sum_{l=0}^\infty \sum_{m=-l}^l \int_0^\infty {d\ponehat \over 4\pi \Ehat_{k_l\ponehat} }  Y_{lm}(\Omega)Y^\ast_{lm}(\Omega') 
 \left(e^{i\beta\Ehat_{k_l \ponehat} (\xonehat-\yonehat)} -{\rm h.c.}\right)\cos\ponehat\xonehat\, \sin\ponehat \yonehat
 \period \label{Ctwo}
\end{align}
Note that for the the exponents  involving $\Ehat_{k_l \ponehat}$ we used the relation (\ref{equalttil1}).  The sum over $m$ can be done by 
 the well-known addition theorem for $Y_{lm}$ namely
\begin{align}
\sum_{m=-l}^lY_{lm}(\Omega) Y^\ast_{lm}(\Omega') = {2l+1 \over 4\pi} P_l(\hat{n}\cdot \hat{n'}) \comma  \label{addthm}
\end{align}
where $P_l(x)$ is the Legendre polynomial  and $\hat{n} = (\sin\theta \cos\varphi, \sin\theta \sin\varphi, \cos\theta)$ denotes the unit vector corresponding to the pair of angles $\Omega(\theta, \varphi)$. 
Furthermore, the integral over $\ponehat$, after rewriting the product of 
 trigonometric functions into a sum of them,  can also be performed using  the  formulas 3.961 of \cite{GR} (with the aid of the relation $\del_z K_0(z) = -K_1(z)$) 
\begin{align}
&\int_0^\infty e^{-b\sqrt{k^2+x^2}}\cos ax dx = {b k \over \sqrt{  a^2+b^2}} K_1(k \sqrt{ a^2+b^2})\comma   \label{GRintone}\\
&\int_0^\infty {x \over \sqrt{k^2+x^2}} e^{-b\sqrt{k^2+x^2}}\sin ax dx = {a k \over \sqrt{ a^2+b^2}} K_1(k \sqrt{a^2+b^2})\comma \label{GRinttwo}
\end{align}
where $K_1(z)$ is the Macdonald function (\ie one of the modified Bessel functions) of order 1 and the both formulas are
 valid for ${\rm Re}\, b >0 \comma 
{\rm Re}\, k>0 $.   In particular, the convergence condition $b>0$ is important 
 since in our case, $b=\pm i \beta (\xonehat -\yonehat)$ and are 
 pure imaginary. Thus, we must regularize them by introducing an infinitesimal 
 positive parameter $\eta>0$ and replace  $b$ by 
\begin{align}
b_- &\equiv -i\beta (\xonehat -\yonehat+i\eta ) \comma \label{bm}\\
b_+ &\equiv +i\beta (\xonehat -\yonehat-i\eta ) \period  \label{bp}
\end{align}

Since the rest of the calculations are somewhat tedious but more or less straightforward, we shall describe  some intermediate steps in the Appendix \ref{app:generalFFO} and 
only list here the important  structures that one will encounter as one proceeds. 
\begin{itemize}
	\item The  terms which contain  $\cos(\xonehat + \yonehat)$ and 
$\sin(\xonehat + \yonehat)$ produced from the product of sines and cosines 
turn out to cancel  completely, because  $\xonehat + \yonehat$ is positive and generically finite  and  the regulator  $\eta$ after performing the integrals can be ignored compared to them. 
\item On the other hand, for the terms containing the difference $\xonehat -\yonehat$, there are two cases. When the difference is finite and hence $\eta$ in $b_\pm$ can be 
 ignored,  all the terms cancel just as in the case above and hence  the commutator vanishes.  

In contrast, when the difference is of order $\eta$ or smaller, then 
 the contribution remains and becomes  proportional to the structure
$K_1(\al k_l \eta)$, with a finite constant $\al$. Now if we first make a cut-off 
 on the angular momentum $l$ so that $k_l=\sqrt{l(l+1)}/2M$ can be large but finite, then $\al k_l \eta \rightarrow 0$ as we send  $\eta \rightarrow 0$. Then, from the behavior of $K_1(z)$ for small $z$, \ie $K_1(z) \simeq 1/z$, we see that the contribution  diverges like $1/\eta$. Thus, we see that as $\xonehat -\yonehat \rightarrow 0$, the commutator diverges as we remove the regulator. 

Together, this is nothing but the behavior of the $\delta$-function $\delta(\xonehat-\yonehat)$ which is proportional to $\delta(\xonetil -\yonetil)$ due to  the relation (\ref{equalttil2}). 
\item In the other limit where $l$ becomes so large that $k_l \eta$ is large, then, 
 $K_1(z)$ damps like $\sim e^{-z}/\sqrt{z}$ and such a region does not contribute. This indicates that we can effectively replace $k_l$ by a large constant independent of $l$. 
\item Then, we are left with the sum over $l$, which produces the angular 
 $\delta$-functions in the manner 
\begin{align}
\sum_{l=0}^\infty {2l+1 \over 4\pi} P_l(\vec{n} \cdot \vec{n'}) 
&= \sum_{l=0}^\infty \sum_{m=-l}^l Y_{lm}(\Omega) Y^\ast_{lm}(\Omega') \nonumber\\
&= \delta(\cos\theta-\cos\theta') \delta(\varphi -\varphi') = {1\over 2\pi} \delta(\vec{n} \cdot \vec{n'} -1) \period
\end{align}
\end{itemize}

Combining, we find  that the commutator is proportional to 
 the desired product of $\delta$-functions and the quantization for an arbitrary 
 FFO with the boundary condition along $\xonehat=0$ in the vicinity of 
 the horizon is  achieved. 

Several remarks are in order:
\begin{itemize}
	\item Although the correct $\delta$-function structure for the canonical 
 commutation relation is confirmed, unfortunately we cannot compute the exact 
normalization constant because the relevant integrals and the infinite sum cannot be performed exactly.  This is regrettable since such a constant must become 
singular as we let the almost lightlike trajectory approach exactly lightlike, and 
 it would be interesting to see how this comes about.  
\item Nevertheless, the fact that the quantization for a general FFO with the boundary condition $\phi(\xonehat=0)=0$ can be carried out as we have shown shows that the number of modes that the FFO sees as he/she passes the horizon {\it does not change} and is naturally {\it half as many as for the case of the 
 two-sided black hole. }
\end{itemize}
\subsection{Quantization of the scalar field by the observer in the \WF frame } 
We now consider the quantization in the \WF frame with the same boundary 
 condition along the slightly time-like line, namely  $(1-\ep)t_M +x^1 =0$. 
Expressed in terms of the \WF variables (see (\ref{relMWF})), this becomes
\begin{align}
z_F (e^{t_F} -\ep \cosh t_F) =0 \period
\end{align}
Since $z_F$ need not vanish, we should set $e^{t_F} = \ep \cosh t_F$. 
This can be easily solved for $t_F$ as 
\begin{align}
t_F  \simeq \half \ln {\ep \over 2}\comma \label{boundarytF}
\end{align}
which is very large and negative.

Now we impose the vanishing condition for $\phi^F$ along this line. 
An advantageous feature of the \WF region is that such a line is, 
 practically,  contained entirely in the flat region. Therefore we can make use 
 of the expression in the flat spacetime and the boundary condition on the field 
$\phi^F(t_F, z_F, \Omega)$ can be written as 
\bal
\label{fieldexpansion}
 \phi^F(z_F,t_F,\Omega)=\sum_{l,m}\int_{-\infty}^{\infty} d\omega  N_\w^F\left(e^{-i\omega t_F} H^{(2)}_{i\omega}(k_lz_F)Y_{lm} (\Omega) a^F_{lm \omega} + \mbox{h.c.}\right)\, .
\eal
 Using the relation 
\ba
 H^{(2)}_{-i\w}=e^{\pi\w}H^{(2)}_{i\w}\, ,
\ea
 we can rewrite \eqref{fieldexpansion} as
\bal
 \phi^F(z_F,t_F,\Omega)& =\sum_{l,m}\int_{0}^{\infty} d\omega  \biggl(N_\w^F e^{-i\omega t_F}H^{(2)}_{i\omega}(k_l z_F)  Y_{lm}(\Omega)a^F_{lm \omega} \no&\qqq\qqq\qq +N_{-\w}^F e^{i\omega t_F}H^{(2)}_{-i\omega}(k_lz_F) Y_{lm}(\Omega) a^F_{lm-\omega}+ \mbox{h.c.}\biggl)\no & =\sum_{l,m}\int_{0}^{\infty} d\omega N_\w^F\left( H^{(2)}_{i\omega}(|k|z_F) Y_{lm}(\Omega)(e^{-i\omega t_F} a^F_{lm \omega} +e^{i\omega t_F} a^F_{lm -\omega})+ \mbox{h.c.}\right)\, .
\eal
 We now impose the boundary condition 
\ba
 \phi^F(z_F,t_F=t^{B}_F,\Omega)=0.\label{tzbc}
\ea
 where $t^{B}_F=\half \ln {\ep \over 2}$, then the modes should satisfy the following relations
\ba
 a^F_{lm \omega}=-e^{2i\w t_F^{B}}a^F_{lm -\omega }\, . \label{bbc}
\ea
 Notice that this argument is valid even when we take the value of $\ep$ very small.

 Thus, the conclusion is that, the boundary condition places relations \eqref{bbc} among  the modes and hence the number of 
 independent modes observed in the \WF frame is halved,  just like the ones in the FFO frame $\hat{a}_{l m,\hat{p}^1}$.
 
\subsection{Quantization of the scalar field by the oberserver in the \WR frame \label{sec:WRscalar} }
Finally, let us consider the quantization by the observer in the \WR frame. 

If we take the same special slightly timelike boundary line which goes through the 
 origin of the coordinate frame $(t_M, x^1)$, this line is outside of the region \WR\!\!. Therefore, there is no boundary condition to impose and the modes which exist
 for the two-sided case  are all present and independent.  

Although it is a valid argument, it certainly depends crucially on the special choice of the boundary line.  Therefore we should also consider the case where the boundary line is slightly shifted  to the positive $x^1$ direction  so that it passes inside \WR very close 
 to its lightlike boundary.  Explicitly, the boundary line is now taken to be along
\begin{align}
t=-{1\over 1-\ep} x^1 + \delta \comma  \label{BCWR}
\end{align}
where $\delta$ is a very small shift.  In this case, the imposition of the boundary condition is meaningful and the argument to follow is of more general validity.  

\begin{figure}[h]
\begin{center}
  \includegraphics[width=4cm]{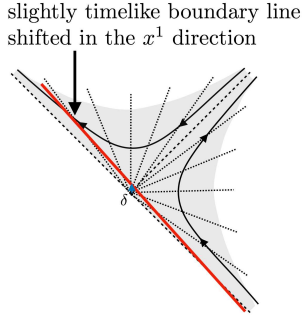}
\caption{A slightly timelike boundary line (shown in red), which is shifted infinitesimally in the positive $x^1$ direction  compared to Figure\  \ref{boundaryline}.\label{shiftedBC}}
 \end{center}
\end{figure}

As discussed in Sec.~3.5, the expansion of $\phi^R(z_R, t_R, \Omega)$ 
near the horizon is given by\footnote{Due to the use of the spherical harmonics 
 instead of the plane wave in the transverse directions, the normalization factor
 $N_\omega$ is slightly different from the one in (\ref{NRomega}).}
\begin{align}
\phi^R(z_R, t_R, \Omega) &= \int_0^\omega d\w N_\omega \sum_{l,m}Y_{lm}(\Omega)
\left( K_{i\omega}(k_l z_R) e^{ -i\omega t_R} a^R_{lm \omega } +{\rm h.c}
\right)\comma  \\
N_\omega &={ \sqrt{\sinh \pi \omega } \over \pi} \period \label{NRomegasph}
\end{align}
Since (\ref{BCWR}) can be rewritten as $x^+=z_R e^{t_R} = \ep t+\delta$, 
$z_R$ is small along the boundary line for finite $t_R$. Now to make use of 
 the form of the $K_{i\omega}(z)$ for small $z$, we make a cut-off for the 
 angular momentum $l$ and consider the states for which $k_l$ is bounded. 
Then, we can use the 
 behavior of $N_\omega K_{i\omega}(y)$ for small $y$, which is given by 
\begin{align}
N_\omega K_{i\omega}(y) &\sim{ \sqrt{\sinh \pi \omega } \over \pi} \times  2\sqrt{{\pi \over \omega \sinh \pi \omega}} 
\cos \left( \omega \ln {y \over 2} -{\rm arg}\, \Ga(i\omega) \right) \nonumber\\
&= {2\over \sqrt{\pi}} {1\over \sqrt{\omega}} \cos \left( \omega \ln {y \over 2} -{\rm arg}\, \Ga(i\omega) \right) \period
\end{align}
Note that this oscillates wildly as $y$ becomes small. 
$y$ here is $k_l z_R$ and as we send the regulator $\ep$ to zero, it becomes 
 of order $\delta$, which is very small.  
 Then,  just as in the case of the quantization in $W_F$,  by 
 imposing a cut-off for the $\omega$-integral, we can apply  the Riemann-Lebesgue  lemma  to conclude that the Fourier type integral tends to vanish and 
 the boundary condition $\phi^R=0$ is automatically satisfied along the matter trajectory.  

Therefore if we exclude the  the highly excited states,  the boundary condition does not impose any relations to the mode operators and the degrees of freedom remain the 
same as in the two-sided case.  For the energetic states, more exact computation 
 is needed to make definite statements. 
\section{Implications on the quantum equivalence principle, the firewall phenomenon and the Unruh effect}
Having analyzed and compared the quantization of a scalar field by different  natural observers in a concrete manner, we now consider the implications of our results. 
\subsection{Quantum equivalence principle and the firewall phenomenon}
One of the clear results is that the degrees of freedom of the modes that the 
 observer sees are in general different, both for the case of the two-sided eternal black hole and the more physical one-sided one.   Explicitly, the FFO and the \WF observers
 see the same number of modes,  while the observer in the \WR frame 
 finds half as many in the two-sided case. On the other hand, in the one-sided case, the number of modes which the FFO and the \WF observers
 see is both halved by the imposition of the boundary condition,  while the number of those seen by  the  \WR observers
 is the same as in the case without the boundary. Thus, in this case number of modes that  FFO, \WF and \WR observers see are identical.

The fact that the sizes of the quantum Hilbert spaces are halved for FFO and \WF observers in the one-sided case and \WR observers in both cases 
 is natural since such observers can only see a part of the spacetime due to 
 the presence of the horizons for them. 

Whether the equivalence principle holds quantum mechanically  is quite a different 
 question.  It asks whether  the FFO, upon crossing the ``horizon," which does not exist for him/her {\it classically}, sees extra or fewer degrees of freedom of 
 the {\it quantum} excitation modes of a field. Our explicit computation shows that,   for both the eternal and the physical black holes, {\it the quantum equivalence principle  holds naturally.} This is essentially due to the fact that no new boundary conditions  for the scalar field appear as seen by an FFO who goes through  the ``horizon". 

Needless to say, this conclusion is valid  under the assumption that the metric 
 of the interior of the \SS black hole is essentially given by the Vaidya-type 
metric. If the interior of the  black hole is such that it cannot be specified 
 just by the information of the metric, the conclusion may differ. 
However, as far as the classical \SS black holes produced by the collapse of 
  matter are concerned, our assumption is conservative and should be reasonable. 

Thus, for a large enough black hole that itself can be treated classically, with a small value of curvature at the horizon, 
 our explicit computations for the quantum effects of the massless scalar field 
 as seen by the  three types of observers should be reliable;  in particular 
 the freely falling observer does not  encounter the so-called firewall phenomenon. 
\subsection{Unruh-like effect near the horizon of a physical black hole \label{sec:Unruh}}
\baselineskip 3.5ex 
The Unruh effect\cite{Unruh} is the simplest example of the non-trivial quantum phenomena due to the difference of the vacua for the relatively accelerated observers. 
In the original case treated by Unruh, a Rindler  observer uniformly accelerated in the  flat  Minkowski space in the positive $x^1$ direction with accleration $a$ (confined to the wedge \WR\!\!), 
 sees in the Minkowski vacuum $\vacM$ a swarm of particles of energy  $ \omega$ with a number density distribution given by 
\begin{align}
{\langle N^R_\omega  \rangle \over {\rm Vol. }}&= {\braM a^{R\dagger}_\omega a^R_\omega \vacM \over {}_{\rm M}\langle 0 | 0\rangle_{\rm M}
{\rm Vol.} }  \propto  {1 \over e^{2\pi \omega/a} -1} \period 
\end{align}
Evidently, this coincides with the thermal distribution of bosons at temperature $a/2\pi$. In fact this computation is {\it truly thermal in nature} since $\ket{0}_M$ is 
an entangled state consisting of the states of \WL as well as of \WR\!\!, and 
one must take a trace over all the states of  \WL to obtain the distribution above. 

Although in this example the background is taken to be the flat spacetime to begin with, one might expect a similar phenomenon to be seen by a stationary observer just outside the horizon of a physical \SS black hole, since the spacetime there is well-approximated by  the right Rindler wedge of a flat Minkowski space. 

However, the analysis cannot be the same  for the following reasons. 
 First, there is no \WL region for the one-sided black hole and hence 
whatever distribution we obtain is {\it not} truly thermal in nature. It simply shows that the 
 concept of ``a particle" depends crucially on the vacuum state, even if it is 
 a pure state. 
The second reason  is the fact that, although the region of our interest 
 is locally a flat Minkowski space,  we must take into account 
 the effect of the boundary condition for the scalar field and its vacuum seen by the FFO,  who corresponds to the Minkowski observer in the Unruh setup.  As discussed in Sec.4.2.3, however, the vacuum $\ket{\hat{0}}_-$  for the  FFO  is {\it not} the genuine Minkowski vacuum.  A related difference is that,  as discussed in Sec.\ref{sec:WRscalar}, 
in our setup the scalar field in the \WR frame is not affected by the boundary 
 condition and hence the number of modes seen in that frame is the {\it same} as that of the FFO. This is in contrast to the case of the flat space, where the number of modes  for the \WR observer is {\it half} that of the Minkowski observer. 

Thus, the question of interest  is what the distribution of the \WR  particles is 
 in the vacuum of the FFO. To answer this, we must express 
 the mode operators  $a^R_{lm \w}$ and their  conjugates of   the \WR observer  in terms of the field  $\phi(\that, \xonehat, \Omega)$ and its modes of  a FFO\footnote{The reason for focusing on the FFO in the $(\that,\xonehat)$ frame 
 is simply that the effect of the boundary condition is the simplest in such a frame.
For the other frames of FFO, one can make a Lorentz transformation for the FFO, 
 {\it with  the boundary condition kept intact.}}.

Unfortunately, in general this computation cannot be performed accurately due to our lack of  knowledge  of the fields outside the approximately flat region. The required calculation  is of the form 
\begin{align}
a^R_{lm\w} &=i  \int d\varphi \int \sin\theta d\theta Y^\ast_{lm}(\Omega)  \int_0^\infty {dz_R \over z_R} f^\ast_{\omega, l} (t_R, z_R) 
 \overleftrightarrow{\del_{t_R}} \phi^{\rm FFO}(\that, \xonehat, \Omega)\comma \label{aRlmomega}
\end{align}
where $f_{\omega, l} (t_R, z_R)Y_{lm}(\Omega) $ is the solution of the equation of motion, corresponding to the mode $a^R_{lm\w}$,  
in the right Rindler wedge in the background of the \SS black hole. 
All we know is that this function takes the form 
$f_{\omega, l}(t_R, z_R) \simeq N_\omega K_{i\omega}(|k_l|z_R) e^{-i\omega t_R}$ in the approximately flat region where $z_R \lsim M$. Therefore, integration 
 over $z_R$, which extends outside such a region, cannot be performed explicitly. 

There is, however, a class of modes for which the computation can be performed 
sufficiently accurately using the function for  the flat space region.  These are the ones with large angular momentum $l$ such that $|k_l|M=\sqrt{l(l+1)}/2 \gg 1$. 
To see this, let us expand the scalar field  into 
 angular momentum eigenstates as $\phi(t, r, \Omega) = \sum_{l,m} \phi_{lm}(r) Y_{lm}(\Omega)$ and write down the equation of motion for 
 $\phi_{lm}(t,r)$ in the \SS metric. It is given by 
\begin{align}
0 &=-{ 1\over 1-{2M \over r}}\del_t^2 \phi_{lm}+ {2(r-M) \over r^2} \del_r \phi_{lm} 
 + \left( 1-{2M \over r}\right) \del_r^2 \phi_{lm}  -{l(l+1) \over r^2} \phi_{lm}
\period 
\label{radialeq}
\end{align}
Now we look at the region  $r \gg M$ where the flat space approximation 
 is no longer valid.  In such a region, writing $\phi_{lm}(t,r) = e^{\pm i \omega t}\phitil_{lm}(r)$, the equation for $\phitil_{lm}(r)$ simplifies to 
 \begin{align}
0 &= \left( \omega^2 + {2 \over r} \del_r + \del_r^2 -{l(l+1) \over r^2} \right) \phitil_{lm}(r) \period
\end{align}
The solution is well-known and is given, with a certain normalization, by 
\begin{align}
\phitil_{lm}(r) &= \sqrt{\omega \over r}\, J_{l+\half} (\omega r) \comma 
\end{align}
where $ J_{l+\half} (\omega r) $ is the Bessel function. 
Its asymptotic form  for large $l$ can be obtained from 
 the formula 10.19.1 of \cite{Olver} as 
\begin{align}
J_{l+\half}(\omega r) \sim {1\over \sqrt{(2l+1)\pi }} e^{-(l+\half) \ln {2l+1 \over e \omega r}} \period
\end{align}
This shows that for $l \gsim \omega r$, this expression is exponentially small in $l$ and contributes negligibly to the integral over $z_R$. 
    {\it Thus, for such modes with high angular momenta, we can effectively need only the function  in the flat region} and the computation is possible. 
Such a calculation is at the same time  self-consistent because $K_{i\omega}(|k_l|z_R)$ damps exponentially for large $|k_l|z_R$ and for large enough $l$ this quantity is  already large for $z_R \simeq M$. Therefore, contribution from $z_R \gsim M$ region is safely neglected. 

To perform the computation of (\ref{aRlmomega}), first  consider the projection of the angular part in (\ref{aRlmomega}) with the use of the  orthogonality of the spherical harmonics. Since $\phi^{\rm FFO}$ contains 
 both $Y_{lm}$ and $Y^\ast_{lm}$, the relevant formulas are  
 $\int d\Omega Y^\ast_{lm}(\Omega) Y_{l'm'}(\Omega) = \delta_{ll'}\delta_{mm'}$ and 
$\int d\Omega Y^\ast_{lm}(\Omega) Y^\ast_{l'm'}(\Omega) =(-1)^m \delta_{ll'}\delta_{m,-m'}$. (The second formula follows from the first by using the relation 
 $Y^\ast_{lm} = (-1)^m Y_{l,-m}$. ) 

Therefore, after the removal of 
 the angular part, what we need to compute is 
\begin{align}
a^R_{lm\omega} &= i \int_0^\infty {dz_R \over z_R} N_\omega K_{i\omega}(|k_l|z) 
 e^{i\omega t_R}  \overleftrightarrow{\del_{t_R}} 
 \int_{-\infty}^\infty {d\ponehat \over \sqrt{4\pi E_{k_l \ponehat}}}
\biggl( e^{-iE_{k_l \ponehat} \that}  
 \ahat^-_{lm \ponehat}\nonumber\\
&\hspace{5cm} + (-1)^m  e^{iE_{k_l \ponehat} \that} 
\ahat^{-\dagger}_{l, -m \ponehat}\biggr) \sin \ponehat \xonehat 
\period \label{aRlm}
\end{align}
To perform the differentiation with respect to $t_R$, we must use the relation between $\that$ and $t_R$ given by the Lorentz transformations
\begin{align}
\that &= \gahat (t_M + \behat x^1) = \gahat z_R ( \sinh t_R  + \behat 
 \cosh t_R )\comma \\
\xonehat &= \gahat (x^1 + \behat t_M)  = \gahat z_R (\cosh t_R + \behat 
\sinh t_R )\comma \\
\behat &= 1-\ep \equiv \tanh \xi \comma \qquad 
\gahat = {1\over \sqrt{1-\beta^2}}
\equiv \cosh \xi \comma 
\end{align}
where we introduced the rapidity variable $\xi$. 
Then, the relevant part of (\ref{aRlm}) can be computed as 
\begin{align}
&e^{i\omega t_R} \overleftrightarrow{\del_{t_R}}\left[ e^{-i E_{k_l \ponehat} \that}  \sin  (\ponehat \xonehat)\right] \nonumber \\
&\quad =e^{i \omega t_R}e^{-iaz_R} \left[ -i (a'z_R+\omega)  \sin b z_R 
+ b' z_R\cos b z_R  \right]\nonumber\\
&\quad= -\half  (a'z+\omega) e^{i\omega t}\left( e^{-i(a-b)z} -e^{-i(a+b)z} \right) 
 + \half b' z e^{i\omega t} \left(e^{-i(a-b)z} + e^{-i(a+b)z} \right) 
\comma 
\end{align}
where 
\begin{align}
a &\equiv E_{k_l \ponehat} \gahat(\sinh t_R + \behat \cosh t_R) \comma 
\qquad 
b \equiv  \ponehat \gahat  (\cosh t_R + \behat \sinh t_R ) \comma \\
a' &\equiv E_{k_l \ponehat} \gahat (\cosh t_R + \behat \sinh t_R) \comma 
\qquad 
b'\equiv  \ponehat \gahat (\sinh t_R + \behat \cosh t_R) \period
\end{align}
These expressions can be further simplified by introducing the parametrization
\begin{align}
E_{k_l\ponehat} &= |k_l| \cosh \uhat \comma \qquad 
 \ponehat = |k_l| \sinh \uhat  \period
\end{align}
Then, we can write 
\begin{align}
a\pm b &= |k_l| \sinh \rho_\pm \comma \qquad 
a'\pm b' = |k_l| \cosh \rho_\pm \comma \\
\rho_\pm &\equiv \xi + t_R \pm \uhat \period
\end{align}

Now we consider the integral over $z_R$ in (\ref{aRlm}). The basic integrals we 
 need are $A_1(c, k)$ and $A_2(c,k)$ given in (\ref{Aoneck}) and (\ref{Atwock}) 
 in Appendix \ref{app:intKiw}. Specifically, the ones we need are 
with $c=\pm (a\pm b)$ and $k=|k_l|$. When these parameters are substituted 
the integrals simplify drastically and we obtain 
\begin{align}
A_1(a\pm b, |k_l|) &= C_\omega \left( e^{\pi \omega/2} e^{-i\omega \rho_\pm}
+ e^{-\pi \omega/2} e^{i\omega \rho_\pm}\right) \comma \\
A_2(a\pm b, |k_l|) &= {\omega C_\omega \over |k_l| \cosh \rho_\pm} 
\left(  e^{\pi \omega/2}
  e^{-i\omega \rho_\pm} - e^{-\pi \omega/2} e^{i\omega \rho_\pm}\right) 
\comma  \\
A_1(-(a\pm b), |k_l|) &=  C_\omega \left( e^{\pi \omega/2} e^{i\omega \rho_\pm}
+ e^{-\pi \omega/2} e^{-i\omega \rho_\pm}\right) \comma \\
A_2(-(a\pm b), |k_l|) &= {\omega C_\omega \over |k_l| \cosh \rho_\pm} 
\left(  e^{\pi \omega/2}
  e^{i\omega \rho_\pm} - e^{-\pi \omega/2} e^{-i\omega \rho_\pm}\right)
\comma 
\end{align}
where 
\begin{align}
C_\omega &\equiv { \pi  \over 2\omega \sinh \pi \omega}  \period
\end{align}
Further,  it is convenient to use the rapidity-based oscillators 
\begin{align}
\ahat^-_{lm\uhat}&\equiv \sqrt{E_{k_l \ponehat}}  \ahat^-_{lm\ponehat} 
\comma 
\end{align}
similarly to (\ref{defaMku}).  

Now, using these formulas, it is straightforward to compute the RHS of the 
formula (\ref{aRlm}) and get the form of $a^R_{lm\omega}$ (and its conjugate)
 in terms of the FFO mode operators $\ahat^-_{lm\rho}$ and $\ahat^{-\dagger}_{lm\rho}$. 
The answers take rather simple forms:
\begin{align}
a^R_{lm\omega} &= {1\over 2i} {e^{-i\omega \xi} \over \sqrt{\pi \sinh\pi\omega}} \int_{-\infty}^\infty d \rho \sin \omega \rho 
\left( e^{\pi \omega/2} \ahat^-_{lm\rho} -(-1)^me^{-\pi \omega/2} \ahat^{-\dagger}_{l, -m\rho} \right) \comma \label{aRahatm}\\
a^{R\dagger}_{lm\omega} &= {1\over 2i} {e^{i\omega \xi} \over \sqrt{\pi \sinh\pi\omega}} \int_{-\infty}^\infty d \rho \sin \omega \rho 
\left((-1)^m e^{-\pi \omega/2} \ahat^-_{l,-m\rho} -e^{\pi \omega/2} \ahat^{-\dagger}_{lm\rho} \right) \period \label{arFFO}
\end{align}
One can  check that they satisfy the correct commutation relation 
 $[a^R_{lm\omega}, a^{R\dagger}_{l'm'\omega'}] = \delta_{ll'}\delta_{mm'}\delta(\omega-\omega') $.

Finally, with the expressions (\ref{arFFO}), we can compute the expectation value of the number operator 
 for the \WR ``particles" in the FFO vacuum $\ket{\hat{0}}_-$. The result is 
\begin{align}
{{}_-\langle \hat{0}| a^{R\dagger}_{lm\omega} a^R_{lm\omega} | \hat{0}\rangle_- \over {}_-\langle \hat{0} | \hat{0} \rangle_- }
&= { 1 \over e^{2\pi \omega} -1}{2\over \pi} \int_{-\infty}^\infty d\rho \sin^2 \omega \rho \period
\end{align}
Several remarks are in order: \\
(i)\ We recognize that the first factor is of the same form as the familiar ``thermal"  distribution.  We emphasize however that in this case it is not genuinely thermal 
 since \WL modes do not exist and hence no tracing over them is involved. 
The fact that the form looks thermal stems from the fact that the expression of 
$a^R_{lm\omega}$ in terms of $\ahat^-_{lm\rho}$ and its conjugate 
 in (\ref{aRahatm}) is essentially the same as (\ref{aRaM}), valid for the 
entire Minkowski space including the region \WL\!\!. 
\\
(ii)\ The last integral represents the cohererent sum over infinite number of rapidities which contribute to the \WR mode.  Although it appears to depend on $\omega$, this factor is divergent and depending on how we cut it off, the $\omega$-dependence will be different.  Moreover, as it becomes clear from the comparison 
 with the usual Unruh effect below, this factor comes from the nature of the boundary  condition along the shock wave, \ie it depends on the interaction between the falling matter and the scalar field.  Therefore, this integral is ambiguous and the form of its  $\omega$ dependence should not be taken seriously.  It indicates, however, 
 that an extra $\omega$-dependence, other than the usual thermal factor, can be 
 possible. \\
(iii)\ As the last remark, note that the dependence on $\xi$, the Lorentz boost 
 parameter, disappeared in the distribution. This is quite natural since the vacuum 
 $\ket{\hat{0}}_-$ should be Lorentz invariant. 
 
In any case, we have found that,  even in the case of the one-sided black hole, 
 the Unruh-like effect does exist. 

It is instructive to compare this with the case of the usual Unruh effect. 
From (\ref{aRaM}), it is easy to find that
\begin{align}
{{}_M\langle 0| a^{R\dagger}_{k\omega} a^R_{k\omega} \ket{0}_M 
\over {}_M\langle 0 \ket{0}_M} &=  \int_{-\infty}^\infty {du \over 4\pi \sinh \pi \omega} 
\int_{-\infty}^\infty du' e^{-i\omega(u-u')} e^{-\pi \omega} 
{{}_M\langle 0|[ a^{M}_{ku}, a^{M\dagger}_{ku'}] \ket{0}_M
\over  {}_M\langle 0 \ket{0}_M }  \nonumber \\
&= {1 \over 2\pi \left( e^{2\pi \omega} -1\right)} {V_{R^2} \over (2\pi)^2} \int_{-\infty}^\infty du \comma 
\end{align}
where ${V_{R^2} \over (2\pi)^2} =\delta^2(k-k)$ is the volume of the two-dimensional space and the divergent integral 
$\int_{-\infty}^\infty du$ counts all the modes with different rapidities making up a \WR  particle wave. Note that in this flat space Unruh effect, the $\omega$ dependence $e^{-i\omega(u-u')}$ cancels  out  due to the appearance of $\delta(u-u')$ coming from the 
commutator $[ a^{R}_{ku}, a^{R\dagger}_{ku'}] $ and we have exactly 
 the thermal form, as is well-known. 
\section{Summary and discussions \label{sec:discussions}}
\subsection{Brief summary}
In this work we have made a detailed  study of the issue  of  the observer dependence for 
the quantization of fields in a curved spacetime, which is 
 one of the crucial problems that one must deal with  whenever one discusses quantum gravity.  Understanding of this issue is particularly important  in cases where an event horizon exists for some of the observers.  Explicitly, we have focused on the quantization  of a scalar field in the most basic such configuration, namely the spacetime  in the vicinity of the horizon of a four-dimensional \SS black hole, including the interior as well as the exterior. 
 Detailed  and comprehensive analyses were performed for the  three typical observers, clarifying how the modes they observe are related. We studied both the two-sided eternal case and the more physical one-sided case produced by the falling 
shell, or a shock wave.  For the latter, 
the effect of  the collapsing matter upon the scalar field outside of the shell is represented by an effective  boundary condition along the shock wave. 

One important conclusion obtained from such explicit calculations  is that 
as long as the interior of a large black hole can be described  more or less by 
a metric like that of Vaidya, 
 the free-falling ovserver sees no change in the Hilbert space structure of the 
quantized field as he/she crosses the horizon. In other words, the equivalence 
 principle holds quantum mechanically as well,  at least in the above sense. 

Another result worth emphasizing is that  in the one-sided case despite the fact that there are no counterparts of the left Rindler modes in the vacuum of the freely falling observer, and hence no tracing procedure over them is relevant,  there still exists  an Unruh-like effect.  Namely, in such a vacuum the number density of the \WR modes contains  the universal factor of ``thermal" distribution in the  frequency $\omega$ (apart from a divergent piece which depends on the interaction between the scalar field and the falling matter.)  

In addition to these results, comprehensive and explicit knowledge of the properties and 
 the relations of the Hilbert spaces for the different observers have been obtained, and we believe this will be of use in better understanding of the quantum properties of 
gravitational physics. 
\subsection{Discussions}
Evidently,  the problem of observer dependence that we studied in the semi-classical  regime in this work is of universal importance in any attempt to understand quantum gravity.  In particular, it would be extremely interesting to see how this 
 problem appears and should be treated in the construction of the ``bulk" from the ``boundary" in the AdS/CFT correspondence, which is anticipated to give 
 important hints for formulating quantum gravity and understanding quantum 
black holes.  Although there have been some attempts to address this question, 
it is not well understood how the change of frame (\ie the choice of ``time") for the quantization,  both in the bulk and the boundary, is expressed and controlled in 
 the AdS/CFT context. The best place to look into would be the AdS$_3$/CFT$_2$ 
 setting, where at least we have some knowledge of how the structure of CFT$_2$ changes under a redefinition of ``time " by a conformal change of variable\cite{NS,GT}. A further advantage  to exploring the observer dependence in AdS$_3$/CFT$_2$ is that AdS$_3$ black holes (\ie, BTZ black holes) are locally equivalent to the pure AdS$_3$ spacetime and we can solve the equations of motion in the black hole spacetime in the same manner as for the pure AdS$_3$.

In this work we have concentrated on the relations between the modes 
seen by different observers and have not touched upon the correlation functions 
 between the fields.   Some two-point correlation functions in the Rindler wedges of the Minkowski space have been studied\cite{Michel:2016qge}, but the most interesting question of whether one can extract physical 
 information from behind the horizon or exchange information 
between different observers by quantum means is yet to be answered. 
We hope to study these and related questions and give a report in the near future. 
\par\bigskip\noindent
{\large\bf Acknowledgment}\par\smallskip\noindent
Y.K. acknowledges  T.~Eguchi for a useful discussion. 
 The research of  K.G is supported in part by  the JSPS Research Fellowship for Young Scientists, while that of 
Y.K. is supported in part by the 
 Grant-in-Aid for Scientific Research (B) 
No.~25287049, both from the Japan 
 Ministry of Education, Culture, Sports,  Science and Technology. 
\appendix
\section{Orthogonality and completeness relations for the modified Bessel functions of imaginary order  \label{app:ortcomp}}
For various basic computations performed in the main text using the expansions in terms of the eigenmodes, the orthogonality and the completeness of the modified Bessel functions  of imaginary order are essential.  In this appendix, we give  some useful  comments on such relations previously obtained in the literature and provide additional information. 
\subsection{Orthogonality \label{app:orthogonality}} 

The orthogonality relations are needed in extracting each mode from the expansion  of the scalar field appropriate for  various coordinate frames. Such relation 
for $K_{i\omega}(x)$ is proven in \cite{Yakubovich, Passian, Szmytkowski} 
and takes the form 
\bal
\int_0^\infty \f{dx}{x}K_{i\w}(x)K_{i\w'}(x)&={1\over \mu(\w)} 
\left(\d(\w-\w') +\delta(\omega+\omega')\right) \comma \label{orthK} 
\eal
where $\mu(\w)$ here and below is given by 
\begin{align}
\mu(\w) &\equiv {2\w \sinh \pi \w \over \pi^2} \period \label{muomega}
\end{align}

The corresponding relations  for the Hankel functions $H^{(i)}_{i\omega}$ for $i=1,2$ have not been explicitly given in the literature but can be derived without difficulty, for example,  by the method described in \cite{Szmytkowski}. 
 The result is 
\begin{align}
\int_0^\infty {dx \over x} H^{(i)}_{i\omega}(x)H^{(i)}_{i\omega'}(x) 
={4e^{\eta_i \pi \w} \over \pi^2 \mu(\w) }  \left(\delta(\omega -\omega') +\delta(\omega+\omega') \right)\quad \eta_1=+1\comma \quad \eta_2=- 1\period  \label{orthH}
\end{align}
\subsection{Completeness \label{app:completeness}}
The completeness relation for the function $K_{i\omega}(x)$ can be written as 
\ba
\int_0^\infty d\w \mu(\w) K_{i\w}(x)K_{i\w}(y)=x\d(x-y)\comma  \label{compK}
\ea
Since $K_{-i\omega} = K_{i\omega}$, we can, if we wish,  extend the range of integration to $[-\infty \le \w \le \infty]$ and multiply the RHS by a factor of 2. 

This relation is equivalent to the inverse of the so-called Kontorovich-Lebedev (KL) transform\cite{KontLeb} below. KL transform $f(\w,y)$ of a function $g(x,y)$ with respect to $x$ (where $y$ is a parameter)  is defined by 
\begin{align}
f(\w,y) &= \mu(\w) \int_0^\infty {dx \over x} K_{i\w}(x) g(x,y) \period 
\label{fomegay}
\end{align}
Then, $g(x,y)$ is obtained in terms of $f(\w, y)$ by the formula
\begin{align}
g(x,y) &= \int_0^\infty d\w K_{i\w}(x) f(\w, y) \period \label{gxy}
\end{align}
If we take $g(x,y) = x\delta(x-y)$, then the formula (\ref{fomegay}) 
gives $f(\w, y) =\mu(\w) K_{i\w}(y)$. Substituting this into (\ref{gxy}) then gives 
\begin{align}
x\delta(x-y) &= \int_0^\infty d\w \mu(\w) K_{i\w}(x) K_{i\w}(y) \comma 
\end{align}
which is precisely the completeness relation (\ref{compK}). 

In fact, without resorting to the KL-formula, there is a rather elementary derivation of (\ref{compK}), starting from the 
 following integral formula\cite{ETII}:
\begin{align}
\int_0^\infty d\omega \cosh a\omega K_{i\omega}(x) K_{i\omega}(y) 
&= {\pi \over 2} K_0( \sqrt{x^2 + y^2 + 2x y \cos a} ) \comma \label{intcosK}
\end{align}
valid for  $x,y>0$, $|{\rm Re}\, a| + |{\rm arg}\, x| < \pi$. (The second condition is stringent. We cannot set $a=\pi$ from the beginning.)  First by differntiating  this with respect to $a$, we get 
\begin{align}
\int_0^\infty d\omega \omega \sinh a \omega K_{i\omega}(x) K_{i\omega}(y) 
&= {\pi x x' \sin a \over 2\sqrt{x^2 + y^2 + 2x y \cos a}} 
K_1 (\sqrt{x^2 + y^2 + 2x y \cos a}) \comma 
\end{align}
where we used the formula $\del_z K_0(z) =-K_1(z)$. 
Now we set $a=\pi -\ep$, where $\ep$ is a positive infinitesimal quantity. Then the RHS becomes 
\begin{align}
{\pi x y \ep \over 2\sqrt{(x-y)^2 + xy \ep^2} } K_1(\sqrt{(x-y)^2 + xy \ep^2}) \period \label{intep} 
\end{align}
For $x-y\ne 0$, this vanishes as $\ep \rightarrow 0$, \ie as $a \rightarrow \pi$. On the other hand, for small $x-y$, 
using the small argument expansion $K_1(z) \simeq  {1\over z}$, (\ref{intep}) becomes 
\begin{align}
{\pi \over 2}{\ep xy  \over (x-y)^2+ \ep ^2 xy  } \period \label{epdelta}
\end{align}
By making a rescaling   $x \rightarrow x/\sqrt{xy}$ and $y \rightarrow y/\sqrt{xy}$  in 
 the well-known representation of the delta function, namely, 
 $\delta(x-y) = (\ep/\pi)/ ((x-y)^2 + \ep^2))$, we readily obtain 
\begin{align}
\delta ((x-y)/\sqrt{xy}) &= \sqrt{xy}\delta(x-y) 
 ={1\over \pi} {\ep xy \over (x-y)^2 +\ep^2  xy } \period
\end{align}
Comparing with (\ref{epdelta}) we  obtain the completeness relation (\ref{compK}). 

The corresponding completeness relations for the Hankel functions are given by 
\begin{align}
\int_0^\infty d\w {\pi^2 \mu(\w)  \over 4 e^{\eta_i \pi \omega} }
H^{(i)}_{i\omega}(x) H^{(i)}_{i\omega}(y) = x\delta(x-y)  \label{compH}
\end{align}
where the sign $\eta_i$ is as defined in (\ref{orthH}). 

\section{Extraction of the modes in various wedges and their relations \label{app:relmodes}}
In this appendix, we provide some details of the computations concerning 
  the extraction of the modes and their relations described in Sec.~2 of the main text. 
\subsection{Klein-Gordon inner products and extraction of the modes\label{app:KG}}
We will be interested in a $d-$dimensional curved space with the metric of the form
\ba
ds^2=-N(x)^2dt^2+g_{ab}dx^adx^b \comma 
\ea
where $N(x)$ is the lapse function. Let $f_A, f_B$ be two independent solutions of the Klein-Gordon equation for this metric.
Define the following current
\ba
J^{\mu}_{f_A,f_B}(x)\equiv f^*_A(x)\overleftrightarrow{\nabla}^\m f_B
\comma 
\ea
which is covariantly conserved $\nabla_{\m}J^{\mu}_{f_A,f_B}(x)=0$. Let $\SG$ be the constant $t$ surface. The conservation property above means that the Klein-Gordon inner product defined by
\bal
(f_A,f_B)_{\rm KG}\equiv i\int_{\SG}d^{d-1}x\f{\s{g}}{N}n_{\m}J^{\mu}_{f_A,f_B}
\comma 
\eal
where $n^\mu$ is the  future directed unit vector normal to  $\SG$, is independen of $t$. This formula is useful in extracting the modes from the field expressed 
 in various coordinates. 
\subsubsection{The right Rindler wedge \label{app:WR}}
Hereafter, we will sed $d=4$. 
In the right Rindler wedge, the metric is given by
\ba
ds^2=-z^2dt^2+dz^2+\sum_{i=2}^3(dx^{i})^2.
\ea
In this case, we can identify  $N=z, g_{zz}=1, g_{ij}=\d_{ij}, \s{g}=1$, and hence  the Klein-Gordon inner product in the right Rindler wedge is defined as
\ba
(f_A,f_B)^R_{\rm KG}=i\int_{0}^{\infty}\f{dz}{z}\int d^{2}x(f^*_A\overleftrightarrow{\pp_{t}}f_B) \period
\ea 
The solutions of the Klein-Gordon equation in this coordinate frame are
\bal
f^R_{k\w }(t,z,x)&=N^R_{\w }K_{i\w}(|k|z)e^{i(kx-\w t)}\comma \no
(N^R_{\w })^2&=\f{\sinh\pi\w}{\pi^2(2\pi)^{2}} \period
\eal
Let us compute the Klein-Gordon inner product of such functions explicitly. We get
\bal
(f^R_{k\w},f^R_{k'\w' })^R_{\rm KG}&=\int_0^\infty \f{dz}{z}\int d^{2}x N^R_{\w }N^R_{\w' }K_{i\w}(|k|z)K_{i\w'}(|k'|z)(\w+\w')e^{i(k-k')x}e^{-i(\w-\w')t}\no
&=(2\pi)^{2}(\w+\w')\d(k-k')N^R_{\w }N^R_{\w' }e^{-i(\w-\w')t}\int_0^\infty \f{dz}{z}K_{i\w}(|k|z)K_{i\w'}(|k'|z)\no
&=\d(k-k')\d(\w-\w') \comma 
\eal
where, getting to the last line,  
 we used the orthogonality of the modified Bessel function (\ref{orthK}) for $\w, \w' >0$. 

Recall that the scalar field in the right Rindler wedge can be expanded as
\bal
\phi^R(t_R,z_R,x)&=\int_0^{\infty}d\w\int d^{2}k \left[f^R_{k\w }(t_R,z_R,x)a^R_{k\w}+{\rm h.c.}\right]\period \label{expandphiR}
\eal
The modes $a^R_{k\w}$ and $a^{R\dagger}_{k\w}$ are extracted using the Klein-Gordon inner product as 
\ba
a^R_{k\w }=(f^R_{k\w },\phi^R)^R_{\rm KG}\comma \qquad a^{R\dg}_{k\w }=-(f^{R*}_{k\w },\phi^R)^R_{\rm KG}\period
\ea
\subsubsection{The future Rindler wedge \label{app:WF}}
In the future Rindler wedge, the metric is given by
\ba
ds^2=-dz_F^2+z_F^2dt_F^2+\sum_{i=2}^3(dx^{i})^2 \period
\ea
In this case, $z_F$ is the time variable, $t_F$ is the  space variable and $N=1, g_{t_Ft_F}=z_F^2, g_{ij}=\d_{ij}, \s{g}=z_F$. The Klein-Gordon inner product in the future Rindler wedge is defined as
\ba
(f_A,f_B)^F_{\rm KG}=i\int_{-\infty}^{\infty}dt_F\int d^{2}x z_F(f^*_A\overleftrightarrow{\pp_{z_F}}f_B) \period
\ea 
Then the solutions of the Klein-Gordon equation which damps at large $|k|z_F$  are
\bal
f^{(2)}_{k\w }(t_F,z_F,x)&=N^F_{\w }H^{(2)}_{i\w}(|k|z_F)e^{i(kx-\w t_F)}
\comma \no
(N^F_{\w })^2&=\f{e^{\pi\w}}{8(2\pi)^{2}} \period
\eal
The Klein-Gordon inner product of these functions is given by 
\bal
&(f^{(2)}_{k\w },f^{(2)}_{k' \w' })^F_{\rm KG}\no &=i\int_{-\infty}^\infty dt\int d^{2}x zN^F_{\w }N^F_{\w' }\left(H^{(1)}_{i\w'}(|k'|z_F)\pp_{z_F} H^{(2)}_{i\w}(|k|z_F)-H^{(2)}_{i\w}(|k|z_F)\pp_{z_F} H^{(1)}_{i\w'}(|k'|z_F)\right) \nonumber \\
&\quad  \cdot e^{-\pi\w}e^{i(k-k')x}e^{-i(\w-\w')t}\no
&=i(2\pi)^{3}\d(k-k')\d(\w-\w')\no
& \quad \cdot z_F N^F_{\w }N^F_{\w' }\left(H^{(1)}_{i\w'}(|k'|z_F)\pp_{z_F} H^{(2)}_{i\w}(|k|z_F)-H^{(2)}_{i\w}(|k|z_F)\pp_{z_F} H^{(1)}_{i\w'}(|k'|z)\right) e^{-\pi\w}\no
&=\d(k-k')\d(\w-\w') \period
\eal
To get to the last line, we used the identity
\ba
H^{(1)}_{i\w}(|k|z)\pp_z H^{(2)}_{i\w}(|k|z)-H^{(2)}_{i\w}(|k|z)\pp_z H^{(1)}_{i\w}(|k|z)=-i\f{4}{\pi z}\period  \label{HoneHtwo}
\ea By similar manipulations, it is easy to get the following inner products
\bal
(f^{(2)*}_{k\w },f^{(2)*}_{k' \w' })^F_{\rm KG}=-\d(k-k')\d(\w-\w') 
\comma \qquad 
(f^{(2)}_{k\w },f^{(1)}_{k' \w' })^F_{\rm KG}=0 \period
\eal
Recall that the scalar field in the future Rindler wedge can be expanded as
\bal
\phi^F(t_F,z_F,x)&=\int_{-\infty}^{\infty}d\w\int d^{2}k \left[f^{(2)}_{k\w }(t_F,z_F,x)a^F_{k\w }+{\rm h.c.}\right] \period  \label{expphiFap}
\eal
Taking the inner product with $f^{(2)}_{k\w }$, we obtain
\ba
a^F_{k\w }=(f^{(2)}_{k\w },\phi^F)^F_{\rm KG}\comma \qquad 
a^{F\dg}_{k\w }=-(f^{(2)*}_{k\w },\phi^F)^F_{\rm KG}\period
\ea
\subsection{Mode operators of \WR and \WF in terms of those of the Minkowski spacetime \label{app:relaWR}}
\subsubsection{ Useful integrals involving $K_{i\w}(z)$ \label{app:intKiw}}
We shall first derive several useful integrals involving $K_{i\w}(z)$, which  play 
 important roles  below in \ref{app:aRaM} and in \ref{sec:Unruh} in the main text.\nxt
\underline{Formula (I)}:\quad The first formula is 
\begin{align}
\int_0^\infty {dz \over z} K_{i\omega}(z) e^{-iz \sinh t -\ep z} 
&= {\pi \over 2\omega \sinh \pi \omega} \left( e^{i\omega t} e^{-\pi \omega/2} + e^{-i\omega t} e^{\pi \omega/2} \right) \comma  \label{Kexpsinh} 
\end{align}
where $\ep$ is an infinitesimal positive parameter, needed to make the integral 
 convergent. 
To prove this formula, we start with the formula 6.795-1 of \cite{GR}, which can be expressed as 
\begin{align}
{\pi \over 2} 
 e^{-z \cosh \tau}&=\int_0^\infty d\w \cos(\omega \tau) K_{i\omega}(z)  \comma \quad 
 \left| {\rm Im}\, \tau \right| <{\pi \over 2} \comma \quad z >0  \period 
\label{GRformula}
\end{align}
By extending the region of $\omega$ to $[-\infty, \infty]$ for convenienece\footnote{This is purely as a mathematical equality. The physical energy $\omega$ is of course non-negative.}, the integral on the RHS can be rewritten as 
\begin{align}
{\rm RHS} &= \half \int_{-\infty}^\infty e^{i\omega' \tau} K_{i\omega'} (z) d\omega'
\period
\end{align}
We now act $\int_0^\infty (dz/z) K_{i\omega}(z)$ on this expression, with $\omega$ non-negative. Then using the orthogonality relation (\ref{orthK}) for $K_{i\omega}(z)$, we get
\begin{align}
\half \int_{-\infty}^\infty d\omega' e^{i\omega' \tau} \int_0^\infty {dz \over z} K_{i\omega}(z) K_{i\omega'}(z)
&= \half \int_{-\infty}^\infty d\omega' e^{i\omega'\tau} {1\over \mu(\omega)} \left( \delta(\omega' -\omega)+ \delta(\omega'+\omega) \right) \nonumber\\&= {1\over 2\mu(\omega)} \left( e^{i\omega \tau} + e^{-i\omega \tau} \right)= {1\over \mu(\omega)} \cos \omega \tau \comma 
\end{align}
where $\mu(\omega)$ is as given in (\ref{muomega}).  Performing the same 
 integral for the LHS as well, (\ref{GRformula}) becomes 
\begin{align}
\int_0^\infty {dz \over z} K_{i\omega} (z) e^{-z \cosh \tau} 
&= {2 \over \pi \mu(\omega)} \cos \omega \tau \period \label{Kexpcosh}
\end{align}
We now make a substitution 
\begin{align}
\tau = t + ((\pi/2) -\ep) i  \comma \label{tau-t}
\end{align}
where  $\ep$ is an infinitesimal  positive quantity. This is legitimate since ${\rm Im}\, \tau$ satisfies the condition for the formula (\ref{GRformula}) to be valid. 
Then by a simple calculation 
 we get $
\cosh \tau = \cosh \left( t + {\pi \over 2 }i -i\ep \right) =  i\sinh t + \ep
$, 
where we reexpressed a positive infinitesimal quantity $\ep \cosh t $ as $\ep$. 
Substituting (\ref{tau-t}) into the RHS of (\ref{Kexpcosh}) and expanding 
$\cos \w \tau$, we obtain the formula (i). 
\nxt
\underline{Formula (II)}\quad The second formula is 
\begin{align}
A_1(c,k) &\equiv \int_0^\infty {dz \over z} K_{i\omega} (kz) e^{-icz-\ep z}
\nonumber\\ &= {\pi 
\over 2\omega \sinh \pi \omega} \biggl\{ e^{\pi \omega/2}  \exp\left( -i\omega \ln  \left( {1\over k}\left(c + \sqrt{ c^2+k^2}\right)\right) \right)  \nonumber\\
&\quad +  e^{-\pi \omega/2}  \exp\left( i\omega \ln  \left( {1\over k}\left(c + \sqrt{ c^2+k^2}\right)\right) \right) \biggr\} \comma  \label{Aoneck}
\end{align}
where $c$ is real and $k$ is  real positive. To prove this formula, 
 we first rescale $z \rightarrow kz$ in formula (I) and then set $c=k\sinh t$. 
Solving $e^t$ in terms of $c$ and substituting into the RHS of formula (I), 
we obtain the integral above. 
\nxt
\underline{Formula (III)} \quad Finally, a formula similar to (II) we need is 
\begin{align}
A_2(c, k) &\equiv\int_0^\infty dz\,  K_{i\omega} (kz) e^{-icz}
\nonumber\\
&={\pi 
\over 2\sinh \pi \omega}{1\over   \sqrt{c^2 + k^2}} 
\biggl\{ e^{\pi \omega/2}  \exp\left( -i\omega \ln  \left( {1\over k}\left(c + \sqrt{ c^2+k^2}\right)\right) \right)  \nonumber\\
&-  e^{-\pi \omega/2}  \exp\left( i\omega \ln  \left( {1\over k}\left(c + \sqrt{ c^2+k^2}\right)\right) \right) \biggr\} \period \label{Atwock}
\end{align}
This formula is obtained simply from $A_1(c,k)$ as $A_2(c,k) = i (\del A_1(c,k)/\del c)$. 
\subsubsection{ $a^R_{k\w}$ in terms of $a^M_{kp^1}$ \label{app:aRaM}}
The free scalar field in the right Rindler wedge can be expanded as  in (\ref{expandphiR}). 
On the other hand, in this region we should be able to express $\phi^R$ 
 in terms of  $\phi^M$ and hence $a^R_{\w k}$ in terms of the Minkowski modes $a^M_{kp^1 }$. 
In the Minkowski spacetime the scalar fields can be written in terms of the coordinate  of \WR as 
\bal
\phi^M(z_R,t_R,x)&=\int \f{dp^1}{\s{2\pi}\s{2E_{k',p^1}}} \int \f{d^2k'}{2\pi} e^{ik'x+ip^{1}x^1-iE_{k',p^1}t_M} a^M_{k'p^1} + \mbox{h.c.}\no
 &=\int \f{dp^1}{\s{2\pi}\s{2E_{k',p^1}}} \int \f{d^{2}k'}{2\pi}\ e^{ik'x+ip^{1}z_R\cosh t_R-iE_{k',p^1}z_R\sinh t_R} a^M_{k'p^1} + \mbox{h.c.} \comma \eal
where in the second line we substituted   $t_M=z_R\sinh t_R, x^1=z_R\cosh t_R$. Thus using the Klein-Gordon inner product we can extract the annihilation operators in the right Rindler coordinate from the expression of the scalar field in the Minkowski spacetime as 
\bal 
 a^{R}_{k\w}&=(f^R_{k\w},\phi^M)^R_{\rm KG}\no
 &=i\int_{0}^{\infty}\f{dz}{z}\int d^{2}x(f^{R*}_{k\w}\overleftrightarrow{\pp_{t_R}}\phi^M)
 \no
 &=i\int_{0}^{\infty}\f{dz_R}{z_R}\int d^{2}x\int \f{dp^1}{\s{4\pi E_{k'p^1}}} \int \f{d^{2}k'}{2\pi}N_\w^R K^*_{i\omega}(|k|z_R)\no &\times \left(e^{-i(kx-\omega t_R)}\overleftrightarrow{\pp_{t_R}}[e^{ik'x+ip^{1}z_R\cosh t_R-iE_{k'p^1}z_R\sinh t_R} a^M_{k'p^1} + \mbox{h.c.}]\right)
 \no
 &=i\int \f{2\pi  dp^1 N_\w^R}{\s{4\pi E_{k'p^1}}} e^{i\omega t_R}\left(\overleftrightarrow{\pp_{t_R}}\int_{0}^{\infty}\f{dz_R}{z_R} K^*_{i\omega}(|k|z_R)e^{-ip^{1}z_R\cosh t_R-iE_{kp^1}z_R\sinh t_R} a^M_{kp^1} \right.\no
 &\left.+\overleftrightarrow{\pp_{t_R}}\int_{0}^{\infty}\f{dz_R}{z_R} K_{i\omega}(|k|z_R)e^{ip^{1}z_R\cosh t_R+iE_{-kp^1}z_R\sinh t_R} a^{M\dg}_{-kp^1}\right) \period \nonumber   
\eal
Let us now use a convenient  parametrization  $E_{kp^1}=|k|\cosh\rho, p^1=-|k|\sinh\rho$,  such that  $E^2_{kp^1} = (p^1)^2 + k^2$ is realized. Then the expression above can be written as  
\bal
a^{R}_{k\w}=i\int \f{ 2\pi dp^1}{\s{2\pi}\s{2E_{k'p^1}}} N_\w^Re^{i\omega t_R}\overleftrightarrow{\pp_{t_R}} \int_{0}^{\infty}\f{dz_R}{z_R} K_{i\omega}(|k|z_R)\left[e^{-i|k|z_R\sinh (t_R+\rho)} a^M_{kp^1} + e^{i|k|z_R\sinh (t_R+\rho)} a^{M\dg}_{-kp^1}\right]  \comma \nonumber
\eal
where we used the property $K_{i\w}(z)=K_{-i\w}(z)$. 

Now by using the formula (I) given in (\ref{Kexpsinh} ), we can perform the 
integral over $z_R$ and get 

\bal
a^{R}_{k\w }&=i\int \f{2\pi dp^1}{\s{2\pi}\s{2E_{k p^1}}} N_\w^R
 {\pi \over 2\omega \sinh \pi \omega} e^{i\omega t_R}\overleftrightarrow{\pp_{t_R}} \biggl[\left( e^{-i\w(t_R+\rho)} e^{\pi \w/2} 
 +e^{i\w(t_R+\rho)} e^{-\pi \w/2} \right)a^M_{kp^1} \no
&\qquad + \left( e^{-i\w(t_R+\rho)} e^{-\pi \w/2} 
 +e^{i\w(t_R+\rho)} e^{\pi \w/2} \right)a^{M\dagger}_{-kp^1} \biggr] \no
&=\int \f{dp^1}{\s{2\pi}\s{2E_{kp^1}}} \f{1}{\s{\sinh\pi\w}}\left(\f{E_{kp^1}-p^1}{E_{kp^1}+p^1}\right)^{\f{-i\w}{2}}\left[e^{\pi\w/2}a^M_{kp^1}+e^{-\pi\w/2}a^{M\dg}_{-kp^1}\right] \period \label{rm}
\eal 
This is the formula quoted in (\ref{WRM}).  Taking the hermitian conjugation 
we obtain the creation operator
\bal
a^{R\dg}_{k\w }=\int \f{dp^1}{\s{2\pi}\s{2E_{kp^1}}} \f{1}{\s{\sinh\pi\w}}\left(\f{E_{kp^1}-p^1}{E_{kp^1}+p^1}\right)^{\f{i\w}{2}}\left[e^{\pi\w/2}a^{M\dg}_{kp^1}+e^{-\pi\w/2}a^{M}_{-kp^1}\right] \period
\eal 
\subsubsection{$a^F_{k\w}$ in terms of $a^M_{kp^1}$\label{app:relaWF} }
As in \WR \,  the free scalar field in  \WF frame  should be describable in terms of 
the Minkowski modes. 
It is  expanded as in (\ref{expphiF}) in terms of 
 the Hankel functions $H^{(2)}_{i\w}(|k|z_F)$, which is recalled in (\ref{expphiFap}) for convenience. 
If we write such a  field in the Minkowski spacetime in terms of the coordinates 
 of \WF,  using the relation $t_M=z_F\cosh t_F, x^1=z_F\sinh t_F$, it reads 
\bal
\phi^M(z_F,t_F,x)&=\int \f{dp^1}{\s{2\pi}\s{2E_{k'p^1}}} \int \f{d^2k'}{2\pi}\ e^{ik'x_F+ip^{1}z\sinh t_R-iE_{k'p^1}z\cosh t_F} a^M_{k'p^1} + \mbox{h.c.} \period
\eal
Using the Klein-Gordon inner product,  we can extract $a^R_{k\w}$ from the 
Minkowski field $\phi^M(z_F,t_F,x)$ as 
\bal
 a^{F}_{k\w}&=(f^F_{k\w},\phi^M)^F_{\rm KG}
 =i\int_{-\infty}^{\infty}zdt_F dx^{2}(f^{F*}_{k\w}\overleftrightarrow{\pp_{z}}\phi^M) \no
 &=i\int_{-\infty}^{\infty}z_Fdt_F dx^2\int \f{dp^1}{\s{2\pi}\s{2E_{k'p^1}}} \int \f{d^2k'}{2\pi}N_\w^F e^{-i(kx_F-\omega t_F)}
\no & \quad \times \left(H^{(2)*}_{i\omega}(|k|z_F)\overleftrightarrow{\pp_{z_F}}[e^{ik'x_F+ip^{1}z_F\sinh t_F-iE_{k'p^1}z_F\cosh t_F} a^M_{k'p^1} + \mbox{h.c.}]\right)
 \no
 &=i\int_{-\infty}^{\infty}z_Fdt_F \int \f{dp^1}{\s{2\pi}\s{2E_{kp^1}}} \f{e^{\pi\w/2}}{2\s{2}}  \biggl(H^{(2)*}_{i\omega}(|k|z)\overleftrightarrow{\pp_{z_F}}\bigl[e^{ip^{1}z_F\sinh t_F-iE_{kp^1}z_F\cosh t_F}e^{i\omega t_F} a^M_{kp^1}
\no
&\quad  +e^{-ip^{1}z_F\sinh t_F+iE_{-kp^1}z_F\cosh t_F}e^{i\omega t_F} a^{M\dg}_{-kp^1}\bigr]\biggr)
\label{aab}
\eal
We now use the following integral representations\cite{Olver}
\bal
{\rm (i)} \left(\f{\a+\b}{\a-\b}\right)^{\nu/2}H^{(1)}_\nu(\s{\a^2-\b^2})&=\f{e^{-\nu\pi i/2}}{\pi i}\int_{-\infty}^{\infty}e^{i\a\cosh\tau+i\b \sinh\tau-\nu \tau}d\tau,\q {\rm Im}(\a\pm\b)>0\comma \no
{\rm (ii)} \left(\f{\a+\b}{\a-\b}\right)^{\nu/2}H^{(2)}_\nu(\s{\a^2-\b^2})&=-\f{e^{\nu\pi i/2}}{\pi i}\int_{-\infty}^{\infty}e^{-i\a\cosh\tau-i\b \sinh\tau-\nu \tau}d\tau,\q {\rm Im}(\a\pm\b)<0.
\eal
Note that the formula (i) can be obtained by analytic continuation $\a\r e^{i\pi}\a,\b\r e^{i\pi}\b$ from the formula (ii). 

For the part of (\ref{aab}) containing $a^M_{kp^1}$, namely  
\ba
H^{(2)*}_{i\omega}(|k|z)\overleftrightarrow{\pp_{z}}\int_{-\infty}^{\infty}dt_Fe^{ip^{1}z_F\sinh t_F-iE_{k'p^1}z_F\cosh t_F}e^{i\omega t_F} a^M_{k'p^1} \comma 
\ea
we can use the formula (ii). On the other hand, for the part containing $a^{M\dagger}_{kp^1}$, \ie 
\ba
H^{(2)*}_{i\omega}(|k|z)\overleftrightarrow{\pp_{z}}\int_{-\infty}^{\infty}dt_Fe^{-ip^{1}z\sinh t_F+iE_{k'p^1}z\cosh t_F}e^{i\omega t_F} a^{M\dg}_{k'p^1} \comma 
\ea
 it is convenient to use the formula (i).  In this way,  we can compute (\ref{aab}) as
\bal
 a^{F}_{\w,k}
&=-z_F \int \f{\pi dp^1}{\s{2\pi}\s{2E_{kp^1}}}  \f{e^{\pi\w/2}}{2\s{2}}\n\biggl(H^{(2)*}_{i\omega}(|k|z)\overleftrightarrow{\pp_{z}}\Bigl[H^{(2)}_{i\w}(|k|z) e^{-\pi\w/2}\left(\f{E_{kp^1}-p}{E_{kp^1}+p}\right)^{-\f{i\w}{2}}a^M_{kp^1}\no
&\qquad -H^{(1)}_{i\w}(|k|z)e^{\pi\w/2} \left(\f{E_{-kp^1}-p}{E_{-kp^1}+p}\right)^{\f{i\w}{2}}a^{M\dg}_{-kp^1}\Bigr]\biggr)
\no
&=i\int \f{dp^1}{\s{2\pi E_{-kp^1}}}\left(\f{E_{kp^1}-p^1}{E_{kp^1}+p^1}\right)^{-\f{i\w}{2}}a^M_{kp^1}\period
\eal
In the last step, we used the identity  (\ref{HoneHtwo}). 

Together with the similar result for $a^{F\dagger}_{k\w}$, we can summarize the results as 
\bal
 a^{F}_{\w,k}&=i\int \f{dp^1}{\s{2\pi E_{kp^1}}}\left(\f{E_{kp^1}-p^1}{E_{kp^1}+p^1}\right)^{-\f{i\w}{2}}a^M_{kp^1} \comma 
\no
 a^{F\dg}_{\w,k}&=-i\int \f{dp^1}{\s{2\pi E_{kp^1}}}\left(\f{E_{kp^1}-p^1}{E_{kp^1}+p^1}\right)^{\f{i\w}{2}}a^{M\dg}_{kp^1} \period
\label{FM}
\eal
This is the relation quoted in (\ref{WFM}) in the main text and its hermitian conjugate. As shown in (\ref{FtransFM}), in terms of the rapidity variable $u$ defined in 
(\ref{rapidityu}), these relations can be interpreted as the Fourier transforms 
 and then it is practically trivial to check the desired commutation relations
\bal
[a^{F}_{k\w},a^{F\dg}_{k'\w'}] &=\d(k-k')\d(\w-\w')
\comma \qquad \mbox{rest $=0$} \period
\eal
\subsection{ Sketch of the proof that $\phi^M(t_M, x^1, x)$
depends only on the modes of \WL (\WR ) for $x^1<0\ (x^1>0)$ \label{app:phiMWLWR}}
In this appendix, we give  a sketch of the proof that the scalar field in 
 the Minkowski space $\phi^M(t_M, x^1, x)$, 
 when expressed in terms of the oscillators of the Rindler wedge \WR and those of  \WL, receive  only the contribution of the former(resp. the latter) in the region \WR  (resp. \WL). 

As in (\ref{expphiM}) in the main text, 
 $\phi^M(t_M, x^1, x)$ is expanded in the plane wave basis as 
\begin{align}
\phi^M(t_M, x^1, x) &= \int_{-\infty}^\infty {dp^1 \over \sqrt{2\pi} \sqrt{2E_{kp^1}}}
\int {d^2k \over 2\pi} e^{ikx + ip^1x^1 -iE_{kp^1}t_M} a^M_{kp^1} + {\rm h.c.} 
\label{phiM}
\end{align} 
Now  substitute the expression of $a^M_{kp^1} = a^M_{ku}/\sqrt{E_{kp^1}}$ in terms of $a^F_{k\w}$ given in (\ref{FtransMF}) and further use the expressions of  $a^F_{k\w}$ and $a^F_{k, -\w}$ in terms of $a^R_{k\w}$ and 
 $a^L_{k\w}$ given in (\ref{relpFRL}) and (\ref{relmFRL}).  This gives $\phi^M$  in terms of the modes of \WR and \WL. After a simple rearrangement we obtain 
\begin{align}
\phi^M(t_M, x^1, x) &= \int{d^2k \over 2\pi \s{4\pi}} e^{ikx} \int_0^\infty {d\w \over \s{2\pi} \s{2\sinh \pi \w}}  \nonumber\\
&\cdot \left( -I(\omega) \left[ e^{\pi \w/2} a^R_{k\w} -e^{-\pi\w/2} a^{L\dagger}_{k\w} \right] + I(-\omega) \left[ e^{-\pi \w/2} a^{R\dagger}_{k\w} -e^{\pi\w/2} a^{L}_{k\w} \right]  \right) + {\rm h.c.}  \comma  \label{phiMRL}
\end{align}
where 
\begin{align}
I(\w) &\equiv \int_{-\infty}^\infty du\, e^{i|k|x^1\sinh u -i|k| t_M\cosh u- i\w u} \period
\end{align}
We must study the conditions under which this integral exists.  First, for $u\rightarrow \infty$, the dominant part of the exponent is $i{ |k|\over 2} e^u (x^1-t_M) $. Thus for the integral to converge in this region, we  need the condition ${\rm Im}\, (x^1+t_M) < 0$. On the other hand for $u\rightarrow -\infty$, the dominant part of the exponent is $-i{|k|\over 2} (x^1+t_M) $ and for the convergence 
  we need ${\rm Im}\, (x^1+t_M) < 0$. These two conditions can be met simultaneously if we make the shift 
\begin{align}
t_M \longrightarrow t_M -i\ep \comma \qquad \ep >0 \period
\end{align}
Then, we can make use of the formula 10.9.16 of \cite{Olver} and get
\begin{align}
I(\omega) &= -i \pi  e^{\pi \omega/2} \left({ t_M -x^1 -i\ep \over t_M +x^1 -i\ep} \right)^{i\omega/2} H^{(2)}_{i\omega}( ((t_M-i\ep)^2 -(x^1)^2)^{1/2}) 
\period
\end{align}
To express $\phi^M(t_M, x^1, x)$ in (\ref{phiMRL}) it is clear that in addition to 
 $I(\w)$ we need 
the integrals  $I(-\w)$, $I(\w)^\ast$ and $I(-\w)^\ast$. To obtain them from $I(\w)$, 
 we need to make use of the well-known relations among the Hankel functions 
(see for example  10.46 and 10.11.9 of \cite{Olver})
\begin{align}
H^{(1)}_{-i\omega}(z) &= e^{-\pi \omega} H^{(1)}_{i\omega}(z) \comma \qquad 
H^{(2)}_{-i\omega}(z) = e^{\pi \omega} H^{(2)}_{i\omega}(z)\comma  \\
{H^{(1)}_{i\omega}(z)}^\ast &= H^{(2)}_{-i\omega}(z^\ast) = e^{\pi \omega} H^{(2)}_{i\omega}(z^\ast) \comma \\
{H^{(2)}_{i\omega}(z)}^\ast &= H^{(1)}_{-i\omega}(z^\ast) = e^{-\pi \omega} H^{(1)}_{i\omega}(z^\ast) \period
\end{align}
We then get
\begin{align}
I(\omega) &= -i \pi  e^{\pi \omega/2} \left({ t_M -x^1 -i\ep \over t_M +x^1 -i\ep} \right)^{i\omega/2} H^{(2)}_{i\omega}( ((t_M-i\ep)^2 -(x^1)^2)^{1/2})\comma  \\
I(-\omega) &= -i \pi  e^{\pi \omega/2} \left({ t_M -x^1 -i\ep \over t_M +x^1 -i\ep} \right)^{-i\omega/2} H^{(2)}_{i\omega}( ((t_M-i\ep)^2 -(x^1)^2)^{1/2}) \comma \\
I(\omega)^\ast &= i \pi  e^{-\pi \omega/2} \left({ t_M -x^1 +i\ep \over t_M +x^1 +i\ep} \right)^{-i\omega/2} H^{(1)}_{i\omega}( ((t_M+i\ep)^2 -(x^1)^2)^{1/2}) \comma \\
I(-\omega)^\ast &= i \pi  e^{-\pi \omega/2} \left({ t_M -x^1 +i\ep \over t_M +x^1 +i\ep} \right)^{i\omega/2} H^{(1)}_{i\omega}( ((t_M+i\ep)^2 -(x^1)^2)^{1/2}) \period
\end{align}

Now rather than displaying the complete expression for $\phi^M(t_M, x^1, x)$, 
 it should suffice to demonstrate that the coefficient of $a^R_{k\w}$ vanishes 
 in \WL region, as the rest of the calculations are entirely similar.  

Let us note that $a^R_{k\w}$ appears in two places, namely $a^F_{k\w}$ part
 of $a^M_{kp^1}$ and $a^{F\dagger}_{-\omega k}$ part of $a^{M\dagger}_{kp^1}$. The total contribution for the coefficient of $a^R_{k\w}$ from these sources  is proportional  to $ -I(\omega) e^{\pi \omega/2} +I(-\omega)^\ast e^{\pi\omega/2}$. 

Consider now the region \WL, where $t_M+x^1 <0$, $t_M-x^1>0$ and of course $x_1 <0$. Thus, apart from the $\pm i\ep$, we have $t_M^2 -(x^1)^2= -z_L^2 <0$ and we must choose the square root branch for the quantity $(-z^2)^{1/2}$. ( Since $z_L=z_R$, we  denote it by $z$ for simplicity hereafater. )  As a concrete choice, let us  take 
the branch cut to be along $[-\infty, 0]$ in the $z$ plane. 
This means that $(-z^2 \pm i\delta)^{1/2} = \pm iz$ for small positive $\delta$.  

First consider the region of \WL where $t_M >0$.  Then, we have  $\delta = \ep t_M$ and  in the expressions of $-I(\w)$ and $I(-\w)^\ast $, we have, respectively,  $H^{(2)}_{i \omega}(-i z)$ and $H^{(1)}_{i\omega} (iz)$.  In this case, 
from the formula 10.11.5 of \cite{Olver}, we have 
\begin{align}
H^{(2)}_{i\omega}(e^{-i\pi}iz) &=  -e^{-\pi \omega} H^{(1)}_{i\omega}(iz) \period
\end{align}
Using this relation, it is easy to see that $-I(\omega) + I(-\omega)^\ast =0$ and the coefficient of $a^R_{i\omega}$ vanishes in \WL, as desired. 

Next, consider the region in \WL where $t_M <0$. Then, we have 
 instead $H^{(2)}_{i \omega}(i z)$ for $-I(\w)$  and $H^{(1)}_{i\omega} (-iz)$
 for $I(-\w)^\ast$. Then, again 
from 10.11.5 of \cite{Olver}, we have 
\begin{align}
H^{(1)}_{i\omega}(e^{i\pi} iz) &= -e^{\pi \omega} H_{i\omega}^{(2)}(iz) 
\comma 
\end{align}
and  $-I(\omega)$ and $I(-\omega)^\ast$  cancel with each other in this case 
 as well. 

Combining, we have shown that  $a^R_{\omega k}$ does not contribute in the expansion in the region \WL.  
\section{Poincar\'e algebra for the various  observers \label{app:Poincare}}
\subsection{Proof of the Poincar\'e algebra in \WF frame \label{app:PoincareWF}}
In this appendix, we shall demonstrate that the generators $M_{01}^F, H^F$, 
and  $P_1^F$  constructed in (\ref{MF}), (\ref{HF}) and (\ref{PoneF}) 
 form the Poincar\'e algebra. 

First, consider the commutator $[H^F, M^F_{01}]$. This can be computed as 
\bal
[H^F,M^F_{01}]&=-\int d^{2}k'\ |k'|\int d^{d-2}k\int d\w'd\w [a^{F\dg}_{k\w'}\cos \left(\f{d}{d\w'}\right)a^F_{k\w'},\w  a^{F\dg}_{k\w}a^F_{k\w}]\no&=-\int d^{2}k\ |k|\int d\w \w a^{F\dg}_{k\w}\left(\cos \left(\f{d}{d\w}\right)(\w a^F_{k\w})-\w\cos \left(\f{d}{d\w}\right)a^F_{k\w}\right)\no
&=\int d^{2}k\ |k|\int d\w a^{F\dg}_{k\w}\sin \left(\f{d}{d\w}\right)a^F_{k\w}\no
&=iP_1^F,\label{commma}
\eal
where in the second line we used the simple identity 
\ba
\left(\f{d}{d\w}\right)^n (\w a_{\w})=n\left(\f{d}{d\w}\right)^{n-1} a_{\w}+\w\left(\f{d}{d\w}\right)^n a_{\w}\period 
\ea

In an entirely similar manner, with $\cos$ and $\sin$ interchanged, $[P_1^F, M_{01}^F] =iH^F$ can be shown. 

Finally, the fact that $[H^F, P_1^F]$ vanishes can be checked as 
\bal
[H^F,P^F_{1}]&=-i\int d^{d-2}k'\ |k'|\int d^{d-2}k|k|\int d\w'd\w  [a^{F\dg}_{k\w'}\cos \left(\f{d}{d\w'}\right)a^F_{k\w'}, a^{F\dg}_{k\w}\sin \left(\f{d}{d\w}\right)a^F_{k\w}]\no&=-i\int d^{d-2}k\ |k|^2\int d\w \w a^{F\dg}_{k\w}\left(\sin \left(\f{d}{d\w}\right)\cos \left(\f{d}{d\w}\right)a^F_{k\w}-\cos \left(\f{d}{d\w}\right)\sin \left(\f{d}{d\w}\right)a^F_{k\w}\right)\no
&=0 \period \label{commmc}
\eal
\subsection{Poincar\'e generators for \WF by the unitary transformation
\label{app:Unitary}} 
Recall that the unitary transformation $U_F$ defined by 
 \ba
U_F=e^{-\f{i\pi}{2}A}\comma \qquad  A=\f{1}{2}\int^{\infty}_{-\infty}d\w a^{F\dg}_{k\w}(-\f{d^2}{d\w^2}+\w^2-1)a^F_{k\w} \comma 
 \ea
converts the mode operator  $a^F_{k\omega}$ into  $a^M_{kp^1}$ in the manner
\ba
 U_Fa^F_{k\w}U_F^{\dg}=i \s{E_{kp^1}}a^M_{kp^1}  \period
 \ea
As an application of this operation, let us show that it transforms $M_{01}^F$ 
 into $M_{01}$,  namely 
 \bal
U_F \left( \int d^2 k \int_{-\infty}^\infty  d\omega\,  \omega a^{F\dagger}_{k\omega} a^F_{k\omega} \right)  U_F^\dg&=i \int d^{2}k \int_{-\infty}^\infty  dp^1\,  E_{kp^1}\ a^{M\dg}_{kp^1}\f{\pp}{\pp p_{1}}a^M_{kp^1} \period 
\eal
First, expand the unitary transformation as the sum of multiple commutators in 
 the usual way:
\bal
U_F M_{01}^FU_F^\dg&=M^F_{01}-\f{i\pi}{2}[A,M^F_{01}]+\f{1}{2!}\left(-\f{i\pi}{2}\right)^2[A,[A,M^F_{01}]]+\ddd\period 
\eal
The single  commutator can be computed as 
\bal
[A,M^F_{01}]&=
[A,\int d^{2}k\int^{\infty}_{-\infty}  d\w  \w a^{F\dg}_{k\w}a^F_{k\w}]\no &=\f{1}{2}\int d^{2}k\int^{\infty}_{-\infty}d\w \int^{\infty}_{-\infty}d\w' [a^{F\dg}_{k\w'}\left(-\f{d^2}{d\w'^2}\right)a^F_{k\w'},\w a^{F\dg}_{k\w}a^F_{k\w}]\no&=
-\f{1}{2}\int d^{2}k\int^{\infty}_{-\infty}d\w\left[a^{F\dg}_{k\w}\left(\f{d^2}{d\w^2}\right)\w a^F_{k\w}-\w a^{F\dg}_{k\w}\f{d^2}{d\w^2}a^F_{k\w}\right]\no
&=-\int d^{2}k\int^{\infty}_{-\infty}d\w a^{F\dg}_{k\w}\f{d}{d\w}a^F_{k\w}\period
\eal
Based on this result, the double commutator is calculated as 
\bal
[A,[A,M^F_{01}]]&=[A,-\int d^{2}k\int^{\infty}_{-\infty} d\w  a^{F\dg}_{k\w}\f{d}{d\w}a^F_{k\w}]\no &=-\f{1}{2}\int d^{2}k\int^{\infty}_{-\infty}d\w\int^{\infty}_{-\infty}d\w' (\w'^2-1)[a^{F\dg}_{k\w'}a^F_{k\w'},a^{F\dg}_{k\w}\f{d}{d\w}a^F_{k\w}]\no&=-\f{1}{2}\int d^{2}k\int^{\infty}_{-\infty}d\w [(\w^2-1)a^{F\dg}_{k\w}\f{d}{d\w}a^F_{k\w}-a^{F\dg}_{k\w}\f{d}{d\w}(\w^2-1)a^F_{k\w}]\no&=\int d^{2}k\int^{\infty}_{-\infty} d\w\,   \w a^{F\dg}_{k\w}a^F_{k\w} \period
\eal
Since this is of the original form of $M_{01}^F$, we see that the rest of the multiple commutators produce  $\int d^{2}k\int^{\infty}_{-\infty}d\w a^{F\dg}_{k\w}\f{d}{d\w}a^F_{k\w}$ and $\int d^{2}k\int^{\infty}_{-\infty} d\w\,   \w a^{F\dg}_{k\w}a^F_{k\w}$ alternately. The coefficients can be easily found in such a way 
 that the series sum up to 
\bal
U_F M_{01}^F U_F^\dg 
&=\cos\f{\pi}{2}\int d^{2}k\int^{\infty}_{-\infty} d\w \,  \w a^{F\dg}_{k\w}a^F_{k\w}+i\sin\f{\pi}{2}\int d^{2}k\int^{\infty}_{-\infty}d\w \, a^{F\dg}_{k\w}\f{d}{d\w}a^F_{k\w}\no
&=i\int d^{2}k\int^{\infty}_{-\infty}d\w \,  a^{F\dg}_{k\w}\f{d}{d\w}a^F_{k\w}
\period
\eal
Making the replacements   $\w\r p^1$ and $a^F_{k\w}\r \s{E_{kp^1}}a^M_{kp^1}$, we obtain the desired result
\bal
U_FM_{01}^FU_F^{\dg}=i \int d^{2}k\int dp^{1}\ E_{kp^1}\ a^{M\dg}_{kp^1}\f{\pp}{\pp p_{1}}a^M_{kp_1}  = M_{01} \period
\eal
\section{Quantization in different Lorentz frames with an almost light-like 
boundary condition \label{app:generalFFO}}
Here we supply some details of the quantization in different Lorentz frames with 
a slightly timelike boundary condition discussed in Sec.~\ref{sec:onesidedFFO}. 

What we shall describe is the computations of the two terms (\ref{Cone}) and 
 (\ref{Ctwo}) which constitute the commutator $\com{\tilde{\pi}(\ttil, \xonetil, \Omega)}{\phi(\ttil, \yonetil, \Omega')}$ given in (\ref{compitilphitil}). 
For the convenience of the reader let us display them again:
\begin{align}
C_1 &= -\gamma\calNtil^2 \sum_{l=0}^\infty \sum_{m=-l}^l \int_0^\infty {d\ponehat \over 4\pi }  Y_{lm}(\Omega)Y^\ast_{lm}(\Omega') 
\left(e^{i\beta\Ehat_{k_l \ponehat} (\xonehat-\yonehat)} +{\rm h.c.}\right)\sin\ponehat\xonehat\, \sin\ponehat \yonehat \comma \label{Coneap} \\
C_2 &= -i\gamma\beta \calNtil^2\sum_{l=0}^\infty \sum_{m=-l}^l \int_0^\infty {d\ponehat \over 4\pi \Ehat_{k_l\ponehat} }  Y_{lm}(\Omega)Y^\ast_{lm}(\Omega') 
 \left(e^{i\beta\Ehat_{k_l \ponehat} (\xonehat-\yonehat)} -{\rm h.c.}\right)\cos\ponehat\xonehat\, \sin\ponehat \yonehat
 \period \label{Ctwoap}
\end{align}
First the sum over $m$ can be performed by the addition theorem for $Y_{lm}$ 
 as already described in (\ref{addthm}). Next, we perform the integral over $\ponehat$. Although the energy dependence in the exponent does not disappear at equal $\ttil$, in contrast to the case for the frame $(\that, \xonehat)$, such an integral 
 can be performed, after expressing the product of trigonometric functions 
 into a sum like $\sin\ponehat \xonehat \sin \ponehat \yonehat 
=\half \left( \cos \ponehat (\xonehat -\yonehat) -\cos \ponehat (\xonehat + \yonehat) \right)$. The relevant formulas were given in (\ref{GRintone}) and (\ref{GRinttwo}),  with appropriate regularizations (\ref{bm}) and (\ref{bp}) for convergence. 

Then, the result for $C_1+C_2$ takes the form 
\begin{align}
C_1+C_2 &=-i{\beta \ga \calNtil^2 \over 8\pi}  \sum_{l=0}^\infty {2l+1 \over 4\pi} P_l(\hat{n}\cdot \hat{n'}) \nonumber\\
 \times \biggl\{&  -{\am+i\eta  \over \sqrt{\am^2+\bem^2 }}K_1(k_l \sqrt{ \am^2+\bem^2 }) 
+{\am-i\eta \over \sqrt{ \am^2+\bep^2 }}K_1(k_l  \sqrt{\am^2+\bep^2  }) \nonumber\\
&+{\am+i\eta \over \sqrt{\ap^2+\bem^2 }}K_1(k_l  \sqrt{\ap^2+\bem^2 }) 
-{\am-i\eta \over \sqrt{\ap^2+\bep^2 }}K_1(k_l  \sqrt{\ap^2+\bep^2 }) 
 \nonumber\\
& + {\ap \over \sqrt{\ap^2+\bem^2 }}K_1(k_l  \sqrt{\ap^2+\bem^2 }) 
-{\ap \over \sqrt{\ap^2+\bep^2 }}K_1(k_l  \sqrt{\ap^2+\bep^2}) \nonumber\\
&-{\am \over \sqrt{\am^2+\bem^2}}K_1(k_l  \sqrt{\am^2+\bem^2 }) 
+{\am \over \sqrt{\am^2+\bep^2}}K_1(k_l  \sqrt{\am^2+\bep^2 })
 \biggr\} \comma 
\end{align}
where 
\begin{align}
a_\pm &\equiv \xonehat \pm \yonehat \comma \qquad 
 b_\pm \equiv \pm i\beta (\xonehat -\yonehat \mp i\eta) \period 
\end{align}

Consider first the four terms in the third and the fourth lines, which contain 
 $\ap^2$ in the square roots of the denominator and in the argument of $K_1$ functions.  Since $\xonehat$ and $\yonehat$ are positive,  $\ap$ is positive and generically finite. Therefore, we can ignore  $\eta$ for these terms. Then, 
$\bep^2=\bem^2$ and hence the two terms in the third line cancel and 
similarly the the two terms in the fourth line cancel.  Therefore these four terms actually do not contribute and we can simplify $C_1+C_2$ to 
\begin{align}
C_1+C_2 &= -i{\beta \ga \calNtil^2 \over 8\pi}  \sum_{l=0}^\infty {2l+1 \over 4\pi} P_l(\hat{n}\cdot \hat{n'}) \nonumber\\
 \times \biggl\{&  -{\am+i\eta  \over \sqrt{\am^2+\bem^2 }}K_1(k_l \sqrt{ \am^2+\bem^2 }) 
+{\am-i\eta \over \sqrt{ \am^2+\bep^2 }}K_1(k_l  \sqrt{\am^2+\bep^2  }) \nonumber\\
&-{\am \over \sqrt{\am^2+\bem^2}}K_1(k_l  \sqrt{\am^2+\bem^2 }) 
+{\am \over \sqrt{\am^2+\bep^2}}K_1(k_l  \sqrt{\am^2+\bep^2 })
 \biggr\} \comma 
\end{align}

To analyze this expression we must distinguish two regions. \nxt
(i)\ If $\am^2$ is finite, then we can again ignore $\eta$ and these four terms 
 cancel in exactly the same fashion. \\
(ii)\ Thus non-vanishing result can possibly be obtained if and only if $|\am| \lsim \eta$.  In such a case, since $\am$ and $b_\pm$ are of the order $\eta$, 
 as long as $k_l$ is not infinite, we can use the approximation $K_1(z) \simeq 
1/z$ and hence each term diverges like  $1/\eta$. 

Combining, this shows that the sum of terms containing $K_1$ function 
behaves precisely like   $\sim  \delta(\xonehat -\yonehat)$. 
The rest of the argument is already given in the main text and 
 the commutator  $\com{\tilde{\pi}(\ttil, \xonetil, \Omega)}{\phi(\ttil, \yonetil, \Omega')}$ in the frame of an arbitrary FFO correctly behaves like the product of appropriate  $\delta$-functions. 

Thus, we can write 
\begin{align}
\com{\tilde{\pi}(\ttil, \xonetil, \Omega)}{\phi(\ttil, \yonetil, \Omega')} &=C_1+C_2 = -i F \gamma \delta(\xonehat -\yonehat) \delta(\cos\theta-\cos\theta')\delta(\varphi -\varphi') \comma 
\end{align}
where we used the relation $\delta(\xonetil-\yonetil) = \ga\delta(\xonehat -\yonehat)$ valid at equal $\ttil$.   $F$ is a constant, which we want to set to unity by adjusting the 
normalization constant $\calNtil$.  To find such $\calNtil$, we need to carry out 
 the integral $i\int d\xonehat  d\cos\theta d \varphi (C_1+C_2)$,  perform the sum over $l$ and set the result to $1$. This unfortunately is quite difficult and we have not been able to find the form of $\calNtil$ explicitly. 
\end{document}